\documentclass[10pt,a4paper,twocolumn,floatfix,groupedaddress,superscriptaddress,prb,aps,showpacs]{revtex4-1}

\usepackage{graphicx}
\usepackage{amsmath}
\usepackage{amssymb}
\usepackage{amsfonts}
\usepackage{dcolumn}
\usepackage{bbm} 

\usepackage{color}
\usepackage{bbding}

\def\XSolidBrush{no}
\def\Checkmark{yes}

\begin{document}
\title{Effective Models for the  Anderson Impurity
and the Kondo Model from Continuous Unitary Transformations}

\author{J\"orn Krones}
\email{joern.krones@tu-dortmund.de}
\affiliation{Lehrstuhl f\"{u}r Theoretische Physik I, 
Technische Universit\"{a}t Dortmund,
 Otto-Hahn Stra\ss{}e 4, 44221 Dortmund, Germany}

\author{G\"otz S. Uhrig}
\email{goetz.uhrig@tu-dortmund.de}
\affiliation{Lehrstuhl f\"{u}r Theoretische Physik I, 
Technische Universit\"{a}t Dortmund,
 Otto-Hahn Stra\ss{}e 4, 44221 Dortmund, Germany}

\date{\textrm{\today}}

\begin{abstract} 
The method of continuous unitary transformations (CUTs) is applied to the Anderson impurity 
and the Kondo model aiming at the systematic derivation of convergent effective models.
If CUTs are applied in a conventional way, diverging differential equations occur.
Similar to poor man's scaling the energy scale, below which the couplings
diverge, corresponds to  the Kondo temperature $T_K$.
We present a way to apply CUTs to the Kondo and to the Anderson impurity model
so that no divergences occur but a converged effective low-energy model
 is derived with small finite parameters at arbitrarily small energies.
The ground state corresponds to a bound singlet
with a binding energy given by the Kondo temperature $T_K$.  
\end{abstract}



\maketitle


\section{Introduction}

We want to apply the approach of continuous unitary transformations (CUT) 
 to two archetypical impurity models which 
exhibit the \emph{Kondo effect}, the Kondo model and the Anderson impurity model 
\cite{kondo64,ander61,hewso93}.
The Kondo effect is one of the fundamental problems of many-body theory
as it appears in a wide range of correlated electron systems, for instance in 
heavy-fermion systems \cite{hewso93,stewa84}, 
in Mott-Hubbard metal-insulators \cite{rozen92,georg96} 
and in nanoscale quantum dots \cite{grobi07}.
In such models, a wide range of energy scales is important \cite{hewso93}: from the
bath electrons' bandwidth $D$, which can be of the order of several eV, 
down to the exponentially small Kondo temperature $T_K$. 

The challenge to treat the exponentially small Kondo energy scale reliably 
has first been solved by the numerical
renormalization group (RG) \cite{wilso75,krish80a,bulla08}. A Bethe ansatz 
solution put the results on a rigorous foundation \cite{tsvel83}.

In recent years, the issue has attracted much attention in the field of 
renormalization approaches. The functional RG approach was applied to the 
Anderson impurity model \cite{hedde04} yielding good results for small and intermediate
interactions, but failing to reproduce the exponentially small Kondo energy
scale in the strong coupling regime. Subsequently, a series of papers 
\cite{karra08,barto09,isido10,freir12,kinza13}
tried different variants of the functional RG approach to reproduce this small 
energy scale. While the successes in the regime of small to intermediate couplings
were very interesting, the strong coupling regime eluded a description by 
functional RG. Only recently, Streib and co-workers provided a functional RG approach
with the correct strong-coupling approach \cite{strei13}. The additional key element in their
study is to use  a large magnetic field as flow parameter which is gradually lowered to zero.
Furthermore, Ward identities and partial bosonization of the spin degrees of freedom
are exploited.

The idea to treat the problem first at large magnetic fields where perturbation
theory is perfectly well-controlled and then reducing the field 
gradually has been put forward by Hewson and collaborators in a series of 
papers \cite{hewso01,hewso06,edwar11}.  Besides this key idea they
rely on perturbation theory in the renormalized effective interaction calling
their approach renormalized perturbation theory (RPT). 
All these intensive studies performed in the last decade illustrate
that the Anderon impurity model in the strong coupling regime represents
a formidable methodological challenge.

In the present article, we want to show that continuous unitary transformations
are also able to treat the strong coupling limit of the Anderson impurity model, i.e.,
the Kondo model with its exponentially small energy scale.
Continuous unitary transformations exhibit an intrinsic energy separation
because processes at higher energies are transformed faster than processes
at lower energies. This feature is similar to standard RG approaches.
The CUT approach can be set up non-perturbatively so that 
it is able to derive effective models even at exponentially low energies.

There are a number of applications of CUTs to the Kondo problem. 
Results of conventional ``poor man's scaling''  \cite{ander70a} could be reproduced by a CUT 
in which  diverging differential equations  occur \cite{kehre06}. This divergence indicates
the Kondo energy scale.  Another application of the CUT
to the Kondo model \cite{vogel05} results in an effective model where the matrix elements of 
the effective interaction still exhibit logarithmic infrared divergences very similar to those found in  a standard perturbative treatment \cite{kondo64}.
Furthermore, there are CUT approaches to the Kondo model which 
succeed in the derivation of a finite, convergent effective models.
But these approaches profit from a detour via the bosonized form of the Kondo model
 \cite{hofst01a,lobas05,sleza03}.

CUTs were also applied to the Anderson impurity model 
\cite{kehre94,kehre94e,kehre96b,staub04,zapal15}. But none of them revealed
the exponential character of the Kondo temperature $T_{\text{K}}$. Nevertheless, an important previous work has been able to retrieve and to improve the Schrieffer-Wolff transformation
\cite{kehre96b}.This approach has been extended recently to impurities which hybridize
with a superconducting environment \cite{zapal15}.

In the present work, we show that CUTs yield the correct low-energy physics of 
the Kondo model and of the Anderson impurity model.
Our approach does not rely on a bosonized reformulation. 
Staying in a purely fermionic description
 leads to convergence problems as one encounters diverging couplings \cite{kehre06}.
But we will show that a change of the reference state during the flow
solves this problem avoiding the diverging couplings and yielding a
finite, convergent effective low-energy model characterized by the exponentially small
energy scale of the Kondo temperature.

This article is set up as follows. In the remainder of this introduction the 
basics of the CUTs are presented. In the next section, the Kondo model is briefly introduced
and its standard treatment by means of CUTs is shown. The resulting flow equations
diverge. Thus, in Sect.\ III a modified approach with a change
of the reference state is introduced which allows us to derive a finite, well-defined
effective model. This model is indeed characterized by the correct exponentially small
energy scale. In Sect.\ IV, the Anderson impurity model is tackled by the standard
CUT which again implies a diverging flow. The corresponding modified flow implying 
convergence is analyzed in Sect.\ V. Finally, the results are summarized in 
Sect.\ VI which also includes an outlook on promising future work.


\subsection{Continuous unitary transformations}

The continuous unitary transformation (CUT), also called flow equation approach, 
is a powerful method of theoretical quantum mechanics aiming at the systematic
derivation of effective low-energy models. 
A CUT transforms a Hamiltonian continuously closer (or even completely) to diagonal form while connecting the transformed to the initial Hamiltonian by a unitary transformation. 
The method was suggested in the mid 90's
\cite{wegne94,glaze93,glaze94} and has been successfully applied to a wide range of
problems in condensed matter physics, for a review see Ref.\ \onlinecite{wegne06}. 
Non-perturbative \cite{wegne94,duffe11,fisch10a,fisch11a,hamer10} as well as perturbative 
\cite{uhrig98c,knett00a} versions have been developed in the course of the last two decades.
CUTs continue to be objects of current research. Only recently, improved versions of the CUT approach have been developed: the enhanced perturbative 
(epCUT), the directly evaluated enhanced perturbative CUT (deepCUT) \cite{krull12}
as well as a graph-theory based version called gCUT \cite{yang11a}.

For a CUT  a continuous parameter $l$ is introduced parametrizing the Hamiltonian
\begin{eqnarray}
\label{flow Hamiltonian}
 H\left(l\right) = U^{\dagger}\left(l\right)H\left(0\right)U\left(l\right).
\end{eqnarray}
Differentiating (\ref{flow Hamiltonian}) with respect to $l$ yields the flow equation
\begin{eqnarray}
\label{flow equation}
 \frac{\partial H\left(l\right)}{\partial l} = \left[\eta\left(l\right),H\left(l\right)\right]
\end{eqnarray}
where the anti-hermitian generator
\begin{equation}
\label{definition generator U}
 \eta\left(l\right) = \frac{\partial U\left(l\right)}{\partial l} U^{\dagger}\left(l\right)
\end{equation}
is introduced. The flow equation (\ref{flow equation}) is the heart of the CUT approach. 
There are numerous ways to choose the generator \cite{wegne94,mielk98,knett00a,knett03a,fisch10a,hamer10}. We will discuss later which kind of generator is employed 
in the present work. 

Upon calculating the commutator between $\eta$ and $H$ generically operator terms will emerge 
which are not present in the original Hamiltonian. Their commutators have to to be computed as well leading to even more terms and so on. This procedcure leads to a proliferating number of emerging terms. Thus, we need to approximate at some point by truncating terms.
Specifically, we employ deepCUT ideas \cite{krull12}, i.e., we will target a specific part of $H$ and determine a system of differential equations which allows us to determine the targeted
quantity correctly  up to a certain order in a small expansion parameter, for instance  the spin-spin interaction $J$ or the hybridization $V$. We  will describe the explicit procedure in the
derivation of the flow equations below.


\section{Kondo model}

First, we present our approach for the Kondo model 
before we will apply it to the Anderson impurity model as well. For this reason,
we briefly review it here.

The Kondo model was introduced by Kondo \cite{kondo64} in order to explain the resistivity minimum
upon lowering the temperature found in metals hosting magnetic impurities. The model describes the interaction of the conduction or bath electrons of the non-magnetic host metal with a localized spin  $\vec{S}_{I}$  of the impurity. The conduction electrons follow the  dispersion $\epsilon^{\phantom{\dagger}}_{\mathbf{k}}$. The interaction is an exchange interaction implying
a spin-spin coupling $J$ between the localized impurity spin and the spins of the 
bath electrons $\vec{s}_b$
\begin{equation}
 \label{Kondohamiltonian}
 H_{\text{K}} = \sum_{\mathbf{k},\sigma} \epsilon^{\phantom{\dagger}}_{\mathbf{k}} 
c^{\dagger}_{\mathbf{k} \sigma} c^{\phantom{\dagger}}_{\mathbf{k} \sigma} 
                  + J\vec{S}_{I} \cdot\vec{s}_b.
\end{equation}
The Hamiltonian is given in second quantization, i.e., 
$c^{\dagger}_{\mathbf{k} \sigma}$ ($c^{\phantom{\dagger}}_{\mathbf{k} \sigma}$) 
creates (annihilates) a bath electron with 
momentum $\mathbf{k}$ and spin $\sigma$ while $\vec{s}_b$ is the bath electrons' spin
\begin{equation}
\vec{s}_b = \frac{1}{N}\sum_{\mathbf{k},\mathbf{k}'}\sum_{\alpha,\beta} c^{\dagger}_{\mathbf{k} \alpha}\vec{\sigma}^{\phantom{\dagger}}_{\alpha\beta}c^{\phantom{\dagger}}_{\mathbf{k}' \beta}
\end{equation}
interacting with the local impurity spin. 
The components of the vector $\vec{\sigma}$ are the Pauli matrices 
$\vec{\sigma}=\sum_{\mu\in x,y,z}\sigma^{\mu}\vec{e}_{\mu}$ as usual.


\subsection{Logarithmic discretization}

\begin{figure}
\includegraphics[width=\columnwidth]{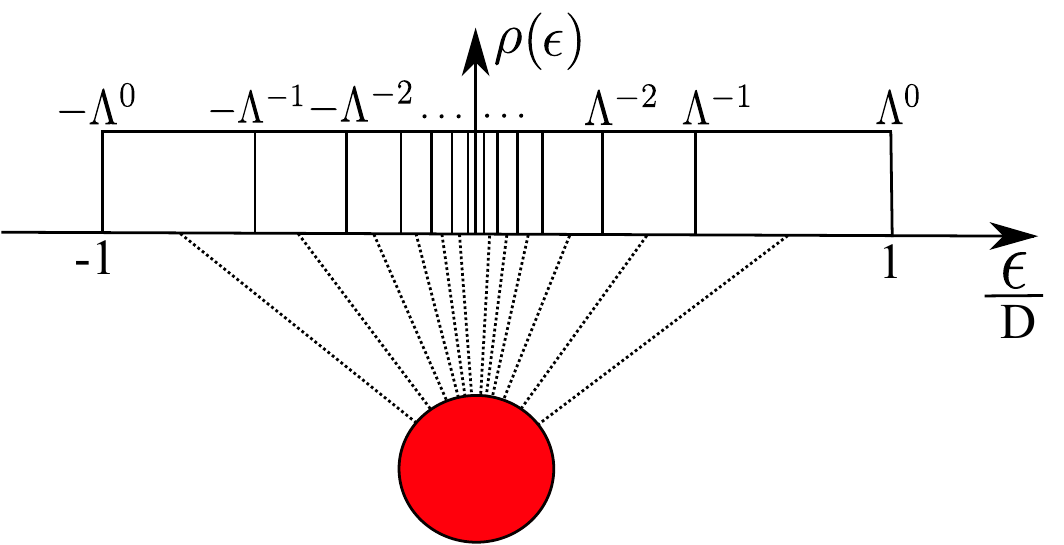}
\caption{(Color online) Logarithmic discretization of the continuum of the bath electrons
 sketched for a flat density of states (DOS)
\mbox{$\rho\left(\epsilon\right)=\rho_0\,\Theta\left(D-\left|\omega\right|\right)$} with \mbox{$\rho_0={1}/(2D)$} and a band width $2D$ coupled to an impurity.}
\label{fig:logarithmic_discretization}
\end{figure}

The Hamiltonian in energy representation is simpler than the initial Hamiltonian. Let us
assume that all parameters are isotropic and thus only depend on the absolute value $\left|\mathbf{k}\right|$ of the momenta. 
Due to this isotropy, it is convenient to introduce spherical coordinates. 
Finally, a substitution $k \rightarrow \epsilon\left(k\right)$
is used. This leads to the continuum energy representation. For
details of the required steps, the reader is referred to Ref.\ \onlinecite{krish80a}.
Numerically, a continuum of operators or states can hardly be 
handled. Thus we use a logarithmic discretization of the energy representation, see,
e.g., Refs.\ \onlinecite{bulla08,krish80a}.

The important energies of the Kondo problem stretch from the bandwidth $D$ 
down to exponentially small energies below the Kondo temperature $T_K$. A linear 
discretization is not suitable in such a problem. Thus one resorts to the logarithmic discretization sketched in Fig.\ \ref{fig:logarithmic_discretization}, for details see
Ref.\  \onlinecite{bulla08}. The continuum of the bath electrons is discretized in exponentially decreasing intervals 
\begin{subequations}
\begin{eqnarray}
\label{Chapter 2 discretized intervals}
 \frac{I_n^{+}}{D}&=&\left[\Lambda^{-n-1},\Lambda^{-n}\right]
\\ 
 \frac{I_n^{-}}{D}&=&\left[-\Lambda^{-n},-\Lambda^{-n-1}\right]
\end{eqnarray}
\end{subequations}
where $\Lambda>1$ determines the discretization  and $n\in \mathbbm{N}$.
The length of the $n$th interval is given by
\begin{equation}
 \frac{d_n}{D}=\left(1-\Lambda^{-1}\right)\Lambda^{-n}.
\end{equation}
In this logarithmic discretization, the higher energies are only covered with low precision while small energy scales are represented with an increasingly higher resolution. More precisely,
the relative precision is the same at all energies, high or low.
In this way, the discretization scheme easily reaches down to exponentially small 
energy scales below the Kondo temperature $T_{K}$. 


\subsection{Discretization of a flat density of states}

In the model (\ref{Kondohamiltonian}) one discretizes the bath electrons' density of states (DOS).
The formal representation is given by
\begin{subequations}
\begin{eqnarray}
\label{discretization parameters flat DOS}
 \epsilon_n^{\pm}&=&\frac{1}{\left|\gamma_n^{\pm}\right|^2} \int_{n,\pm} \epsilon\,\, \rho\left(\epsilon\right)\,\,\text{d}\epsilon
\\ 
\left|\gamma_n^{\pm}\right|^2 &=&  \int_{n,\pm} \rho\left(\epsilon\right)\text{d}\epsilon
\end{eqnarray}
\end{subequations}
where we integrate over the intervals $I^{\pm}_n$ 
according to
\begin{eqnarray}
 \int_{n,+} = \int_{D\Lambda^{-n-1}}^{D\Lambda^{-n}}
\quad , \quad
 \int_{n,-} = \int_{-D\Lambda^{-n}}^{-D\Lambda^{-n-1}}.
\end{eqnarray}
Then, the discretized Hamiltonian reads
\begin{eqnarray}
\nonumber
  H_{\text{K}} &=& \sum_{n,\sigma} \epsilon_{n} :c^{\dagger}_{n \sigma} c^{\phantom{\dagger}}_{n \sigma}:
\\
\label{Kondo Hamiltonian energy representation}
&+& \sum_{\mu}\sum_{\alpha,\beta,n,m} J_{nm}\sigma^{\mu}_{\alpha\beta}
S_I^{\mu}:c^{\dagger}_{n \alpha}c^{\phantom{\dagger}}_{m \beta}:
\end{eqnarray}
where 
\begin{equation}
 \label{starting values Kondo model}
J_{nm}=J\gamma_n\gamma_m. 
\end{equation}
The colons denote that the operators are normal-ordered with respect to the Fermi sea
of the bath electrons
\begin{equation}
 :c^{\dagger}_{n \alpha}c^{\phantom{\dagger}}_{m \beta}
:= c^{\dagger}_{n \alpha}c^{\phantom{\dagger}}_{m \beta} - 
\langle \text{FS}|c^{\dagger}_{n \alpha}c^{\phantom{\dagger}}_{m \beta}|\text{FS}\rangle.
\end{equation}
For a flat density of states
\begin{equation}
 \rho\left(\omega\right) = \rho_0\,\Theta\left(D - \left|\omega\right|\right)
\quad,  \quad \rho_0= 1/(2D)
\end{equation}
we can easily calculate the parameters (\ref{discretization parameters flat DOS})
\begin{subequations}
\begin{eqnarray}
\label{discretized parameters flat DOS}
\frac{\epsilon_n^{\pm}}{D} &=& \pm\frac{1}{2}\left(1 + \Lambda^{-1}\right)\Lambda^{-n}
\\ 
\left|\gamma_n^{\pm}\right|^2 &=& \frac{1}{2}\left(1 - \Lambda^{-1}\right)\Lambda^{-n}.
\end{eqnarray}
\end{subequations}


\subsection{Diagonalization of the spin-spin interaction}

In order to diagonalize the spin-spin interaction we introduce the generator
\begin{equation}
 \label{Kondo Hamiltonian generator}
 \eta = \sum_{\mu}\sum_{\alpha,\beta,n,m} \eta_{nm}\sigma^{\mu}_{\alpha\beta}S_I^{\mu}:c^{\dagger}_{n \alpha}c^{\phantom{\dagger}}_{m \beta}:
\end{equation}
which has the same structure as the corresponding term in the Hamiltonian \eqref{Kondohamiltonian}.
Specifically, we use the sign generator
\begin{equation}
 \label{eq:sign_generator}
 \eta_{nm} = \text{sgn}\left(\epsilon_{n} - \epsilon_{m}\right)J_{nm}.
\end{equation}
Without further approximations this choice of the generator would lead to an effective model in which the spin-spin interaction is diagonalized within degenerate subspaces, for
a proof of this statement see Ref.\ \onlinecite{fisch10a}. 
As soon as approximations are used, this
statement does not  hold true necessarily and the resulting
flow equations might even diverge. Nevertheless, despite the approximations,  the CUT approach commonly yields sensible results if the approximations are physically justified.

Calculating the commutator between $\eta$ from (\ref{Kondo Hamiltonian generator}) and $H_{K}$ from (\ref{Kondo Hamiltonian energy representation}), terms emerge which so far did
not appear in (\ref{Kondo Hamiltonian energy representation}). For instance, 
quartic operators in the fermionic bath operators occur. We discard them \emph{after} normal-ordering with respect to the reference state which is the Fermi sea of the fermionic bath 
so far. Due to the normal-ordering  feedback to the spin-spin interaction
in order $J^2$ is properly captured. 
Furthermore, bilinear hopping terms occur which we discard as well
because they only weakly renormalize the single-particle energies $\epsilon_n$
in order $J^2$. The remaining terms of the commutator are compared to the derivative of
$H_K$ leading to  the flow equation (\ref{flow equation})
\begin{eqnarray}
\label{DEQ poor mans scaling with CUT Kondo}
  && \partial_{l} J_{nm} = -\left|\epsilon_{n} - \epsilon_{m}\right|J_{nm}
\\ \nonumber
 && \quad - \sum_{x}\left(\text{sgn}\left(\epsilon_{n}-\epsilon_{x}\right) - 
\text{sgn}\left(\epsilon_{x} - \epsilon_{m}\right)\right)\left(1-2\theta_x\right)J_{nx}J_{xm}   
\end{eqnarray}
where
\begin{eqnarray}
 \theta_x = \langle \text{FS}|c^{\dagger}_{x \sigma}c^{\phantom{\dagger}}_{x \sigma}|\text{FS} \rangle
\end{eqnarray}
stems from the normal-ordering with respect to the non-interacting Fermi sea.
The linear term in (\ref{DEQ poor mans scaling with CUT Kondo}) is a 
generic term for a generator of the form (\ref{eq:sign_generator}).
Usually, it implies exponential convergence at large $l$. 
Solving (\ref{DEQ poor mans scaling with CUT Kondo}) numerically, however, reveals that the flow (\ref{DEQ poor mans scaling with CUT Kondo}) diverges 
at some flow parameter $l_{0}$ which is related to the energy scale $T_{K}=l_{0}^{-1}$, see 
Figs.\ \ref{fig:ROD_Kondo_no_change_of_ref_state} and \ref{fig:Tk_Kondo_N80_L2}.


\subsection{Residual off-diagonality (ROD)}

\begin{figure}
\includegraphics[width=\columnwidth]{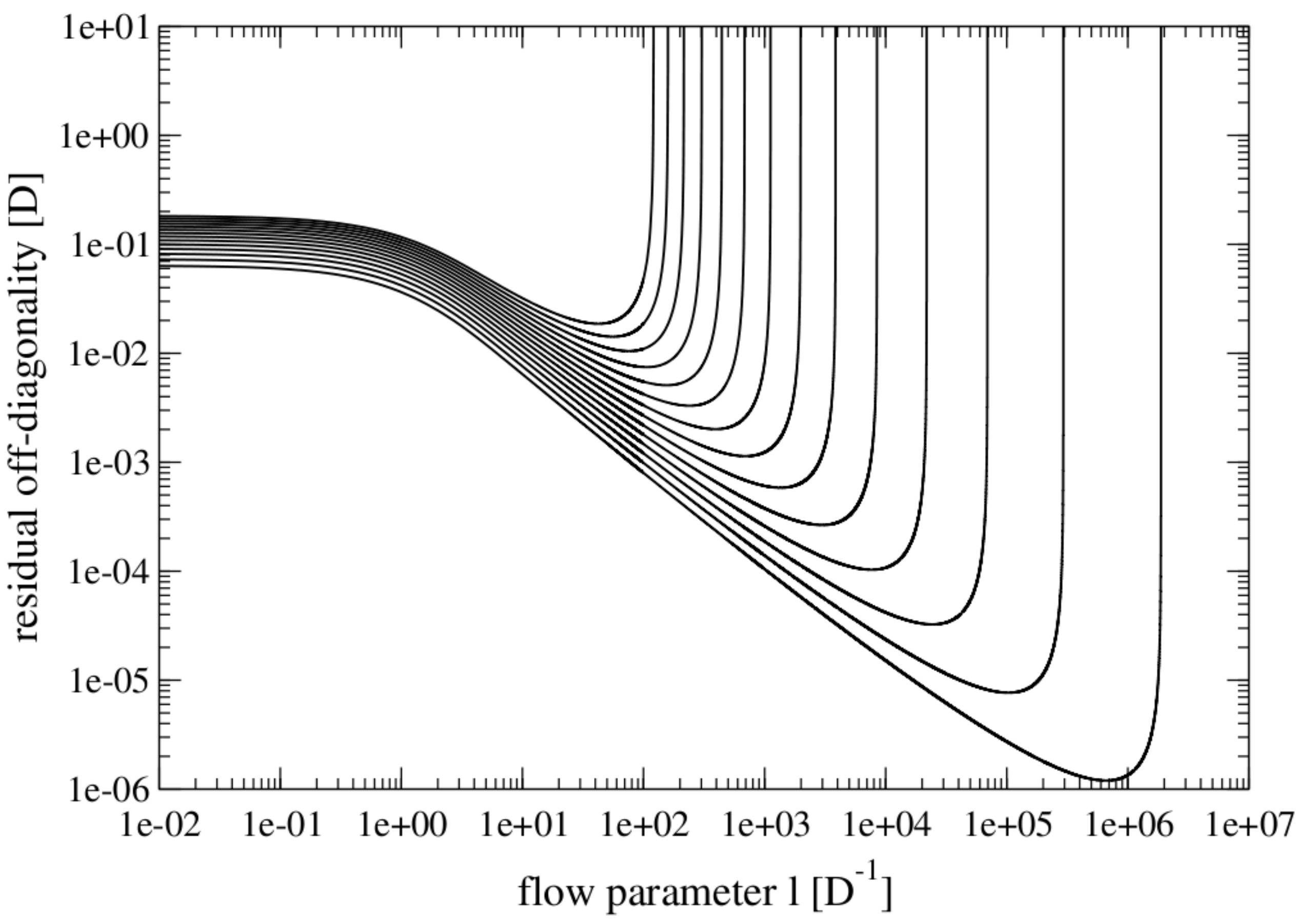}
\caption{(Color online) Residual off-diagonality (ROD) (\ref{ROD for DEQ without change of ref state Kondo model}) of the flow equation (\ref{DEQ poor mans scaling with CUT Kondo}) 
for the Kondo model with \mbox{$N=80$}, \mbox{$\Lambda=2$} and from left to right: 
\mbox{$2\rho_0J=0.2$}, $0.19$, $0.18$, ..., $0.09$, $0.08$, $0.07$.}
\label{fig:ROD_Kondo_no_change_of_ref_state}
\end{figure}

\begin{figure}
\includegraphics[width=\columnwidth]{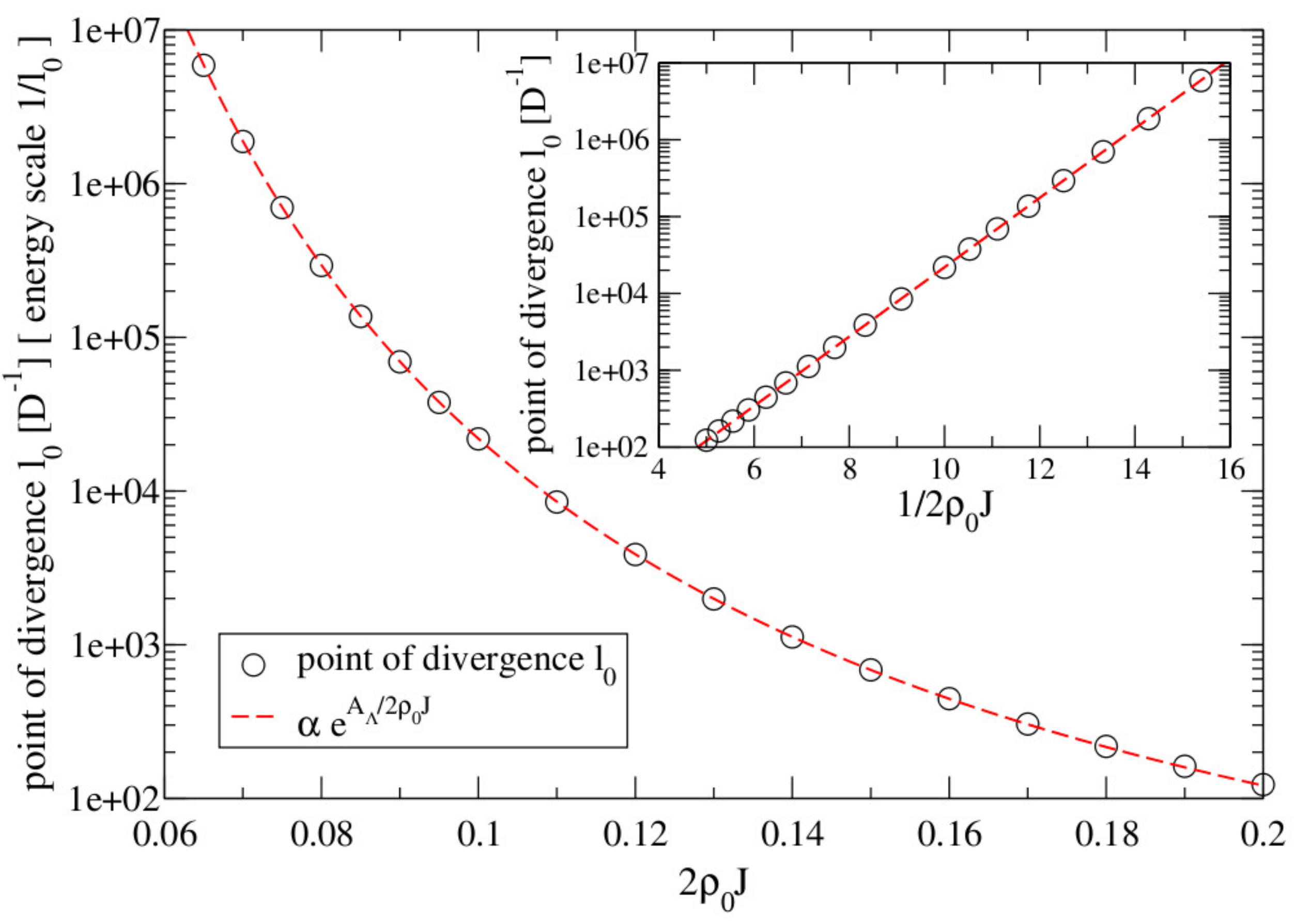}
\caption{(Color online) Flow parameter $l_{0}$ at which the flow equation diverges, 
cf.\ Fig.\ 
\ref{fig:ROD_Kondo_no_change_of_ref_state},  for the Kondo model with \mbox{$N=80$} and \mbox{$\Lambda=2$} in a logarithmic plot vs.\ $2\rho_0J$. The inset shows the exponential character 
\mbox{$l_{0}\propto \exp\left(A_{\Lambda}/2\rho_{0}J\right)$} in a logarithmic plot as
function of $1/2\rho_{0}J$. The factor $A_{\Lambda}$ is given in 
Eq.\ (\ref{discretization effect on Tk}) and stems from the discretization.}
\label{fig:Tk_Kondo_N80_L2}
\end{figure}

In Fig.\ \ref{fig:ROD_Kondo_no_change_of_ref_state} 
we show the residual off-diagonality (ROD) which is defined by
\begin{eqnarray}
\label{Chapter 4 definition ROD}
\text{ROD}^2 := \sum_{n: h_{n}\in \eta} \left|h_{n}\right|^2,
\end{eqnarray}
where $h_{n}$ denotes the coefficients in the Hamiltonian. We sum over the square of the absolute value of all coefficients which contribute to the generator \cite{fisch10a}.
In this way, the decrease of the ROD measures the convergence of the
CUT as function of $l$. 
In the case of the flow equation (\ref{DEQ poor mans scaling with CUT Kondo}) 
the ROD is given by
\begin{eqnarray}
\label{ROD for DEQ without change of ref state Kondo model}
\text{ROD}^2 = \sum_{\underset{n\neq m}{n,m}} \left|J_{nm}\right|^2.
\end{eqnarray}
In Fig.\ \ref{fig:ROD_Kondo_no_change_of_ref_state} one clearly sees that the flow
first appears to converge properly, indicated by a decreasing ROD. But there is
a value $l_0$ of the flow parameter $l$ at which the ROD changes its behavior and rises again.
It even diverges quickly beyond $l_0$.

\begin{figure}
\includegraphics[width=\columnwidth]{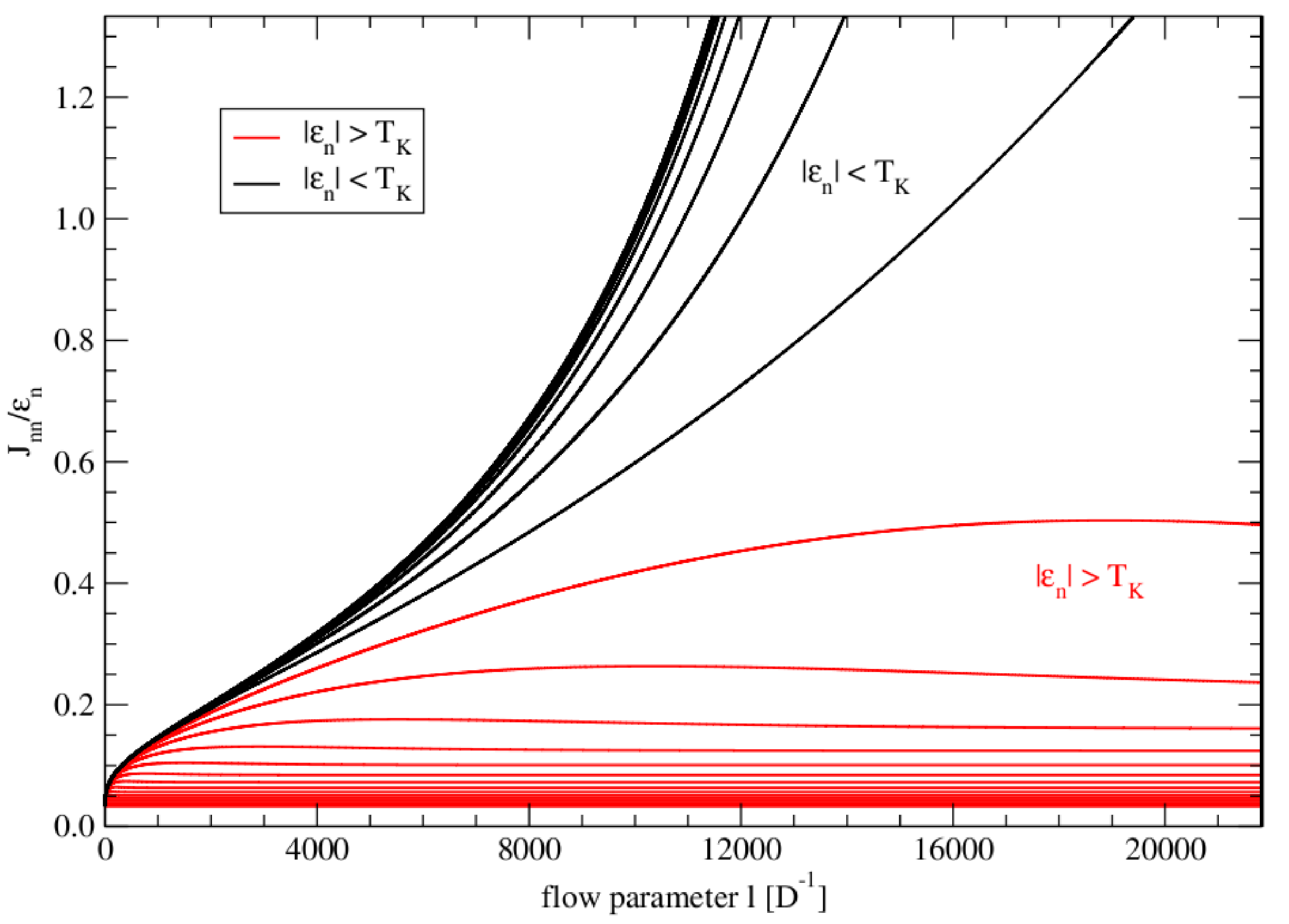}
\caption{(Color online) Flow of the diagonal spin-spin exchange interactions 
$\frac{J_{nn}\left(l\right)}{\left|\gamma_n\right|^2}$ of the Kondo model with 
\mbox{$N=80$}, \mbox{$\Lambda=2$} and \mbox{$2\rho_0J=0.1$}.
The Kondo temperature $T_{K}$ is taken from the inverse of the point of divergence $l_0$. 
The black lines show the exchange couplings with an index $n$ for which
\mbox{$\left|\epsilon_n\right|<T_K$} holds while the red lines show the couplings 
with an index $n$ for which \mbox{$\left|\epsilon_n\right|>T_K$} holds. 
Note that only couplings with $\left|\epsilon_n\right|<T_K$ diverge.}
\label{fig:J_nn_kondo_model_N80_L2_no_change_of_ref_state}
\end{figure}

In Fig.\ \ref{fig:Tk_Kondo_N80_L2} the inverse energy scale $l_{0}$ is analyzed at which the flow 
equation diverges. We find an exponential behavior of the form 
\begin{equation}
  l_{0} = T_{K}^{-1} \propto \text{e}^{\frac{A_{\Lambda}}{2\rho_{0}J}}
\end{equation}
where the prefactor
\begin{equation}
  \label{discretization effect on Tk}
A_{\Lambda}=\frac{1}{2}\frac{\Lambda+1}{\Lambda-1}\ln\Lambda
\end{equation}
takes a well-known discretization effect \cite{krish80a} into account which is
independent of the applied method NRG, CUT, or others.

This result is very similar to the outcome of 
Anderson's ''poor man's scaling'' \cite{ander70a}. 
The resulting differential equations diverge at the Kondo temperature $T_{K}$ which
we derived here in leading order in $J$. Truncating the flow equations in higher orders would 
provide higher order contributions to the Kondo temperature \cite{kehre06}.

Fig.\ \ref{fig:J_nn_kondo_model_N80_L2_no_change_of_ref_state} depicts the flow of the 
relative exchange couplings $J_{nn}/\left|\gamma_n\right|^2$ for various values of $n$.
Only couplings $J_{nn}$ with an index $n$ for which $\left|\epsilon_n\right|<T_K$ diverge in contrast to the couplings with indices corresponding to $\left|\epsilon_n\right|>T_K$ 
which converge towards a finite value.
This observation is very interesting because it clearly shows that the 
spin-spin interaction only plays a dominant role \emph{below} the Kondo energy scale.
To our knowledge, it has not been derived before that only the exhange couplings to levels
\emph{below} the Kondo energy diverge while the one to levels above this scale
stay finite.


\section{Modified approach - Change of the reference state during the flow}

Here we present a modification of the above approach which avoids the occurring divergences. The 
caveat of the above approach is the chosen reference state, i.e., the state to which the
CUT aims to map the ground state. So far the ground
state of the diagonal part
\begin{equation}
 H_{D,\text{old}} = \sum_{n,\sigma} \epsilon_{n} c^{\dagger}_{n \sigma} c^{\phantom{\dagger}}_{n \sigma}
\end{equation}
was taken as the reference state leaving
the spin of the impurity free. Thus the reference state is two-fold degenerate.
 Here we include the diagonal spin-spin interactions
\begin{equation}
 H_{K, \text{diag}} = 
\sum_{\mu}\sum_{\alpha,\beta,n} J_{nn}\sigma^{\mu}_{\alpha\beta}S_I^{\mu}:c^{\dagger}_{n \alpha}c^{\phantom{\dagger}}_{n \beta}:
\end{equation}
into the diagonal Hamiltonian
\begin{equation}
\label{diagonal Hamiltonian new operator basis for the Kondo model}
 H_{D,\text{modified}} = H_{D,\text{old}} + H_{K,\text{diag}}. 
\end{equation}
The key idea is to take the ground state of Eq. (\ref{diagonal Hamiltonian new operator basis for the Kondo model}) as reference state. In view of the divergence of the spin-spin couplings
at low energies it is indeed highly plausible, if not compulsory,
 that these couplings must be included
in the determination of the reference state because it should be close to the
true ground state. Moreover, it is known that the ground state of the Kondo model 
consists of a singlet state which implies that the impurity spin is correlated with
a spin from the bath electrons.

In order to choose the ground state of (\ref{diagonal Hamiltonian new operator basis for the Kondo model}) as the reference state we have to understand how it depends on the diagonal spin-spin interactions $J_{nn}$. 
In principle, it seems that we are facing a many-body problem again that
is almost as difficult as the original Hamiltonian. 
But we can find the ground state 
\eqref{diagonal Hamiltonian new operator basis for the Kondo model} by a much simpler consideration. 
If the couplings $J_{nn}$ are small enough, for instance during the early stages of the flow,
the ground state is the Fermi sea. 
The reason is that the spin-spin interactions do not have any effect on the
Fermi sea because all bath sites are either empty or doubly occupied so that there 
is no spin present. In order to have any effect, a spin in the bath
must be created by either adding a fermion above the Fermi level or removing one
from below the Fermi level. This costs energy.
Then, the energy gain due to the spin-spin interaction must compensate this energy loss.
This can only happen if the couplings 
$J_{nn}$ are large enough relative to the energies $\epsilon_n$.

The couplings $J_{nn}$ increase during the flow and thus the concomitant energy gain
increases compared to the energy loss. The energy balance depends on the site $n$ and there
will be one specific site where the energy balance favors the singlet formation.
The other sites remain in a Fermi sea, unaffected by the spin-spin interaction.

\begin{figure}
\includegraphics[width=\columnwidth]{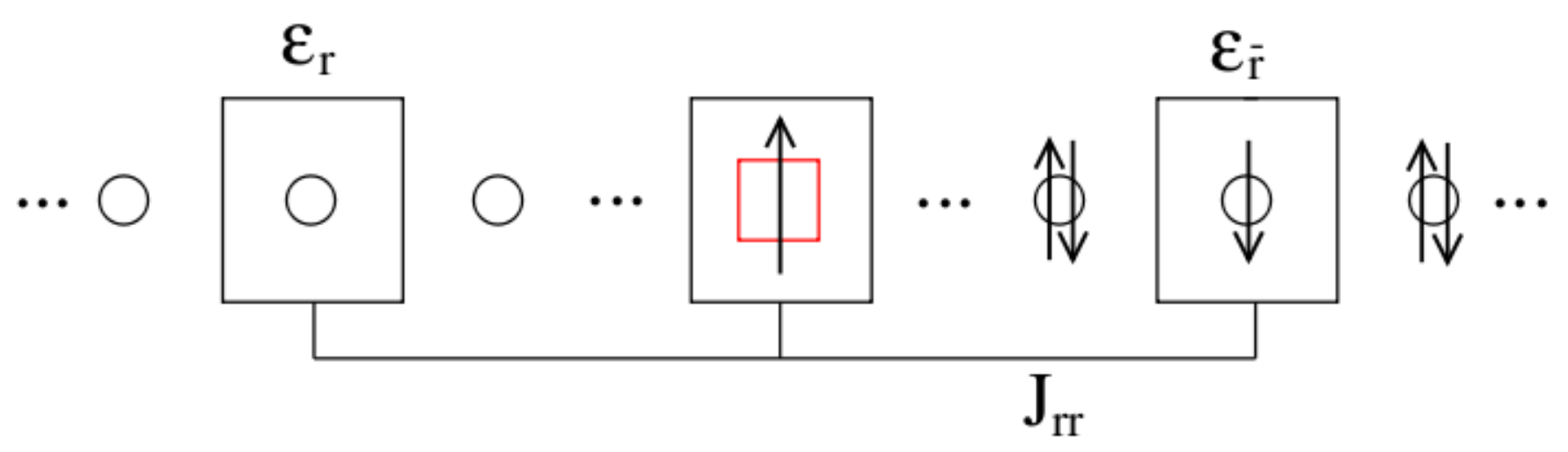}
\includegraphics[width=\columnwidth]{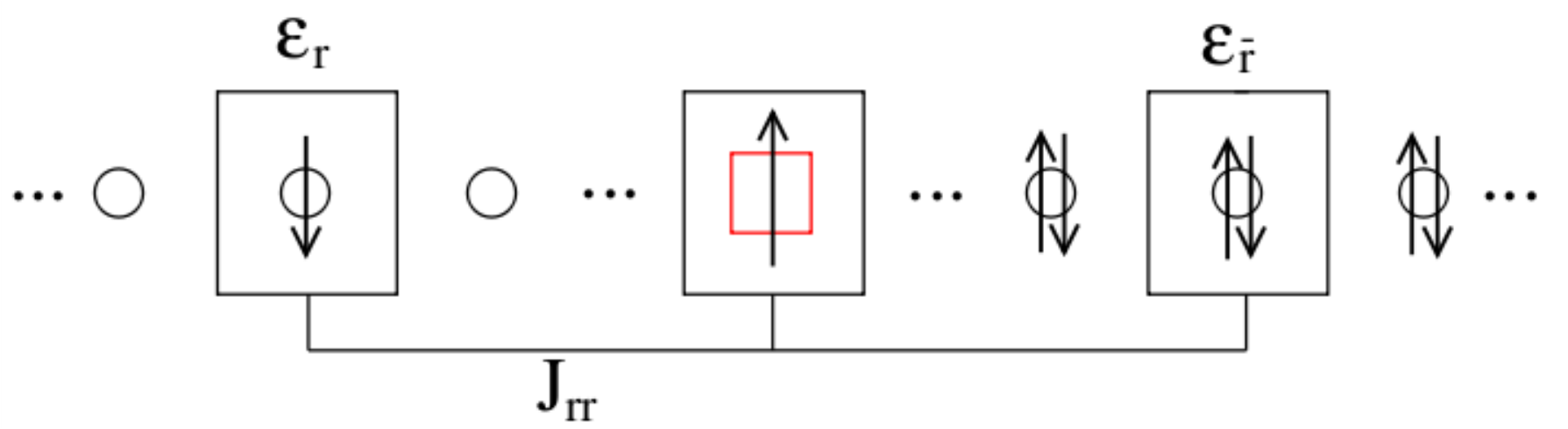}
\caption{(Color online) Part  $H_r$ of the Hamiltonian comprising the first sites for which it becomes energetically favorable to create a spin from the Fermi sea so that the spin-spin exchange is non-trivial. Due to particle-hole symmetry removing a particle from a negative level implies the same energy loss as adding a particle to the corresponding positive level.}
\label{fig:Hr_new_reference_state_pos}
\end{figure}

In order to understand how the ground state changes one only has to focus on the impurity and the specific sites  where it becomes energetically favorable to create a spin.
We include three sites in our analysis because if the creation of
a spin is favorable at the negative level $\epsilon_{\bar{r}} = -\epsilon_{r}$
by removing a particle then the same holds true at the positive level $\epsilon_{r}$ 
by adding a particle due to particle-hole symmetry. 
Thus, we consider
\begin{eqnarray}
\nonumber
  H_r&=&\sum_{\sigma} \epsilon_{r} \left(c^{\dagger}_{r \sigma}c^{\phantom{\dagger}}_{r \sigma} 
	- c^{\dagger}_{\bar{r} \sigma}c^{\phantom{\dagger}}_{\bar{r} \sigma}\right)
	\\
	\label{Chapter 7 H_r for the Kondo and Anderson model}
	&+& J_{rr}\sum_{\mu}\sum_{\alpha,\beta}\sigma_{\alpha\beta}^{\mu}S_I^{\mu}
	\left(c^{\dagger}_{r \alpha}c^{\phantom{\dagger}}_{r \beta}
	 +c^{\dagger}_{\bar{r} \alpha}c^{\phantom{\dagger}}_{\bar{r} \beta}\right) 
\end{eqnarray}
where $\bar{r}$ labels operators acting on the site with energy $\epsilon_{\bar{r}}=-\epsilon_r$. 
This problem can be solved by exact diagonalization and we find 
$32$ eigenstates.

In order to test the modified approach keeping the numerical calculation effort minimum, we 
neglect some of these eigenstates and keep the following
\begin{itemize}
\item[1.] The energetically low-lying ones which are influenced by the spin-spin coupling, namely the singlet state $|s^{\pm}\rangle$ and the triplet states $|t^{\pm}_{i}\rangle$.
\item[2.] The Fermi sea $|\text{FS}, \sigma\rangle$ because the reference state is changed if the singlet states is lowered below the Fermi sea, i.e., these states compete to be the ground state.
\item[3.] The state $|\tilde{\sigma}\rangle$ (cf.\ Eq.\ (\ref{used_states_adapted_operator_basis})) because it may also become the ground state.
\end{itemize}
These are $12$ states (cf.\ Eqs.\ (\ref{used_states_adapted_operator_basis})) out of the $32$ eigenstates of the Hamiltonian (\ref{Chapter 7 H_r for the Kondo and Anderson model}).  
In essence, we neglect all states with an energy larger than the energies of the triplet states.

The modified approach does not rely on this approximation, but using the complete
adapted operator basis would be less transparent and the computational effort would
increase significantly. Moreover, we will see that this choice of kept states
yields the expected energy scales. Nevertheless, it will be an interesting issue
 to implement the flow equations for the complete set of states to study 
the influence of the reduction of the number of kept states or to study whether
an even stricter truncation is sufficient as well.
In Appendix \ref{Appendix Basis states for the new operator basis}  all eigenstates of the Hamiltonian (\ref{Chapter 7 H_r for the Kondo and Anderson model}) are listed for 
completeness. The kept states are
\begin{subequations}
\label{used_states_adapted_operator_basis}
\begin{eqnarray}
&&|s^{-} \rangle = \frac{1}{\sqrt{2}}\left(|0,\uparrow,\downarrow \rangle - |0,\downarrow,\uparrow \rangle \right)
\\
&&|t^{-}_1 \rangle = |0,\uparrow,\uparrow \rangle
\\ 
&&|t^{-}_2 \rangle = \frac{1}{\sqrt{2}}\left(|0,\uparrow,\downarrow \rangle + |0,\downarrow,\uparrow \rangle \right)
\\
&&|t^{-}_3 \rangle = |0,\downarrow,\downarrow \rangle
\\ 
&&|s^{+} \rangle = \frac{1}{\sqrt{2}}\left(|\!\downarrow,\uparrow,\uparrow\downarrow \rangle - |\uparrow,\downarrow,\uparrow\downarrow \rangle \right)
\\ 
&&|t^{+}_1 \rangle = |\uparrow,\uparrow, \uparrow\downarrow  \rangle
\\ 
&&|t^{+}_2 \rangle = \frac{1}{\sqrt{2}}\left(|\!\downarrow,\uparrow,\uparrow\downarrow \rangle + |\uparrow,\downarrow,\uparrow\downarrow \rangle \right)
\\ 
&&|t^{+}_3 \rangle = |\!\downarrow,\downarrow,\uparrow\downarrow \rangle. 
\\ 
&&|\text{FS},\uparrow \rangle = |0,\uparrow,\uparrow\downarrow \rangle \phantom{\frac{1}{\sqrt{6}}}
\\
&&|\text{FS},\downarrow \rangle = |0,\downarrow,\uparrow\downarrow \rangle  \phantom{\frac{1}{\sqrt{6}}}
\\ 
&&|\tilde{\uparrow} \rangle = \frac{1}{\sqrt{6}}\left[|\uparrow,\uparrow,\downarrow \rangle-2 |\uparrow,\downarrow,\uparrow \rangle+ |\!\downarrow,\uparrow,\uparrow \rangle\right] 
\\ 
&&|\tilde{\downarrow} \rangle = \frac{1}{\sqrt{6}}\left[|\!\downarrow,\downarrow,\uparrow \rangle -2 |\!\downarrow,\uparrow,\downarrow \rangle+  |\uparrow,\downarrow,\downarrow \rangle\right]. 
\end{eqnarray}
\end{subequations}
The notation encodes the states $|{r}, {d}, {\bar{r}} \rangle$ 
where ${r}$  represents the state of positive level, ${\bar{r}}$ the corresponding state
at the negative level and ${d}$ the state of the impurity.
The states $|s^{-}\rangle$ and $|t_{i}^{-}\rangle$ refer to the singlet and triplet states formed with the negative level at $\epsilon_{\bar{r}}=-\epsilon_r$ 
while the states $|s^{+}\rangle$ and $|t_{i}^{+}\rangle$ refer to the singlet and triplet states formed with the positive level at $\epsilon_{r}$.
The states $|\text{FS},\sigma \rangle$ are the Fermi sea and a spin $\sigma$ at the impurity 
while the states $|\tilde{\sigma} \rangle$ refer to a state with an effective spin $\frac{1}{2}$.

The eigenvalues of the states in Eq.\ (\ref{used_states_adapted_operator_basis}) read
\begin{subequations}
\label{change of reference state full model eigenvalues}
\begin{eqnarray}
E_{s^{\pm}} =  -\frac{3J_{rr}}{2} - \epsilon_r
\, &,& \quad
E_{t^{\pm}_i} = \frac{J_{rr}}{2} - \epsilon_r 
\\ 
E_{\text{FS},\sigma} = -2\epsilon_r
\, &,& \quad
E_{\tilde{\sigma}}  = -2J_{rr}.
\end{eqnarray}
\end{subequations}
The indices $s^{\pm}$ and $t^{\pm}_i$ refer to the singlet and triplet states formed with the negative level at $\epsilon_{\bar{r}}=-\epsilon_r$ 
and the positive level at $\epsilon_{r}$ in Eq.\ (\ref{used_states_adapted_operator_basis}) while the indices $\text{FS},\sigma$ refer to the 
Fermi sea and $\tilde{\sigma}$ to the states with effective spin $\frac{1}{2}$ in Eq.\ (\ref{used_states_adapted_operator_basis}).

\begin{figure}
\includegraphics[width=\columnwidth]{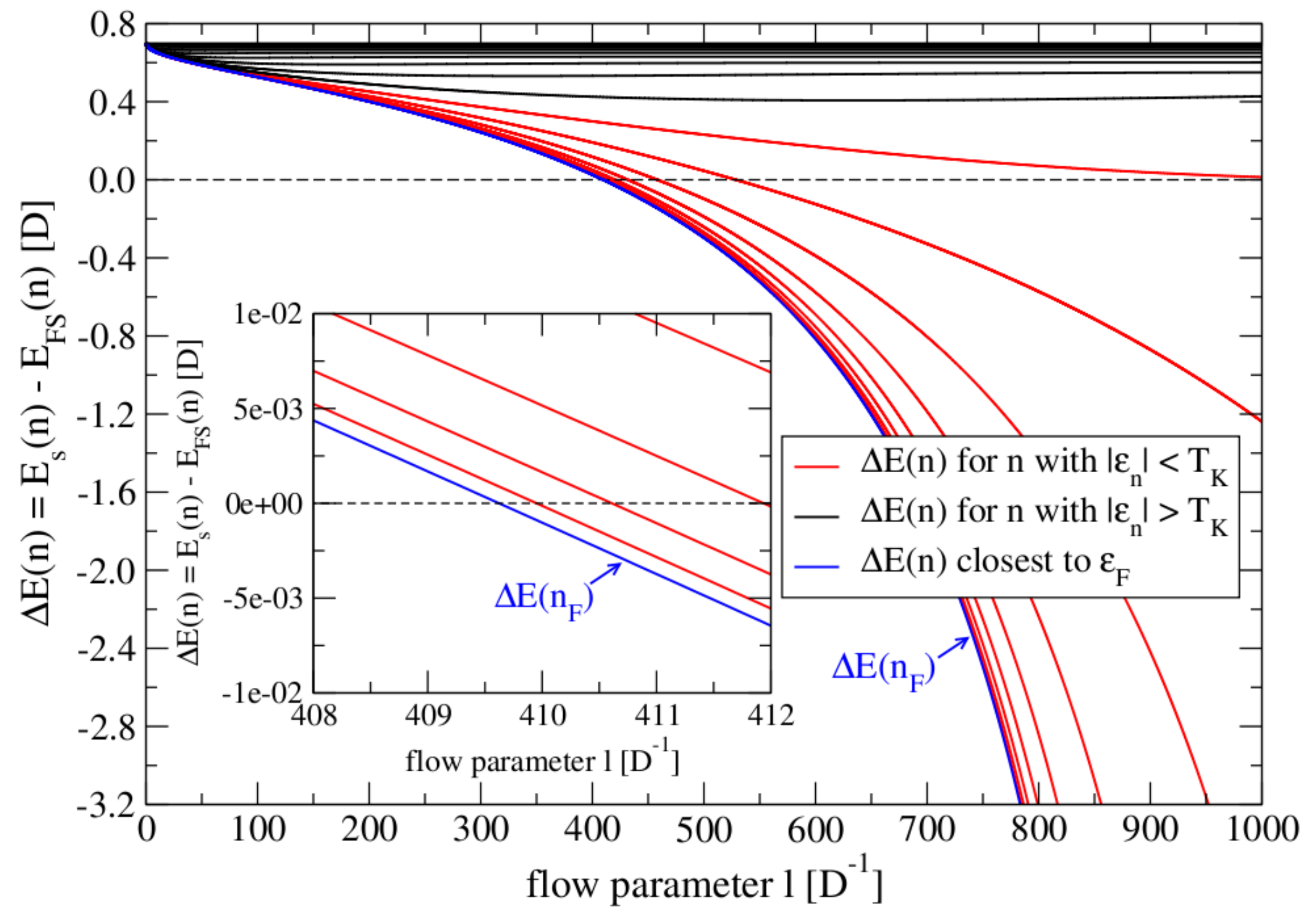}
\caption{(Color online) Energy difference between the singlet state and the Fermi sea $\Delta E\left(n\right) = E_s\left(n\right) - E_{\text{FS}}\left(n\right)$ 
at site $n$ from (\ref{change of reference state full model eigenvalues}) for the Kondo model with $N=40$, $\Lambda=2$ and $2\rho_0J=0.14$. The sites $n$ belong to the negative energy levels
and the absolute values of the energies are decreasing from top to bottom. 
The Kondo temperature  $T_{K}$ is taken from the inverse flow parameter $l_{0}$ where the flow
starts to diverge. The black lines mark the flow for the $n$ for which \mbox{$\left|\epsilon_n\right|<T_{K}$} while the red lines mark the flow for the $n$ for 
which\mbox{ $\left|\epsilon_n\right|>T_{K}$}. At the value of $l$ where one 
$\Delta E\left(n\right)$ vanishes the reference state is changed. 
The blue line shows $\Delta E\left(n_F\right)$ at the site closest to the Fermi level. The inset shows that the site closest to the Fermi level forms the singlet first.}
\label{fig:flow_of_Delta_E_N40_L2_J014_Kondo}
\end{figure}

The parameter regimes with their respective ground states are given by
\begin{subequations}
\begin{eqnarray}
 J_{rr}<\frac{2\epsilon_r}{3} && \quad  \text{ground state:} \quad |\text{FS},\sigma \rangle
\\
\frac{2\epsilon_r}{3}<J_{rr}<2\epsilon_r &&\quad \text{ground state:} \quad |s^{\pm} \rangle
\\
J_{rr}>2\epsilon_r &&\quad \text{ground state:} \quad |\tilde{\sigma} \rangle.
\end{eqnarray}
\end{subequations}
As soon as the point 
\begin{eqnarray}
\label{ratio at which the singlet forms}
J_{rr}\left(l_{0}\right)=\frac{2\epsilon_r}{3}
\end{eqnarray} 
is reached the ground state switches from the Fermi sea to the singlet state.
To be precise, both of them are doubly degenerate because the singlet
can be formed with site $r$ or $\overline{r}$ due to particle-hole symmetry.
The singlet states are no Slater determinants and thus Wick's theorem cannot be applied. But the remaining bath, of course, remains a Slater determinant and Wick's theorem
and the usual normal-ordering can be used as before.

At the point where the singlet states become energetically favorable, we choose the singlet
states as the reference states and the states (\ref{used_states_adapted_operator_basis}) as a new basis. This means that first,  we solve the flow (\ref{DEQ poor mans scaling with CUT Kondo}) in the conventional operator basis with the starting values (\ref{starting values Kondo model}). 
We track the ratio $J_{nn}/\epsilon_{n}$ at each site and as soon as the ratio 
reaches $2/3$ for some $n$ we change the reference state. 

Fig.\ \ref{fig:flow_of_Delta_E_N40_L2_J014_Kondo} depicts the energy difference
\begin{eqnarray}
\label{energy difference of the singlet and the Fermi sea at site n}
\Delta E\left(n\right) = E_s\left(n\right) - E_{\text{FS}}\left(n\right)
\end{eqnarray}
between the energy of the singlet state $E_s$ and the energy of the Fermi sea $E_{\text{FS}}$ at a site $n$ for the Kondo model. At the first value of $l$ where
this difference vanishes, the reference state is changed. 
The couplings $J_{nn}$ only diverge for indices $n$ with $\left|\epsilon_n\right|<T_K$ (cf.\ 
Fig.\ \ref{fig:J_nn_kondo_model_N80_L2_no_change_of_ref_state}), i.e., only couplings below the Kondo energy scale become large enough to make a change of reference state possible.
Thus, the approach ensures intrinsically that the singlet will form below the Kondo energy scale.
The results displayed in Fig.\ \ref{fig:J_nn_kondo_model_N80_L2_no_change_of_ref_state} show that the smaller the 
energy $\epsilon_n$, the faster the ratio $J_{nn}/\epsilon_{n}$ increases. 
Thus, the singlet forms at the lowest energy scale in the system. 
In a continuum of states this would be infinitesimally close to  the Fermi level  
$\epsilon_{F}$. 
In the logarithmically discretized numerical treatment this is the bath site with the
lowest energy.

Once we changed the reference state we compute the flow as discussed below in 
a modified operator basis. We do not allow for further changes of the reference state
which may occur in principle. But we will show below that the chosen switched reference state
ensures a convergent flow. To achieve convergence is the primary goal
of our present study. In addition, the singlet reference states acquire
a binding energy equal to the Kondo energy scale in the course of the flow.
Thus there is no indication of a need to change the reference state further.
Moreover, each change of reference state is cumbersome to implement so that we
have to leave a comprehensive discussion of this point to future research.


\subsection{Effective model and the modified operator basis}

In the next step we determine the effective Hamiltonian and the modified 
flow equations due to the changed reference state. 
We denote the site which is part of the modified operator basis by ${r}$ 
and the flow parameter at which the reference state is changed by $l_0$. 
When the flow parameter reaches the point $l=l_1$, the effective Hamiltonian is still
of the form
\begin{eqnarray}
\nonumber
 H\left(l_1\right) &=& \sum_{n,\sigma}\epsilon_n :c^{\dagger}_{n \sigma} c^{\phantom{\dagger}}_{n \sigma}:
  \\ 
	&+& \sum_{\mu}\sum_{n, m}\sum_{\alpha, \beta} J_{nm}\left(l_1\right) 
	\sigma^{\mu}_{\alpha\beta}S_I^{\mu}:c^{\dagger}_{n \alpha}c^{\phantom{\dagger}}_{m \beta}: .
\end{eqnarray}
The sites denoted by $r$ and $\bar{r}$, where $\epsilon_{\bar{r}}=-\epsilon_r$, form the singlet state with the impurity. Thus we treat them separately. 
Next we introduce the modified operator basis pertaining to these three sites which is
adapted to the changed reference state
\begin{eqnarray}
\label{operator-basis full model normal-ordering}
 \hat{O}_{kq}=| k \rangle\langle q | - \langle \hat{O}_{kq} \rangle
\end{eqnarray}
where $k$ and $q$ denote the basis states from (\ref{used_states_adapted_operator_basis}). 
The subtraction of the expectation values stands for normal-ordering.
Because the states $|s^{-}\rangle$ and $|s^{+}\rangle$ are degenerate we cannot use a single reference state but a reference ensemble \cite{reisc04}
\begin{eqnarray}
\label{expectation value new operator basis}
\langle \hat{O} \rangle = \frac{1}{2}\left(\langle s^{-}| \hat{O} |s^{-} \rangle+ \langle s^{+}| \hat{O} |s^{+} \rangle\right).
\end{eqnarray}

The operator basis from (\ref{operator-basis full model normal-ordering}) is normal-ordered with respect to this reference ensemble.
Due to the normal-ordering (\ref{operator-basis full model normal-ordering}), 
hopping terms will
generically emerge in the course of the flow which eventually change the energies
$E_k$. 
But all effects on $E_{k}$ stemming from the normal-ordering are at least of order 
$J^3$ because all terms arising in this way are of order $J^2$ and they 
need at least one more commutation to act on $E_{k}$ which increases the order 
in powers of $J$ at least by one.
We focus on orders up to $J^2$ in the local energies and thus neglect the normal-ordering 
from (\ref{operator-basis full model normal-ordering}) using 
$\hat{O}_{kq}=| k \rangle\langle q |$ instead.

We expand all terms in the modified operator basis yielding the Hamiltonian
in the form
\begin{eqnarray}
\nonumber
\bar{H}\left(l_0\right) &=&\sum_{n\neq \pm r,\sigma} \epsilon_{n} :c^{\dagger}_{n \sigma}c^{\phantom{\dagger}}_{n \sigma}:
   + \sum_{k,q} E_{kq} | k\rangle\langle q |
\\ \nonumber
&+& \sum_{k,q}\sum_{n,m\neq \pm r}\sum_{\alpha,\beta} J^{kq\alpha\beta}_{nm} 
|k\rangle\langle q|:c^{\dagger}_{n \alpha}c^{\phantom{\dagger}}_{m \beta}:
\\
&+&\sum_{k,q,\sigma}\sum_{n\neq \pm r} \Gamma^{kq\sigma}_{n}
\left(|k \rangle\langle q |  c^{\phantom{\dagger}}_{n \sigma} + c^{\dagger}_{n \sigma}|q \rangle\langle k |  \right).
\label{hamiltonian new operator basis full model}
\end{eqnarray}
The basis states and the modified operators were chosen such that terms acting only on the sites 
$r$ or $\bar{r}$ are diagonal. Hence we know by construction that the $E_k$ are given by the eigenvalues in (\ref{change of reference state full model eigenvalues}) where 
$J_{rr}=J_{rr}\left(l_1\right)$ while the starting values for the other coefficients are determined from
\begin{subequations}
\begin{eqnarray}
J^{kq\alpha\beta}_{nm}\left(l_1\right)&=&J_{nm}\left(l_1\right)
\sum_{\mu}\sigma^{\mu}_{\alpha\beta} \langle k |S_I^{\mu}|q \rangle
\\ \nonumber 
\Gamma^{kq\sigma}_{n}\left(l_1\right) 
&=&J_{nr}\left(l_1\right)\sum_{\mu,\alpha}\sigma^{\mu}_{\alpha\sigma}\langle k |S_I^{\mu}c^{\dagger}_{r \alpha}|q \rangle
\\             
&+& J_{n\bar{r}}\left(l_1\right)\sum_{\mu,\alpha}\sigma^{\mu}_{\alpha\sigma}\langle k |S_I^{\mu}c^{\dagger}_{\bar{r} \alpha}|q \rangle
\\
E_{kq}\left(l_1\right)
&=& 
   2\Re\left(\sum_{\mu}\sum_{\alpha, \beta}J_{r\bar{r}}\sigma^{\mu}_{\alpha\beta}\langle k|S_I^{\mu} c^{\dagger}_{\bar{r} \alpha}c^{\phantom{\dagger}}_{r \beta}| q \rangle\right).  
\end{eqnarray}
\end{subequations}

The structure of the generator after changing the reference ensemble is given by
\begin{eqnarray}
\nonumber
\eta&=&\sum_{k,q}\sum_{\alpha,\beta}\sum_{n,m\neq \pm r} \eta^{kq\alpha\beta}_{nm,J} |k\rangle\langle q|:c^{\dagger}_{n \alpha}c^{\phantom{\dagger}}_{m \beta}:
\\ \label{Chapter 7 generator for the full problem new operator basis}
&+&\sum_{k,q,\sigma}\sum_{n\neq \pm r} \eta^{kq\sigma}_{n,\Gamma}
\left(|k \rangle\langle q |  c^{\phantom{\dagger}}_{n \sigma} - \text{h.c.}  \right)
 + \sum_{k,q} \eta^{E}_{kq} | k\rangle\langle q |. \qquad
\end{eqnarray}
We want to eliminate all terms that couple to the reference ensemble. Thus we choose the coefficients of the generator of the form
\begin{subequations}
\begin{eqnarray}
\eta^{E}_{kq} &=& \!\!\left\{\!\!\begin{array}{cc} \text{sgn}\left(E_k - E_q\right)E_{kq} & \text{if} \, k, q = s^{\pm} \\ 0 & \text{otherwise} \end{array} \right.
\\ 
\eta^{kq\alpha\beta}_{nm,J} &=& \!\!\left\{\!\!\begin{array}{cc} \text{sgn}\left(E_k-\!E_q+
\!\epsilon_{n}-\!\epsilon_{m}\right)
J^{kq\alpha\beta}_{nm} & \text{if} \, k, q = s^{\pm} \\ 0 & \text{otherwise} \end{array} \right. \quad
\\
\eta^{kq\sigma}_{n,\Gamma} &=& \!\!\left\{\!\!\begin{array}{cc} 
\text{sgn}\left(E_k-E_q-\epsilon_{n}\right)\Gamma^{kq\sigma}_{n} & \text{if}  
\, k, q = s^{\pm} \\ 0 & \text{otherwise} \end{array} \right.,
\end{eqnarray}
\end{subequations}
where $k,q = s^{\pm}$ means that $k$ or $q$ are in one of the singlet states.

Summarizing, we only include terms which couple to the singlet states.
We emphasize that this implies that terms which couple to the triplet states formed from the
impurity spin and a spin on a bath site at the Fermi level are not eliminated because they are not included in the generator. Thus, the 
reference state, which becomes the ground state in the course of the flow, consists of a singlet 
and a Fermi sea. 
But this does not imply that the complete effective model is reduced to a singlet and a free
fermionic bath. This fact makes an exhaustive analysis of the effective model challenging.

In order to obtain the modified flow equation (\ref{flow equation}), we commute the generator 
(\ref{Chapter 7 generator for the full problem new operator basis}) 
with the Hamiltonian (\ref{hamiltonian new operator basis full model}). We truncate terms which have a quartic structure in the fermionic bath operators.
The resulting flow equation reads
\begin{subequations}
\label{flow_kondo_modified}
\begin{eqnarray}
\nonumber
\partial_l E_{kq} 
&=&\left(E_{qq} - E_{kk}\right)\eta^{E}_{kq}
 + \sum_{p\neq q}\eta^{E}_{kp} E_{pq} 
 - \sum_{p\neq k}\eta^{E}_{pq} E_{kp}
\\ \nonumber
&+&\sum_{p}\sum_{\alpha,\beta}\sum_{n,m\neq \pm r} \eta^{kp\alpha\beta}_{nm,J} J^{pq\beta\alpha}_{mn}\theta_{n}\left(1-\theta_{m}\right)
\\ \nonumber
&-&\sum_{p}\sum_{\alpha,\beta}\sum_{n,m\neq \pm r} \eta^{pq\alpha\beta}_{mn,J} J^{kp\beta\alpha}_{nm}\theta_{n}\left(1-\theta_{m}\right)
\\ \nonumber
&-&\sum_{n\neq \pm r}\sum_{p,\gamma} \left(\eta^{pk\gamma}_{n,\Gamma}\Gamma^{pq\gamma}_{n} + \eta^{pq\gamma}_{n,\Gamma}\Gamma^{pk\gamma}_{n}\right)\theta_{n}
\\ 
&+&\sum_{n\neq \pm r}\sum_{p,\gamma} \left(\eta^{kp\gamma}_{n,\Gamma}\Gamma^{qp\gamma}_{n} + \eta^{qp\gamma}_{n,\Gamma}\Gamma^{kp\gamma}_{n}\right)\left(1-\theta_{n}\right) \qquad
\label{flow equation change of ref state full model E}
\end{eqnarray}

\begin{eqnarray}
\nonumber
\partial_l \Gamma^{kq\sigma}_{n}
&=&\left(\epsilon_{n} + E_{qq} - E_{kk}\right)\eta^{kq\sigma}_{n,\Gamma}
\\ \nonumber
&+&\sum_{p\neq q} \left(\eta^{kp\sigma}_{n,\Gamma} E_{pq} - \eta^{E}_{pq} \Gamma^{kp\sigma}_{n}\right)
\\ \nonumber
&-&\sum_{p\neq k} \left(\eta^{pq\sigma}_{n,\Gamma} E_{kp} - \eta^{E}_{kp} \Gamma^{pq\sigma}_{n}\right)
\\ \nonumber 
&+&\sum_{p}\sum_{x\neq \pm r,\gamma}
\left(\eta^{kp\gamma}_{x,\Gamma}J^{pq\gamma\sigma}_{xn} - \eta^{pq\gamma\sigma}_{xn,J} \Gamma^{kp\gamma}_{x}\right)\left(1-\theta_{x}\right)
\\ \label{flow equation change of ref state full model G}
&+&\sum_{p}\sum_{x\neq \pm r,\gamma}
\left(\eta^{pq\gamma}_{x,\Gamma}J^{kp\gamma\sigma}_{xn} - \eta^{kp\gamma\sigma}_{xn,J} \Gamma^{pq\gamma}_{x} \right)\theta_{x} \quad
\end{eqnarray}

\begin{eqnarray}
\nonumber
\partial_l J^{kq\alpha\beta}_{nm}
&=&\left(\epsilon_{m} - \epsilon_{n} + E_{qq} - E_{kk}\right)\eta^{kq\alpha\beta}_{nm,J}
\\ \nonumber
&+&\sum_{p\neq q} \left(\eta^{kp\alpha\beta}_{nm,J}E_{pq} - \eta^{E}_{pq} J^{kp\alpha\beta}_{nm} \right)
\\ \nonumber
&-&\sum_{p\neq k} \left(\eta^{pq\alpha\beta}_{nm,J}E_{kp} - \eta^{E}_{kp} J^{pq\alpha\beta}_{nm} \right)
\\ \nonumber
&+&\sum_{x\neq \pm r,\gamma,p}\left(\eta_{nx,J}^{pq\alpha\gamma} J_{xm,J}^{kp\gamma\beta} - \eta_{xm,J}^{kp\gamma\beta} J_{nx}^{pq\alpha\gamma}\right)
\theta_{x}
\\ \nonumber
&+&\sum_{x\neq \pm r,\gamma,p}\left(\eta_{nx,J}^{kp\alpha\gamma} J_{xm,J}^{pq\gamma\beta} - \eta_{xm,J}^{pq\gamma\beta} J_{nx}^{kp\alpha\gamma}\right)
\left(1-\theta_{x}\right)
\\ \nonumber
&-&\sum_{p}\left(\eta^{pk\alpha}_{n,\Gamma}\Gamma^{pq\beta}_{m}+\eta^{pq\beta}_{m,\Gamma}\Gamma^{pk\alpha}_{n}\right)
\\ \label{flow equation change of ref state full model J}
&-&\sum_{p}\left(\eta^{qp\alpha}_{n,\Gamma}\Gamma^{kp\beta}_{m}+\eta^{kp\beta}_{m,\Gamma}\Gamma^{qp\alpha}_{n}\right).
\end{eqnarray}
\end{subequations}
where $\Theta_x:=\langle c^{\dagger}_{x \sigma} c^{\phantom{\dagger}}_{x \sigma} \rangle$ 
is the expectation value with respect to the Fermi sea. It occurs upon the normal-ordering
of the fermionic bath operators with respect to the Fermi sea.

\subsection{Results of the modified flow}

\label{ssec:modified_flow}

\begin{figure}
\includegraphics[width=\columnwidth]{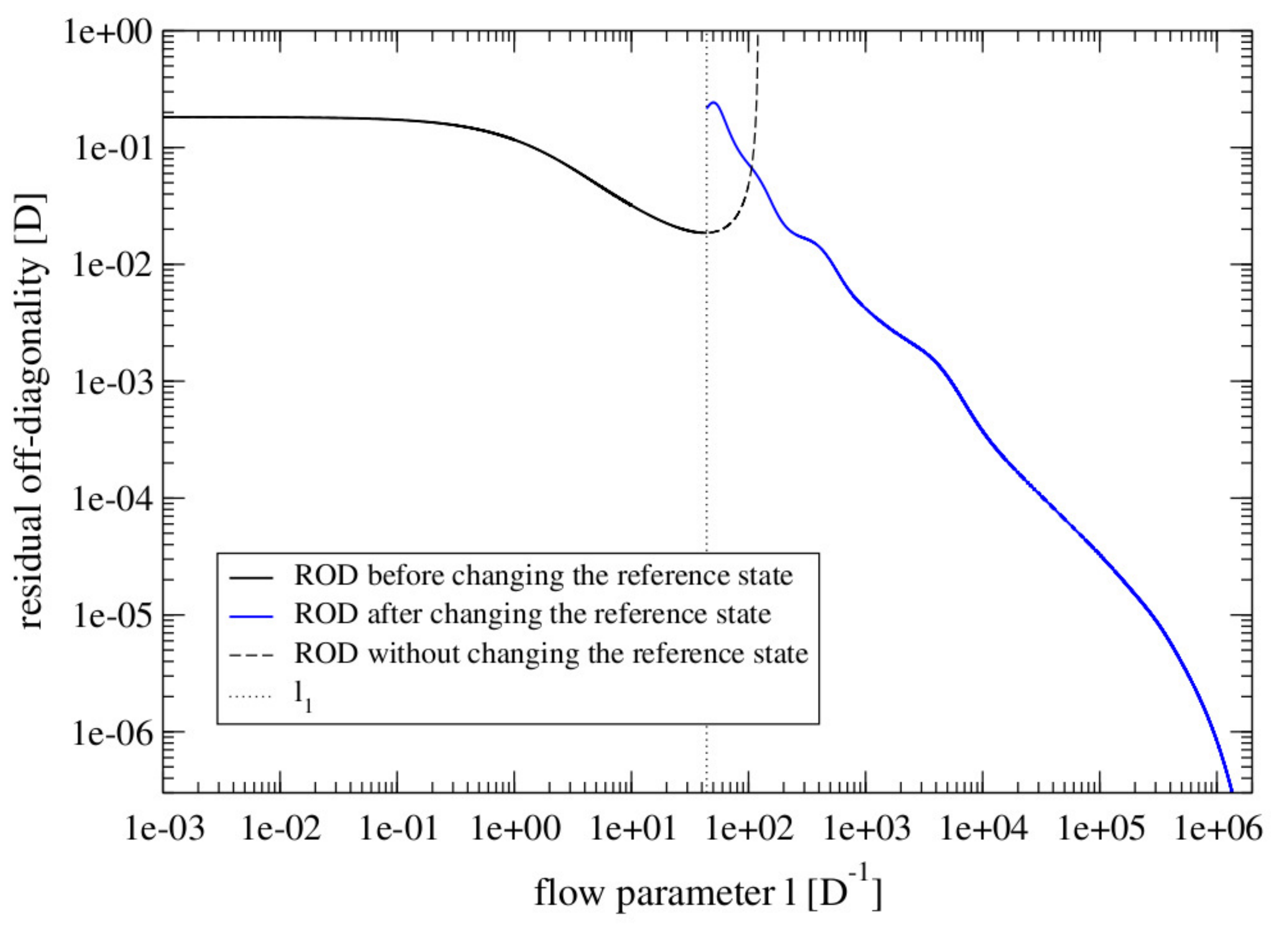}
\caption{(Color online) ROD for the Kondo model with \mbox{$N=40$}, 
\mbox{$\Lambda = 2$} and \mbox{$2\rho_0J=0.2$} obtained from the modified approach where the 
reference state is changed during the flow at $l_{1}$. 
First, the Fermi sea is the reference state and (\ref{DEQ poor mans scaling with CUT Kondo}) 
is solved with the initial values (\ref{starting values Kondo model}). 
After changing the reference state the modified flow \eqref{flow_kondo_modified} is solved.
The modified flow converges in contrast to the original CUT. The ROD for the original CUT 
(\ref{DEQ poor mans scaling with CUT Kondo}) without changing the reference state is 
denoted by the dashed line.}
\label{fig:ROD_Kondo_N40_L2_J02}
\end{figure}

The number of indices is very large and one should reduce the
number of differential equations by exploiting symmetries.
A lot of combinations of the $k$ and $q$ indices, for instance, do not occur due to
spin conservation which reduces the numerical effort.
In the following paragraphs the results obtained by the modified approach are presented.
Fig.\ \ref{fig:ROD_Kondo_N40_L2_J02} depicts the ROD of the Kondo model in the modified approach. 
The aim is to verify that the flow \eqref{flow_kondo_modified}  converges. 

We start from the Kondo Hamiltonian and solve the original flow (\ref{DEQ poor mans scaling with CUT Kondo}) 
with the initial values (\ref{starting values Kondo model}).
This implies that the Fermi sea is the reference state. 
For small $l$ the ratio $J_{nn}/\epsilon_{n}$ is significantly smaller than $2/3$,
cf.\ (\ref{ratio at which the singlet forms}), and the flow proceeds as long as this holds true. 
At some value $l_1$ the spin-spin coupling $J_{nn}$ becomes 
large enough so that $J_{nn}/\epsilon_{n}=2/3$ is fulfilled. 
As soon as this happens, we change the reference state and rewrite the Hamiltonian 
in the form (\ref{hamiltonian new operator basis full model}) 
with the modified operator basis (\ref{used_states_adapted_operator_basis}). 
Then, we use the modified flow \eqref{flow_kondo_modified} and 
continue with the flow starting at $l_1$. 
All sets of differential equations are solved by a $4$th-order Runge-Kutta algorithm. 

For small $l$ the ROD is the same as for the original flow (\ref{DEQ poor mans scaling with CUT Kondo}) because the reference state is not changed yet. Once we switch to the modified reference state, the ROD changes discontinuously because the generator is changed so that 
other types of terms are included in the ROD. One may wonder why the ROD increases upon changing the reference state although we aim at eliminating less terms than before. Recall that  
we only rotate away terms that couple to the singlet states. But one must also bear in mind that we include completely different types of terms in the generator after the change to the
modified operator basis. 
In particular, terms that are diagonal in the fermionic bath operators are then included which were not included before. For instance, we may inspect terms of the form
\begin{eqnarray}
 J^{t^{\pm}_i,s^{\pm},\alpha\beta}_{nn} | t^{\pm}_i\rangle\langle s^{\pm}|:c^{\dagger}_{n \alpha}c^{\phantom{\dagger}}_{n \beta}:
\end{eqnarray}
with initial values at $l_1$ that are proportional to $J_{nn}\left(l_1\right)$. 
Such terms were not eliminated by the CUT in the conventional, original operator basis. Thus, 
at $l=l_1$ these terms are large compared to the terms in the generator before the reference state is changed which have been suppressed by a factor 
$\exp\left(-|\epsilon_n - \epsilon_m|l_1\right)$. 
Once the reference state is changed, these diagonal terms in the fermionic bath operators are
included in the ROD. Thus, the ROD increases abruptly upon switching the reference state and using the modified generator.

The dashed line in Fig.\ \ref{fig:ROD_Kondo_N40_L2_J02} shows the behavior of the ROD if the reference state is not changed. 
In this case the ROD diverges at $l_0^{-1}$ corresponding to the Kondo temperature $T_K$. 
We conclude that the modified flow equation \eqref{flow_kondo_modified} is indeed able to prevent this divergence leading  to an effective Hamiltonian
with finite couplings even at the Fermi level $\epsilon_F=0$.

\begin{figure}
\includegraphics[width=\columnwidth]{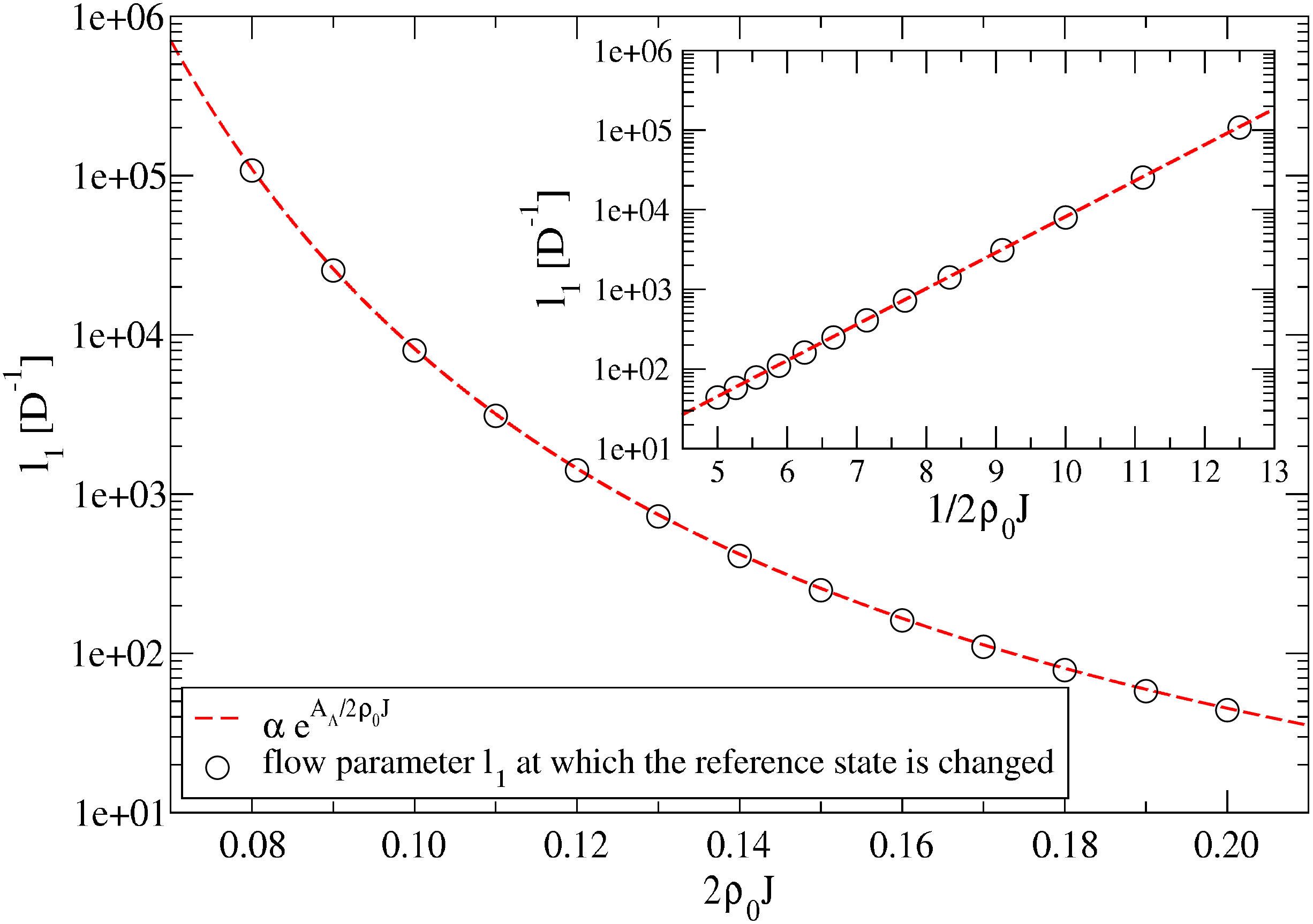}
\caption{(Color online) Inverse energy scale $l_1$ at which the reference state is changed for the Kondo model with \mbox{$\Lambda=2$} and \mbox{$N=52$}.
The inverse energy scale $l_1$ is increasing proportionally to the inverse of the Kondo temperature 
\mbox{$T_K^{-1}\propto\exp\left(A_{\Lambda}/2\rho_{0}J\right)$}. 
The factor $A_{\Lambda}$ from (\ref{discretization effect on Tk}) is due to the discretization.}
\label{fig:energy_scale_change_of_ref_state_N52_L2_l0_Kondo_alternative}
\end{figure}

We succeeded to provide a method that yields an effective Hamiltonian with small finite
parameters for $\epsilon\rightarrow\epsilon_F$. 
With the conventional, original  approach we found the Kondo energy scale only as the point at
 which the running couplings diverge \cite{kehre06}. 
Fig.\ \ref{fig:energy_scale_change_of_ref_state_N52_L2_l0_Kondo_alternative} depicts the inverse energy scale given by the flow parameter $l_1$ at which the reference state is changed for the Kondo model.  We retrieve the exponential energy scale
\begin{eqnarray}
l_1^{-1} \propto \text{e}^{-\frac{A_{\Lambda}}{2\rho_0J}}
\end{eqnarray}
where the factor $A_{\Lambda}$ is given by (\ref{discretization effect on Tk}) taking
discretization effects into account \cite{krish80a}.
Thus, we confirm that the energy scale at which the reference state is changed is proportional to the Kondo temperature $T_K$, at least at the level of accuracy of the present study. 

\begin{figure}
\includegraphics[width=\columnwidth]{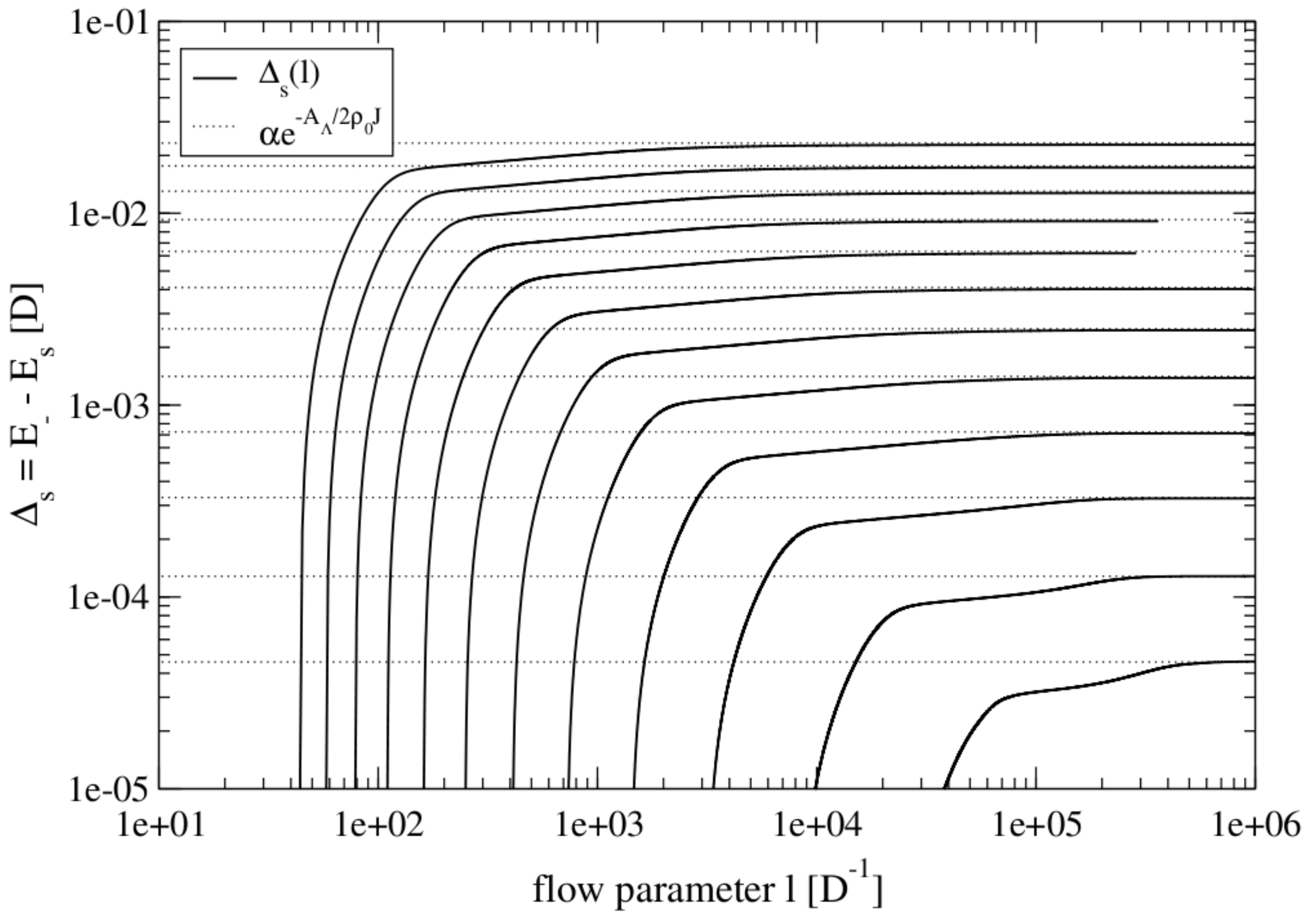}
\caption{(Color online) Flow of \mbox{$\Delta_{s}\left(l\right) = E_{-}\left(l\right) - E_{s}\left(l\right)$} with $E_{-}$ and $E_{s}$ 
from (\ref{Chapter 7 effective Hamiltonian without interaction part energy levels}) 
for the Kondo model with $N=40$ and $\Lambda=2$.
The quantity $\Delta_{s}$ converges to the binding energy of the Kondo singlet for $l\rightarrow\infty$.}
\label{fig:E_singlett_m_E_FS_N40_L2_Kondo}
\end{figure}

In addition, there is another very interesting energy scale in the effective model
which is the \emph{binding energy} of the singlets. 
This energy is the energy by which the singlets are separated from the Fermi sea 
states with a localized spin on the impurity. Note that this does not imply
that the system is gapped because the other fermionic sites still exist
and represent massless excitations in the continuum limit.

We can identify the local binding energy easily as
the energy difference $\Delta_s$ of the two lowest energies $E_{s}$ and $E_{-}$ in the diagonal part (\ref{Chapter 7 effective Hamiltonian without interaction part energy levels}) 
of the Hamiltonian (\ref{Chapter 7 effective Hamiltonian without interaction part}) 
where the $E_{kq}$ are diagonalized first, cf.\ also \ref{change of reference state full model eigenvalues}). Hence we consider
\begin{eqnarray}
\label{Delta_s}
 \Delta_s= E_{-} - E_{s}
\end{eqnarray}
at $l\to\infty$.

If we omit the remaining interactions at finite $l$, we can define an approximative binding energy of the singlet state by the energy difference between the two 
lowest energy levels
\begin{eqnarray}
\Delta_s\left(l\right) = E_{-}\left(l\right) - E_{s}\left(l\right)
\end{eqnarray}
from the effective Hamiltonian
\begin{eqnarray}
\nonumber
 H_{\text{eff}}&=&\sum_{n,\sigma}\epsilon_n :c^{\dagger}_{n \sigma} 
c^{\phantom{\dagger}}_{n \sigma}:
                + \sum_{k}E_{kk}| k \rangle\langle k| 
\\ \label{Chapter 7 effective Hamiltonian without interaction part}
               &+&g\sum_{\sigma}\left(| \text{FS},\sigma \rangle\langle \tilde{\sigma}| + | \tilde{\sigma} \rangle\langle \text{FS},\sigma|\right)
\end{eqnarray}
where we have first diagonalized the non-diagonal terms proportional  to $E_{kq}$. 
Diagonalizing the terms $|k \rangle\langle q|$ yields the eigenvalues
\begin{subequations}
\begin{eqnarray}
\nonumber
&&E_{s^{\pm}} 
\, , \quad  
E_{t^{\pm}_{1,2,3}}
\\ \label{Chapter 7 effective Hamiltonian without interaction part energy levels}
&&E_{\pm} = \frac{1}{2}\left[E_{\text{FS},\sigma} + E_{\tilde{\sigma}} \pm \sqrt{\left(E_{\text{FS},\sigma} - E_{\tilde{\sigma}}\right)^2 + 4g^2}\right] \quad
\end{eqnarray}
\end{subequations}
where the lowest-lying level is the energy of the singlet $E_s$ and the second lowest-lying is the energy level $E_{-}$ which belongs to a linear
combination of the Fermi sea and the $|\tilde{\sigma}\rangle$-states, see also (\ref{change of reference state full model eigenvalues}). 

\begin{figure}
\includegraphics[width=\columnwidth]{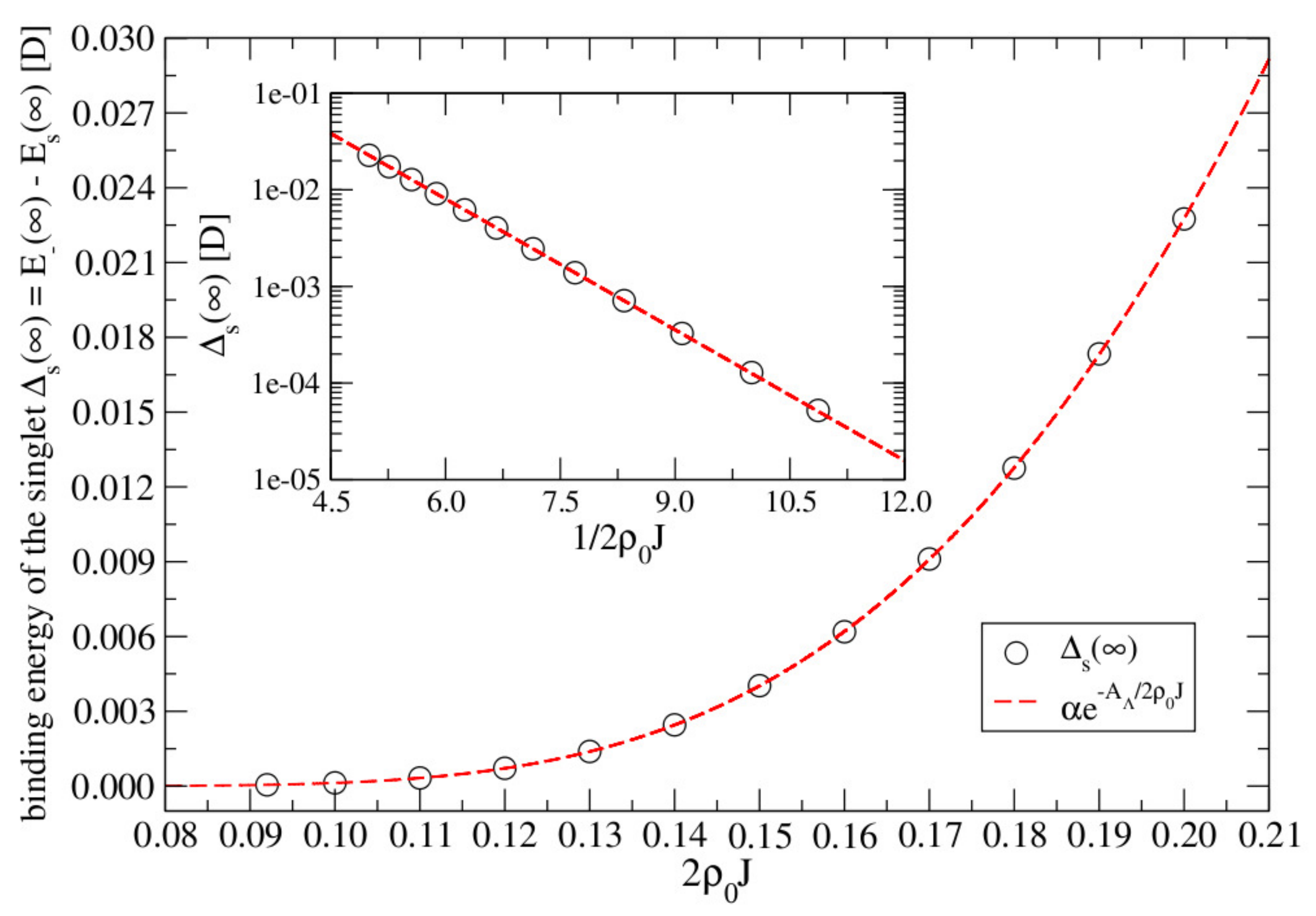}
\caption{(Color online) Binding energy of the singlet state \mbox{$\Delta_{s}\left(\infty\right) = E_{-}\left(\infty\right) - E_{s}\left(\infty\right)$} with 
$E_{-}$ and $E_{s}$ from Eqs.\ (\ref{Chapter 7 effective Hamiltonian without interaction part energy levels}) and (\ref{change of reference state full model eigenvalues}) 
for the Kondo model with \mbox{$N=40$} and \mbox{$\Lambda=2$}. The binding energy shows the generic exponential behavior 
\mbox{$\Delta_{s}\left(\infty\right)\propto\exp\left(A_{\Lambda}/2\rho_{0}J\right)$}. 
The factor $A_{\Lambda}$ from (\ref{discretization effect on Tk}) is an effect of the discretization.}
\label{fig:E_singlett_m_E_FS_N40_L2_Kondo_vs_J}
\end{figure}

Fig.\ \ref{fig:E_singlett_m_E_FS_N40_L2_Kondo} depicts the corresponding flow of $\Delta_s(l)$.
The energy necessary to break up the singlet ground state is given by the difference
\begin{eqnarray}
\Delta_s\left(\infty\right) = E_{-}\left(\infty\right) - E_{s}\left(\infty\right)
\end{eqnarray}
where $E_{s}$ is the singlet energy, which is the lowest lying state, and $E_{-}$ is given by (\ref{Chapter 7 effective Hamiltonian without interaction part energy levels}), 
which is the first excitation above the singlet state \emph{involving} the impurity. 
We draw the reader's attention to the fact that $\Delta_s$ can only be interpreted as the binding energy for $l\rightarrow\infty$ because only then the ground state of the 
effective model will be the singlet state. For smaller $l$ there are still interaction terms present that act on the singlet state which vanish in the limit $l\rightarrow\infty$.

Upon increasing $l$, $\Delta_s\left(l\right)$ increases rapidly until it converges towards the 
binding energy of the singlet state $\Delta_s\left(\infty\right)$. The binding energy is analyzed in Fig.\ \ref{fig:E_singlett_m_E_FS_N40_L2_Kondo_vs_J}.
For $l\rightarrow\infty$ we find an exponential behavior of the form
\begin{eqnarray}
\label{exponential behaviour binding energy Kondo}
\Delta_s\left(\infty\right)\propto\text{e}^{- \frac{A_{\Lambda}}{2\rho_0J}}
\end{eqnarray}
where the discretization factor $A_{\Lambda}$ is given by (\ref{discretization effect on Tk}).

Hence, we conclude that the modified CUT approach 
not only yields an effective Hamiltonian with finite couplings,
but also results in a model in which  the 
Kondo energy scale is already manifest in the diagonal part. 
It is no longer  hidden within the intricate interplay of different physical processes.

Recall that earlier CUT approaches led to diverging couplings \cite{kehre06} or to
an effective model where the parameters exhibited logarithmic infrared divergences very similar to those found by a standard perturbative treatment \cite{kondo64}.
Alternatively, the detour via a bosonized form of the Kondo model was taken
 \cite{hofst01a,lobas05,sleza03}. The mapping of the fermionic Kondo model
to the bosonized one is systematically controlled only in the wide band limit.


\section{Anderson impurity model}

In the last two sections we considered the Kondo model first in the
standard CUT akin to poor man's scaling and second with a change of
reference state. The first approach yields diverging couplings while
the second provides a well-defined effective model with finite couplings.
In the present section and in the next one we will extend these treatments
to the Anderson impurity model.


\subsection{Parametrization of the Anderson impurity model}

We consider the Anderson impurity model in its standard form \cite{ander61,hewso93}
\begin{eqnarray}
\nonumber
 H &=& \sum_{\mathbf{k},\sigma} \epsilon^{\phantom{\dagger}}_{\mathbf{k}} c^{\dagger}_{\mathbf{k} \sigma} c^{\phantom{\dagger}}_{\mathbf{k} \sigma} 
+\sum_{\sigma} \epsilon_d d^{\dagger}_{\sigma} d^{\phantom{\dagger}}_{\sigma} + U d^{\dagger}_{\uparrow}d^{\dagger}_{\downarrow} d^{\phantom{\dagger}}_{\downarrow}d^{\phantom{\dagger}}_{\uparrow} 
\\\label{SIAM}
&+&\sum_{\mathbf{k},\sigma}  \left(V^{\phantom{\dagger}}_{\mathbf{k}}d^{\dagger}_{\sigma} c^{\phantom{\dagger}}_{\mathbf{k} \sigma} 
+ V^{*\phantom{\dagger}}_{\mathbf{k}} c^{\dagger}_{\mathbf{k} \sigma}d^{\phantom{\dagger}}_{\sigma}  \right).
\end{eqnarray}
If we discretize it in the energy representation employing the
formulae (\ref{discretization parameters flat DOS}) we obtain
\begin{eqnarray}
\nonumber
 H&=&\sum_{s=\pm}\sum_{n,\sigma} \epsilon_n^s c^{\dagger}_{n \sigma,s}c^{\phantom{\dagger}}_{n \sigma,s}
   + \epsilon_d\sum_{\sigma}d^{\dagger}_{\sigma}d^{\phantom{\dagger}}_{\sigma}
	 +  Ud^{\dagger}_{\uparrow}d^{\dagger}_{\downarrow}d^{\phantom{\dagger}}_{\downarrow}d^{\phantom{\dagger}}_{\uparrow}
\\ \label{discretization flat DOS}
	&+& \sum_{s=\pm}\sum_{n,\sigma} V \gamma_n^s \left(c^{\dagger}_{n \sigma,s} d^{\phantom{\dagger}}_{\sigma} + d^{\dagger}_{\sigma}c^{\phantom{\dagger}}_{n \sigma,s} \right) .
\end{eqnarray}

The Schrieffer-Wolff transformation to a Kondo-type model has been realized
already in the early days of CUTs \cite{kehre96b} and extended recently to
superconducting hosts \cite{zapal15}. Here, we  use a slightly 
different approach to eliminate the hybridization elements $V_{n}$.
We consider a discretized  flat density of states (DOS) (\ref{discretization flat DOS}) and 
choose the ground state of
\begin{equation}
\label{initial diagonal anderson hamiltonian for schrieffer wolff trafo}
 H_D = \sum_{n, \sigma} \epsilon_n c^{\dagger}_{n \sigma} c^{\phantom{\dagger}}_{n \sigma} + \epsilon_d \sum_{\sigma}n_{d,\sigma} + Un_{d,\uparrow}n_{d,\downarrow}
\end{equation}
as the reference state where $n_{d,\sigma}$ is the occupation operator of the
$d$-level
\begin{eqnarray}
 n_{d,\sigma} = d^{\dagger}_{\sigma}d^{\phantom{\dagger}}_{\sigma}.
\end{eqnarray}

In the present article, we restrict ourselves to the particle-hole symmetric cases.
Then, the singly occupied impurity state is the lowest-lying eigenstate of (\ref{initial diagonal anderson hamiltonian for schrieffer wolff trafo}) 
and thus there are two degenerate reference states with a spin degree of freedom at the impurity.
Hence we have to use a reference ensemble for the impurity operators and the normal-ordering scheme employed is defined by
\begin{eqnarray}
 \langle \uparrow | :\hat{A}: | \uparrow \rangle + \langle \downarrow | :\hat{A}: | \downarrow \rangle = 0.
\end{eqnarray}

In order to be able to use sign generators similar to \eqref{eq:sign_generator},
it must be evident which change of energy they induce. The sign of this energy change
determines the sign of the term in the generator. To this end, we introduce
an operator basis whose terms imply a unambiguous change of local energy on
the impurity. The chosen operator basis has already been used successfully before
in the derivation of generalized $t$-$J$ models from Hubbard models \cite{hamer10}.
Its terms are shown in Tab.\ \ref{tab:one}. 

\begin{table}[hbt]
\begin{tabular}{| c | c |}
\hline
bosonic operators & fermionic operators \\
\hline
$\mathbbm{1}\,\,\,$                                                        
& $F^{\phantom{\dagger}}_{1,\uparrow}=\left(1-n_{\text{d},\downarrow}\right)d^{\phantom{\dagger}}_{\uparrow}$ 
\\ 
$n_{z}=n_{\text{d},\uparrow}-n_{\text{d},\downarrow}\quad\quad$                             
& $F^{\phantom{\dagger}}_{1,\downarrow}=\left(1-n_{\text{d},\uparrow}\right)
d^{\phantom{\dagger}}_{\downarrow}$ 
\\ 
$d^{\dagger}_{\uparrow}d^{\phantom{\dagger}}_{\downarrow}$           
& $F^{\phantom{\dagger}}_{2,\uparrow}=n_{\text{d},\downarrow}d^{\phantom{\dagger}}_{\uparrow}\,\,\,\,\,\quad\quad$ 
\\
$d^{\dagger}_{\downarrow}d^{\phantom{\dagger}}_{\uparrow}$           
& $F^{\phantom{\dagger}}_{2,\downarrow}=n_{\text{d},\uparrow}d^{\phantom{\dagger}}_{\downarrow}\,\,\,\,\,\quad\quad$ 
\\ 
$d^{\phantom{\dagger}}_{\downarrow}d^{\phantom{\dagger}}_{\uparrow}$ 
& $F^{\dagger}_{2,\uparrow}=n_{\text{d},\downarrow}d^{\dagger}_{\uparrow}\,\,\,\,\,\quad\quad$  
\\
$d^{\dagger}_{\uparrow}d^{\dagger}_{\downarrow}$           
& $F^{\dagger}_{2,\downarrow}=n_{\text{d},\uparrow}d^{\dagger}_{\downarrow}\,\,\,\,\,\quad\quad$  
\\
$\bar{n}=n_{\text{d},\uparrow}+n_{\text{d},\downarrow}-\mathbbm{1}\,$                    
& $F^{\dagger}_{1,\uparrow}=\left(1-n_{\text{d},\downarrow}\right)d^{\dagger}_{\uparrow}$  
\\
$\hat{D}=2n_{\text{d},\uparrow}n_{\text{d},\downarrow}-\bar{n}\quad$                        
& $F^{\dagger}_{1,\downarrow}=\left(1-n_{\text{d},\uparrow}\right)d^{\dagger}_{\downarrow}$  
\\
\hline
\end{tabular}
\caption{Impurity operator basis with $n_{\text{d},\sigma} = d^{\dagger}_{\sigma} 
d_{\sigma}$.}
\label{tab:one}
\end{table}

The reason for this choice of operator basis becomes evident 
upon inspecting the local impurity configurations which are connected by these operators. 
The energy difference between the empty and the singly occupied state is different from the 
energy difference between the singly and the doubly occupied state.
Thus, there is no unique energy change induced by the operator $d^{\dagger}_{\sigma}$
because it connects the empty to the singly occupied state and the singly occupied to the doubly occupied state.

The projected operator $F^{\dagger}_{1,\sigma}=\left(1-n_{\bar{\sigma}}\right)d^{\dagger}_{\sigma}$ only connects the empty to the singly occupied state while the projected operator 
$F^{\dagger}_{2,\sigma}=n_{\bar{\sigma}}d^{\dagger}_{\sigma}$ only connects the singly occupied to
the doubly occupied state. Thus, there are unambiguous energy differences induced by the
projected operators reading
\begin{subequations}
\begin{eqnarray}
\label{Chapter 7 energy difference impurity operator}
 &&\Delta {E_1} = \epsilon_{\text{d}} = \tilde{\epsilon}_{\text{d}} - \tilde{U}
\\ 
 &&\Delta {E_2} = \epsilon_{\text{d}} + U = \tilde{\epsilon}_{\text{d}} + \tilde{U}
\end{eqnarray}
\end{subequations}
where the coefficients $\tilde{U}$ and $\tilde{\epsilon}_{\text{d}}$ are the coefficients of the Anderson impurity Hamiltonian (\ref{initial Hamiltonian Chapter 7}) 
expressed in the operator basis given in Tab.\ \ref{tab:one}.
The values of the coefficients are given in 
(\ref{starting values for the flow schrieffer-wolff transformation}).

The Anderson impurity Hamiltonian expressed in the projected impurity operator basis takes
the form
\begin{eqnarray} 
\label{initial Hamiltonian Chapter 7}
 H= H_D + H_R
\end{eqnarray}
with the diagonal part $H_D$ and the hybridization part $H_R$
\begin{subequations}
\begin{eqnarray} 
\label{H_D Anderson model old operator basis}
 H_D &=& \sum_{n, \sigma} \epsilon_n :c^{\dagger}_{n \sigma} c^{\phantom{\dagger}}_{n \sigma}: +
 \tilde{\epsilon}_d \bar{n} + \tilde{U}\hat{D}
\\ \nonumber
 H_R &=&\sum_{n, \sigma} V_n\left(F^{\dagger}_{1 \sigma}c^{\phantom{\dagger}}_{n \sigma} + 
c^{\dagger}_{n \sigma}F^{\phantom{\dagger}}_{1 \sigma}\right)
    \\ 
		&+& \sum_{n, \sigma} \Gamma_n\left(F^{\dagger}_{2 \sigma}c^{\phantom{\dagger}}_{n \sigma} + 
		c^{\dagger}_{n \sigma}F^{\phantom{\dagger}}_{2 \sigma}\right).
\end{eqnarray}
\end{subequations}
The coefficients in the projected operator basis are given by
\begin{eqnarray}
\label{starting values for the flow schrieffer-wolff transformation}
 V_n = V\gamma_n
\, , \,\,\,
 \Gamma_n = V\gamma_n
\, , \,\,\,
 \tilde{\epsilon}_d = \epsilon_d + \frac{U}{2}
\, , \,\,\,
 \tilde{U} = \frac{U}{2}
\end{eqnarray}
with the parameters $\epsilon_n$ and $\gamma_n$ from (\ref{discretized parameters flat DOS}).
The fermionic bath operators are still normal-ordered with respect to the Fermi sea.


\subsection{Elimination of the hybridization}

We want to eliminate the hybridization elements and analyze the spin-spin interaction
induced thereby. This amounts up to the Schrieffer-Wolff transformation
realized by CUTs \cite{kehre96b} or the systematic derivation of 
$t$-$J$ models from Hubbard models by CUTs \cite{hamer10}. 
We choose the generator
\begin{eqnarray}
\nonumber 
 \eta &=&\sum_{n, \sigma} \eta^{V}_n\left(F^{\dagger}_{1 \sigma}c^{\phantom{\dagger}}_{n \sigma} - c^{\dagger}_{n \sigma}F^{\phantom{\dagger}}_{1 \sigma}\right)
      \\
			\nonumber 
			&+&\sum_{n, \sigma} \eta^{\Gamma}_n\left(F^{\dagger}_{2 \sigma}c^{\phantom{\dagger}}_{n \sigma} - c^{\dagger}_{n \sigma}F^{\phantom{\dagger}}_{2 \sigma}\right)
      \\ 
			\label{Chapter 7 generator Schrieffer-Wolff}
			&+&\sum_{n, m, \sigma}
			\eta^{t}_{nm}:c^{\dagger}_{n \sigma} c^{\phantom{\dagger}}_{m \sigma}:.
\end{eqnarray}

For the flow equation (\ref{flow equation}) we commute the generator (\ref{Chapter 7 generator Schrieffer-Wolff}) with the Hamiltonian (\ref{initial Hamiltonian Chapter 7})
which generates terms not present in the initial Hamiltonian reading
\begin{subequations}
\begin{eqnarray}
\label{new emerging terms schrieffer-wolff trafo with CUT}
H_{t} &=& \sum_{n,m,\sigma} t_{nm}:c^{\dagger}_{n \sigma} c^{\phantom{\dagger}}_{m \sigma}:
\\\nonumber  
H_{J} &=&
 \sum_{n,m,\sigma}J^{\uparrow\downarrow}_{nm}\,d^{\dagger}_{\sigma}d^{\phantom{\dagger}}_{\bar{\sigma}}c^{\dagger}_{n\bar{\sigma}}c^{\phantom{\dagger}}_{m\sigma}
+\sum_{n,m,\sigma}J^{n_z}_{nm\sigma}\,\,n_z:c^{\dagger}_{n \sigma} c^{\phantom{\dagger}}_{m \sigma}:
\\ \nonumber
&+&\sum_{n,m,\sigma}J^{\bar{n}}_{nm}\,\,\bar{n}:c^{\dagger}_{n \sigma} c^{\phantom{\dagger}}_{m \sigma}:
\\ 
&+&\sum_{n,m}J^{\pm}_{nm}
\left(d^{\dagger}_{\uparrow}d^{\dagger}_{\downarrow}c^{\phantom{\dagger}}_{n\downarrow}
c^{\phantom{\dagger}}_{m\uparrow}+
c^{\dagger}_{m\uparrow}c^{\dagger}_{n\downarrow}d^{\phantom{\dagger}}_{\downarrow}
d^{\phantom{\dagger}}_{\uparrow}\right).
\end{eqnarray}
\end{subequations}
All these emerging terms are of order $V^2$ and coincide with terms in the arising 
in the standard Schrieffer-Wolff transformation \cite{schri66}.
We aim at computing  the couplings $J^{(i)}_{nm}$ from 
(\ref{new emerging terms schrieffer-wolff trafo with CUT}) in order $V^2$
and thus the commutators 
\begin{eqnarray}
\label{relevant commutators schrieffer-wolff}
\left[\eta_R,H_D + H_R\right]
\quad \text{and} \quad
\left[\eta_t,H_D\right]
\end{eqnarray}
are needed; all other commutations yield terms of order $V^3$ or higher.

Aiming at $H_J$ from (\ref{new emerging terms schrieffer-wolff trafo with CUT}) in order $V^2$ 
we neglect all terms which contribute to $H_J$ in order $V^3$.
We point out that this implies also to neglect terms in order  $V^2$ which are not 
of the form $H_J$ and will influence $H_J$ only in order  $V^3$ or higher.
This argument applies to  the flow of $H_D$ and the emerging hopping terms $t_{nm}$
because they lead to corrections to $H_J$ of order $V^3$ and higher only. 
In this way, the flow equations simplify to
\begin{subequations}
\label{flow equation schrieffer-wolff transformation sign generator simplified}
\begin{eqnarray}
\partial_l V_n &=& \eta^{V}_{n}\left(\epsilon_{n}-\tilde{\epsilon}_d+\tilde{U}\right) 
\phantom{\sum_{n}}
 \\
 \partial_l \Gamma_n &=& \eta^{\Gamma}_{n}\left(\epsilon_{n}-\tilde{\epsilon}_d-\tilde{U}\right) \phantom{\sum_{n}}
\\ 
\partial_l J^{\uparrow\downarrow}_{nm}&=&\eta^{\Gamma}_{n}\Gamma_{m}+\eta^{\Gamma}_{m}\Gamma_{n}- \eta^{V}_{n}V_{m}-\eta^{V}_{m}V_{n}\phantom{\sum_{n}}
\\ 
\partial_l J^{\pm}_{nm}&=&\eta^{\Gamma}_{n}V_{m}+\eta^{\Gamma}_{m}V_{n} -\eta^{V}_{n}\Gamma_{m}-\eta^{V}_{m}\Gamma_{n}
\end{eqnarray}
\end{subequations}
where only $J^{\uparrow\downarrow}_{nm}$ has to be known due to spin-rotation symmetry
\begin{eqnarray}
\label{spin rotation symmetry}
 \sigma J^{n_z}_{nm\sigma} &=& \frac{1}{2}J^{\uparrow\downarrow}_{nm}. 
\end{eqnarray}

The operator $F^{\dagger}_{1\sigma}c^{\phantom{\dagger}}_{n \sigma}$ promotes the empty impurity state to the singly occupied one while annihilating a particle with energy $\epsilon_n$ in the bath. This leads to a change of energy 
\begin{eqnarray}
 \Delta E_{1,n} = \tilde{\epsilon}_d - \tilde{U} - \epsilon_n.
\end{eqnarray}
The operator $F^{\dagger}_{2\sigma}c^{\phantom{\dagger}}_{n \sigma}$ promotes the singly occupied to the doubly occupied impurity level and annihilates a particle with energy $\epsilon_n$ in the bath. This implies a change of energy 
\begin{eqnarray}
 \Delta E_{2,n} =  \tilde{\epsilon}_d + \tilde{U} - \epsilon_n.
\end{eqnarray}
Thus, the sign generator takes the form
\begin{subequations}
\begin{eqnarray}
 \eta^{V}_{n} &=& -\text{sgn}\left(\epsilon_n - \tilde{\epsilon}_d + \tilde{U}\right)V_n
\\ 
 \eta^{\Gamma}_{n} &=& -\text{sgn}\left(\epsilon_n - \tilde{\epsilon}_d - \tilde{U}\right)\Gamma_n.
\end{eqnarray}
\end{subequations}


\subsection{Diagonalization of the induced spin-spin interaction}

In addition, we want to diagonalize the induced spin-spin interaction at the same time as it is generated upon eliminating the hybridization.
For this reason, we add the following terms to the generator
\begin{eqnarray}
\nonumber
\eta_{J} &=&
 \sum_{n,m,\sigma}\eta^{\uparrow\downarrow}_{nm}d^{\dagger}_{\sigma}
d^{\phantom{\dagger}}_{\bar{\sigma}}c^{\dagger}_{n\bar{\sigma}}
c^{\phantom{\dagger}}_{m\sigma}
+\sum_{n,m,\sigma}\eta^{n_z}_{nm\sigma}n_z:c^{\dagger}_{n \sigma} 
c^{\phantom{\dagger}}_{m \sigma}:
\\ \nonumber
&+&\sum_{n,m}\eta^{\pm}_{nm}
\left(d^{\dagger}_{\uparrow}d^{\dagger}_{\downarrow}
c^{\phantom{\dagger}}_{n\downarrow}
c^{\phantom{\dagger}}_{m\uparrow}-
c^{\dagger}_{m\uparrow}c^{\dagger}_{n\downarrow}
d^{\phantom{\dagger}}_{\downarrow}d^{\phantom{\dagger}}_{\uparrow}\right)
\\ \label{generator induced spin-spin interaction}
&+&\sum_{n,m,\sigma}\eta^{\bar{n}}_{nm}\bar{n}:c^{\dagger}_{n \sigma} 
c^{\phantom{\dagger}}_{m \sigma}:.
\end{eqnarray}
These terms lead in the sign generator to terms of the same kind with the
prefactors
\begin{subequations}
\begin{eqnarray}
\eta^{\uparrow\downarrow}_{nm} &=& \text{sgn}\left(\epsilon_n - \epsilon_m\right) J^{\uparrow\downarrow}_{nm}
\\ 
\eta^{n_z}_{nm} &=& \text{sgn}\left(\epsilon_n - \epsilon_m\right) J^{n_z}_{nm}
\\ 
\eta^{\bar{n}}_{nm\sigma} &=& \text{sgn}\left(\epsilon_n - \epsilon_m\right) J^{\bar{n}}_{nm}
\\ 
\eta^{\pm}_{nm} &=& -\text{sgn}\left(\epsilon_n + \epsilon_m\right) J^{\pm}_{nm}.
\end{eqnarray}
\end{subequations}

\begin{figure}
\includegraphics[width=\columnwidth]{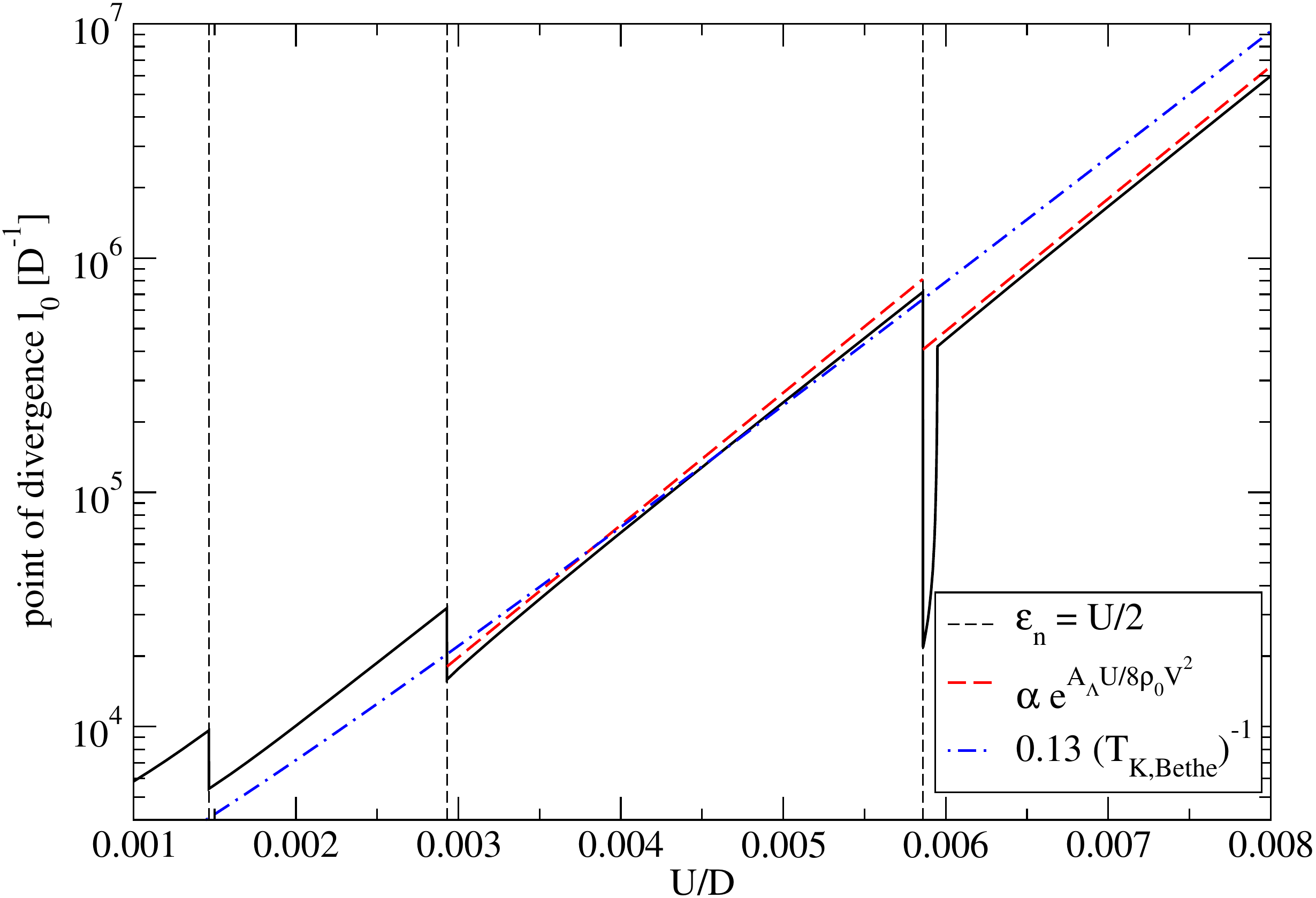}
\caption{(Color online) Flow parameter $l_{0}$ at which the flow equation diverges (compare Fig. \ref{fig:ROD_Kondo_no_change_of_ref_state}) for the Anderson impurity model with 
\mbox{$N=60$}, \mbox{$\Lambda=2$} and \mbox{$V/D=0.01414$} in a logarithmic plot vs. $U/D$. The inverse energy $l_{0}$ shows the generic exponential behavior
\mbox{$l_{0}\propto\exp\left(A_{\Lambda}U/8\rho_{0}V^2\right)$} of the Kondo temperature $T_{K}$ of the Anderson impurity model. The factor $A_{\Lambda}$ is given by 
(\ref{discretization effect on Tk}).
The dashed-dotted curve (blue) depicts the Bethe ansatz result \eqref{eq:bethe_aim}.}
The discontinuous behavior is a result of the discretization and happens every time when $U/2$ crosses an energy level $\epsilon_n$ (dotted vertical lines). 
For decreasing $\Lambda$ the
discontinuities become smaller (compare Fig. \ref{fig:Tk_Anderson_N200_L12}).
\label{fig:Tk_Anderson_N60_L2}
\end{figure}

We only track terms that act in lowest order $J^2$ on the spin-spin interaction
neglecting higher order contributions in $J$.  Hence the commutators 
$\left[\eta_{J}, H_D + H_J\right]$ are needed. Among the resulting terms only the terms 
are kept that act on $H_J$. All other terms are neglected because their feedback on the spin-spin interaction is at least of order $J^3$.
Calculating the commutators and comparing the coefficients in the flow equation 
(\ref{flow equation}) yields additional terms to the flow equation 
(\ref{flow equation schrieffer-wolff transformation sign generator simplified})
so that one arrives at
\begin{subequations}
\label{DEQ Anderson poor mans scaling with CUT}
\begin{eqnarray}
\partial_l J^{n_z}_{nm\sigma}
&=&\left(\epsilon_{m}-\epsilon_{n}\right)\eta^{n_z}_{nm\sigma}
\nonumber
\\
&-&\frac{1}{2}\sum_{x}\sigma\left(\eta^{\uparrow\downarrow}_{nx}J^{\uparrow\downarrow}_{xm}-
\eta^{\uparrow\downarrow}_{xm}J^{\uparrow\downarrow}_{nx}\right)\left(1-2\theta_{x}\right)
\quad
\\ 
 \partial_l J^{\uparrow\downarrow}_{nm}
&=&\left(\epsilon_{m}-\epsilon_{n}\right)\eta^{\uparrow\downarrow}_{nm}
\nonumber
\\
&+&\sum_{x}\sigma \left(\eta^{n_z}_{xm\sigma}J^{\uparrow\downarrow}_{nx}
- \eta^{n_z}_{nx\sigma}J^{\uparrow\downarrow}_{xm}\right)\left(1-2\theta_{x}\right)
\nonumber
\\
&+&\sum_{x}\sigma \left(\eta^{\uparrow\downarrow}_{xm} J^{n_z}_{nx\sigma}
- \eta^{\uparrow\downarrow}_{nx} J^{n_z}_{xm\sigma}\right)\left(1-2\theta_{x}\right)
\quad
\\ 
\partial_l J^{\bar{n}}_{nm}
&=&\left(\epsilon_{m}-\epsilon_{n}\right)\eta^{\bar{n}}_{nm}
\nonumber
\\
&-&\frac{1}{2}\sum_{x}\left(\eta^{\pm}_{xn}J^{\pm}_{xm}+\eta^{\pm}_{xm}J^{\pm}_{xn}\right) 
\left(1-2\theta_{x}\right)
\\ 
\nonumber
 \partial_l J^{\pm}_{nm} 
&=& \left(\epsilon_{n}+\epsilon_{m}\right)\eta^{\pm}_{nm}
\\ \nonumber
&+&\sum_{x}\left(\eta^{\bar{n}}_{xm\uparrow}  J^{\pm}_{nx}+\eta^{\bar{n}}_{xn\downarrow}
J^{\pm}_{xm}\right)\left(1-2\theta_{x}\right)
\\
&-&\sum_{x}\left(\eta^{\pm}_{nx} J^{\bar{n}}_{xm\uparrow}+\eta^{\pm}_{xm} 
J^{\bar{n}}_{xn\downarrow}\right)\left(1-2\theta_{x}\right)
\end{eqnarray}
\end{subequations}
where $\sigma$ labels the spin if it is used as index
while it takes the values $\sigma=\pm 1$ as a coefficient. The occupation number
\begin{eqnarray}
\label{expectation value normal-ordering}
 \theta_x = \langle c^{\dagger}_{x \sigma} c^{\phantom{\dagger}}_{x \sigma} \rangle
\end{eqnarray}
is calculated with respect to the Fermi sea and results from the normal-ordering of the fermionic bath operators.

\begin{figure}
\includegraphics[width=\columnwidth]{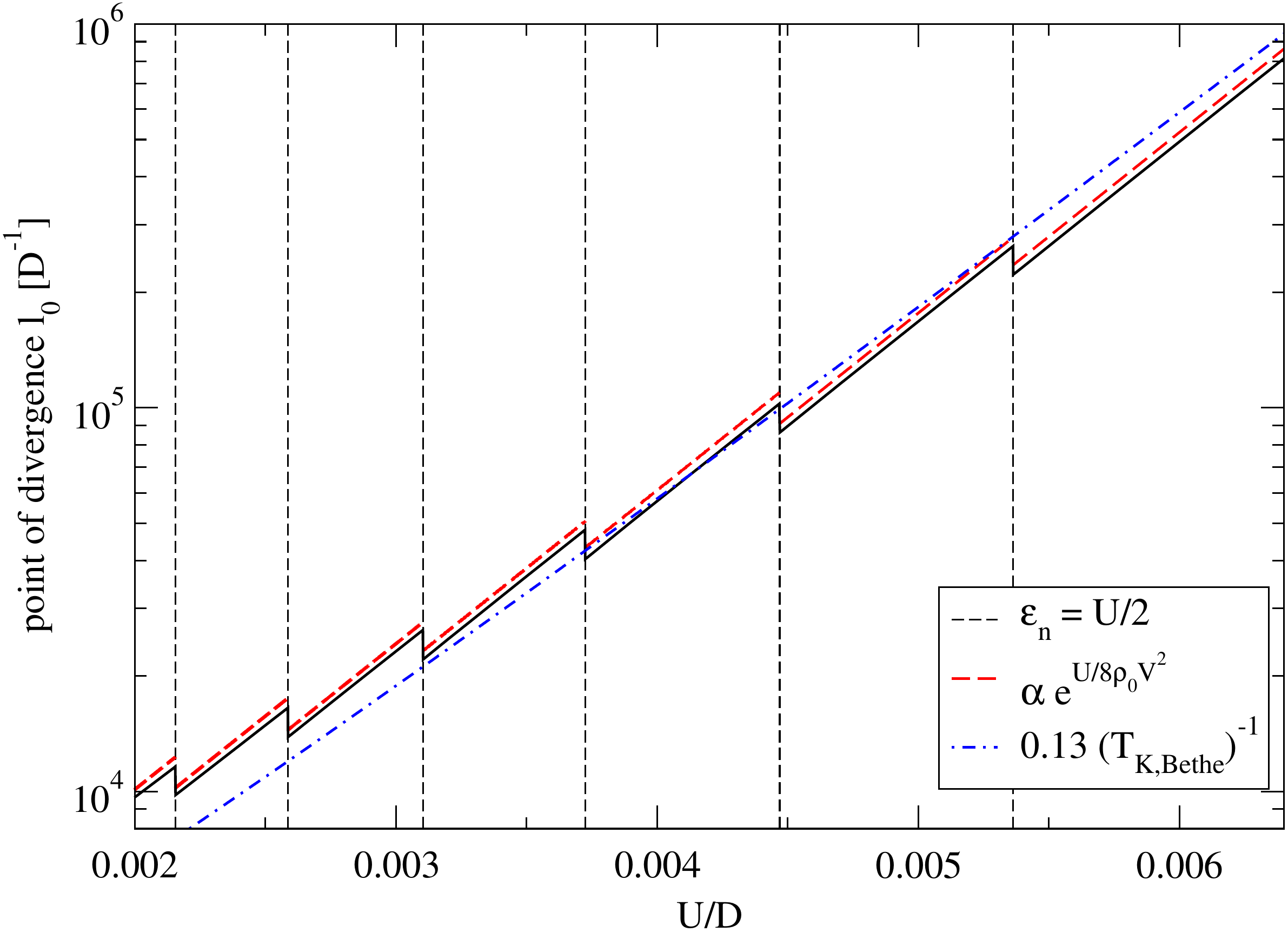}
\caption{(Color online) Flow parameter $l_{0}$ at which the flow equation diverges 
(cf.\ Fig.\ \ref{fig:ROD_Kondo_no_change_of_ref_state}) 
for the Anderson impurity model with \mbox{$N=200$}, \mbox{$\Lambda=1.2$} and \mbox{$V/D=0.01414$} in a logarithmic plot vs. $U/D$. The inverse energy $l_{0}$ 
displays the generic exponential character 
\mbox{$l_{0}\propto\exp\left(A_{\Lambda}U/8\rho_{0}V^2\right)$} 
of the Kondo temperature $T_{K}$ of the Anderson impurity model. 
The discretization factor $A_{\Lambda}$ is given by 
(\ref{discretization effect on Tk}). The dashed-dotted curve (blue)
depicts the Bethe ansatz result \eqref{eq:bethe_aim}.
The discontinuous behavior is a result of the discretization and occurs each time when 
$U/2$ crosses an energy level $\epsilon_n$ 
(dotted vertical lines).}
\label{fig:Tk_Anderson_N200_L12}
\end{figure}

Taking a closer look at the flow equation (\ref{DEQ Anderson poor mans scaling with CUT}) reveals that $J^{n_z}$ and $J^{\uparrow\downarrow}$ only influence each other. They do not couple to 
$J^{\pm}$ or $J^{\bar{n}}$ which also only influence each other.
The spin-rotation symmetry
\begin{eqnarray}
\label{Chapter 7 spin-rotation symmetry}
 \sigma J^{n_z}_{nm\sigma} = \frac{1}{2}J^{\uparrow\downarrow}_{nm}
\end{eqnarray}
holds true during the whole flow which simplifies the flow equation for 
$J^{\uparrow\downarrow}_{nm}$ to
\begin{eqnarray}
\nonumber
\partial_l J^{\uparrow\downarrow}_{nm}
=&-&\left(\epsilon_{n}-\epsilon_{m}\right)\eta^{\uparrow\downarrow}_{nm}
\\ 
&-& \sum_{x}\left(\eta^{\uparrow\downarrow}_{nx}J^{\uparrow\downarrow}_{xm} - \eta^{\uparrow\downarrow}_{xm}J^{\uparrow\downarrow}_{nx}\right)\left(1-2\theta_{x}\right).
\label{DEQ poor mans scaling with CUT Anderson}
\end{eqnarray}
This differential equation is the same as Eq.\ (\ref{DEQ poor mans scaling with CUT Kondo}) which is the flow equation for the diagonalization of the spin-spin interactions in the Kondo model. 
Recall that in the Anderson impurity model we aim at eliminating the charge fluctuations induced by the hybridization $V_{nm}$ and diagonalizing the induced spin-spin interaction $J_{nm}$ 
simultaneously. We emphasize that this is not in one-to-one correspondence
to  applying a Schrieffer-Wolff transformation first and then diagonalizing the effective Kondo Hamiltonian.

\begin{figure}
\includegraphics[width=\columnwidth]{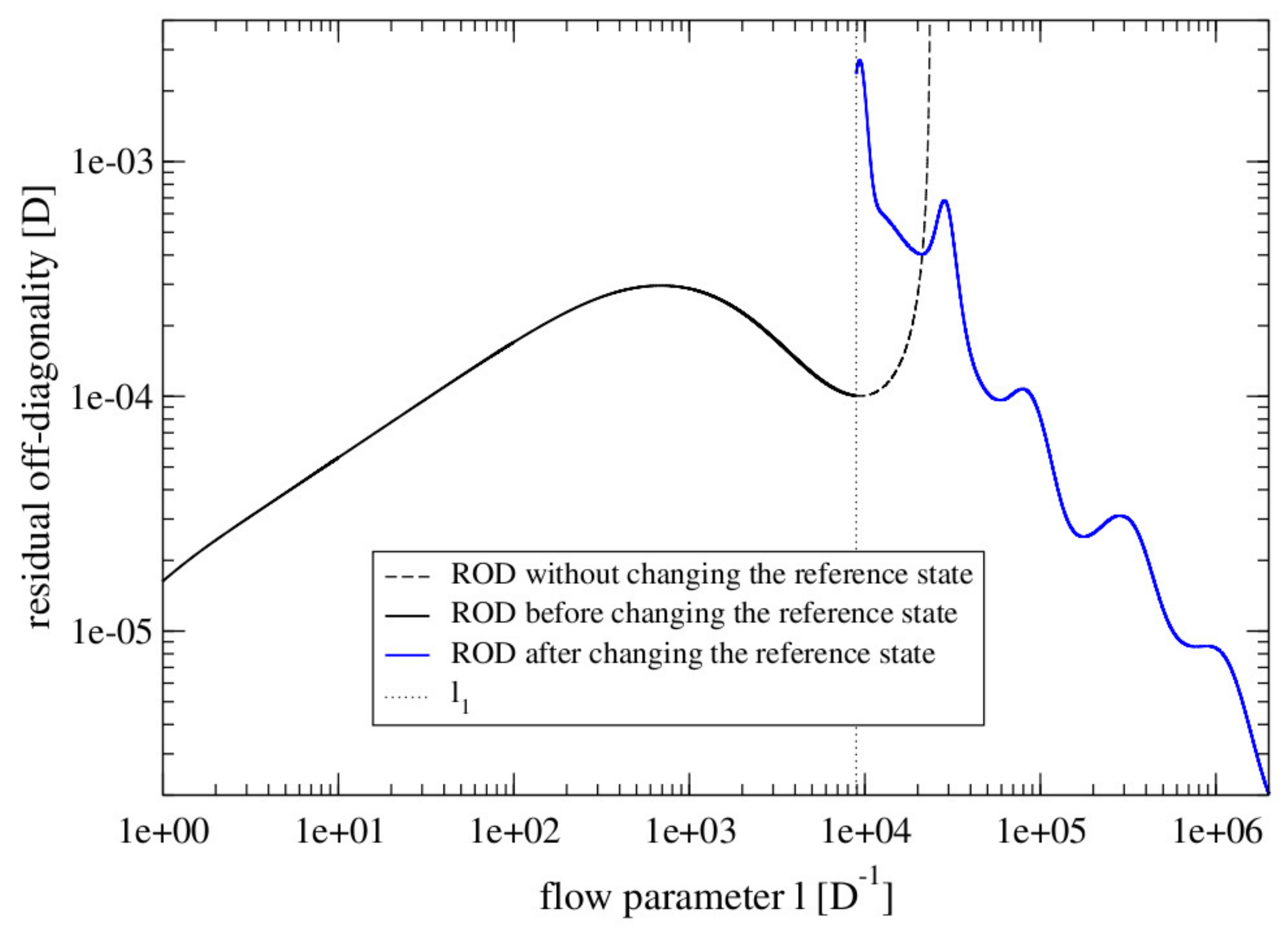}
\caption{(Color online) ROD for the Anderson impurity model with 
\mbox{$N=52$}, \mbox{$\Lambda=2$}, \mbox{$U/D=3.2\cdot 10^{-3}$} , and \mbox{$V/D=0.01414$}. 
The first part of the flow only shows the ROD of (\ref{DEQ poor mans scaling with CUT Anderson}) for the induced spin-spin interaction $J_{nm}^{\uparrow\downarrow}$ from (\ref{new emerging terms schrieffer-wolff trafo with CUT}). The diagonalization of the induced spin-spin interaction leads to the divergence depicted as dashed black line.
After changing the reference state we solve the flow
(\ref{flow_kondo_modified}) which turns out to converge in contrast to the original flow 
(\ref{DEQ poor mans scaling with CUT Anderson}).}
\label{fig:ROD_Anderson_N52_L2_V01414_U12em3}
\end{figure}

In order to determine the Kondo energy scale of the Anderson impurity 
Hamiltonian with CUTs we combine 
(\ref{flow equation schrieffer-wolff transformation sign generator simplified}),
and (\ref{DEQ Anderson poor mans scaling with CUT}). 
This flow equation also leads to divergence on an energy scale 
that depends on the parameters $U$ and $V$. 
Fig.\ \ref{fig:Tk_Anderson_N60_L2} shows the point of divergence $l_0$ as function of
the interaction $U$. 
There are certain values of $U$ for which discontinuous jumps occur. In the intervals between
the discontinuities we find the correct exponential behavior of the Kondo temperature 
$T_{k}^{-1}\propto\exp(A_{\Lambda}\frac{U}{8\rho_0V^2})$ where $A_{\Lambda}$ from 
(\ref{discretization effect on Tk}) captures the
influence of the discretization on the Kondo temperature, cf.\ Ref.\ \onlinecite{krish80a}. 
In order to show that the overall behavior is indeed the correct one, Fig.\
\ref{fig:Tk_Anderson_N60_L2} also depicts the Bethe ansatz result \cite{tsvel83,hewso93}
\begin{equation}
\label{eq:bethe_aim}
T_{\rm K, Bethe} = U \sqrt{\frac{\Delta}{2A_\Lambda U}} 
\exp\left(-\frac{\pi A_\Lambda U}{8\Delta}+\frac{\pi\Delta}{2A_\Lambda U}\right)
\end{equation}
where we use the hybridization $\Delta=\pi V^2/(2D)$ as usual shorthand. Moreover
we introduced the discretization factor $A_\Lambda$ whereever the ratio $U/\Delta$ occurs.

The origin of the discontinuities is the discretization of the bare energy levels. 
Each time the interaction $\frac{U}{2}$ 
crosses an energy level $\epsilon_n$, one sign in the generator 
\begin{subequations}
\label{generator turn away hybridization Schrieffer-Wolff}
\begin{eqnarray}
 \eta_{n}^{V} &=& -\text{sgn}\left(\epsilon_n + U/2\right)V_n
\\ 
 \eta_{n}^{\Gamma} &=& -\text{sgn}\left(\epsilon_n - U/2\right)\Gamma_n
\end{eqnarray}
\end{subequations}
is changed discontinuously implying a discontinuity in all other quantities as well.

In Figs.\ \ref{fig:Tk_Anderson_N60_L2} and \ref{fig:Tk_Anderson_N200_L12} the dashed vertical lines show the values of the interaction where $\frac{U}{2}=\epsilon_n$. One clearly sees that 
the discontinuities  occur indeed exactly when
$\frac{U}{2}$ crosses an energy level $\epsilon_n$. 
In Fig.\ \ref{fig:Tk_Anderson_N200_L12} the discretization parameter $\Lambda$ is decreased and thus more energy levels lie in the considered interval. As a result more discontinuities
occur, but the weight $\left|\gamma_n\right|^2$ 
carried by the respective energy levels decreases 
so that the induced jumps become smaller. Thus, in the limit of $\Lambda\to 1$ the curve would not
display jumps anymore.

Summarizing this section, we succeeded in eliminating the hybridization in the Anderson impurity model by means of a continuous unitary transformation and the thus induced spin-spin interaction
until a small energy scale (large values of the flow parameter $l$) where the flow
diverges. This energy scale turns out to be the Kondo energy scale $T_K$, capturing the 
correct exponential behavior in $U$
\begin{eqnarray}
T_K = l_0^{-1} = C\left(U\right)\exp\left(-A_{\Lambda} \frac{U}{8\rho_0V^2}
\right)
\end{eqnarray}
where $C\left(U\right)$ stems from the discretization and describes the discontinuous behavior observed in Figs.\ \ref{fig:Tk_Anderson_N60_L2} and \ref{fig:Tk_Anderson_N200_L12}; 
$C(U)$  is constant in each interval between two discontinuities. The factor $A_{\Lambda}$ 
captures the discretization corrections in the exponent \cite{krish80a}.
To our knowledge, the correct exponential scale has not yet been found by a CUT so far.
Still, we do not obtain a finite effective model, but a divergent flow.


\section{Modified approach to the Anderson impurity model}

\begin{figure}
\includegraphics[width=\columnwidth]{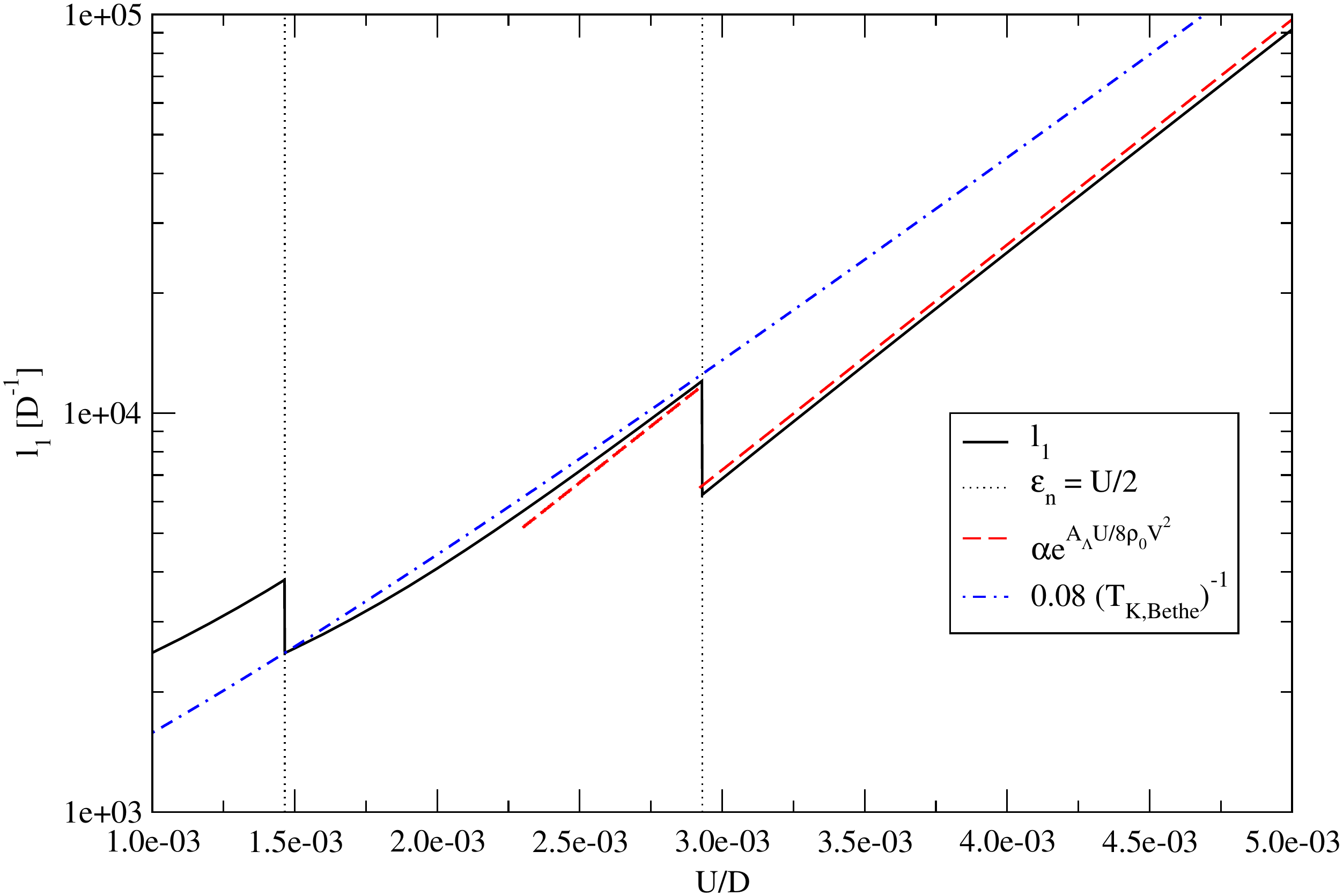}
\caption{(Color online) Inverse energy scale $l_1$ at which the reference state is changed for the Anderson impurity model with 
\mbox{$N=52$}, \mbox{$\Lambda=2$} and 
\mbox{$V/D=0.01414$}.
The inverse energy scale $l_1$ is increasing proportional to the inverse of the Kondo temperature \mbox{$T_K^{-1}\propto\exp\left(A_{\Lambda}U/8\rho_{0}V^2\right)$}. 
The dashed-dotted curve (blue) depicts the Bethe ansatz result \eqref{eq:bethe_aim}.
The factor $A_{\Lambda}$ from (\ref{discretization effect on Tk}) takes the discretization into account. The discontinuities stem also from the discretization of the energy. }
\label{fig:energy_scale_change_of_ref_state_N52_L2_V01414_l0_Anderson_alternative}
\end{figure}

Here, we apply the modified approach to the Anderson impurity model. First we follow the procedure of the last section and start from the Anderson impurity Hamiltonian in the form 
(\ref{initial Hamiltonian Chapter 7}). The hybridization is eliminated 
by using (\ref{flow equation schrieffer-wolff transformation sign generator simplified}). Simultaneously, the thus induced spin-spin interaction is diagonalized by using 
(\ref{DEQ poor mans scaling with CUT Anderson}). For small values of the flow parameter $l$ 
the spin-spin exchange couplings $J_{nm}^{\uparrow\downarrow}$ are generated in the process of
the elimination of the hybridization elements.

Beyond some value of $l$ the hybridization elements become negligible. Then the backaction of the induced spin-spin interaction on itself in (\ref{DEQ poor mans scaling with CUT Anderson}) 
is the driving effect in the flow equation which leads to diverging couplings (dashed black line in Fig.\ \ref{fig:ROD_Anderson_N52_L2_V01414_U12em3}).
At an even larger value $l_1$ the formation of a singlet state with the impurity becomes energetically favorable. This is determined just as in the case of the Kondo model referring
to the coupling $J_{nm}^{\uparrow\downarrow}$ instead of $J_{nm}$. To this end, we
consider only the singly occupied impurity state because $l_1$ is much larger than $1/U$
so that the charge fluctuations on the impurity will not play any significant role 
at this stage of the flow. Hence the couplings $J^{\pm}_{nm}$ and $J^{\bar{n}}_{nm}$ 
do not need to be considered beyond $l_1$.
Using the spin-rotation symmetry 
(\ref{Chapter 7 spin-rotation symmetry}) for $J^{n_z}_{nm}$ and 
$J^{\uparrow\downarrow}_{nm}$ leads to an effective Kondo Hamiltonian 
with the couplings $J_{nm}=J^{\uparrow\downarrow}_{nm}$. Subsequently, 
we can follow the same procedure as in the Kondo model, i.e., we change the reference state
to a mixture of singlets between the impurity spin and bath fermions at the Fermi level
and we expand the effective Kondo Hamiltonian
in the modified operator basis (\ref{used_states_adapted_operator_basis}) and 
use the modified flow equation \eqref{flow_kondo_modified}.

Fig.\ \ref{fig:ROD_Anderson_N52_L2_V01414_U12em3} shows the ROD of the Anderson impurity model obtained with this modified approach. We only plot the ROD of the spin-spin couplings
\begin{eqnarray}
\label{ROD for DEQ without change of ref state Anderson model}
\text{ROD}^2 = \sum_{\underset{n\neq m}{n,m}} \left|J^{\uparrow\downarrow}_{nm}\right|^2.
\end{eqnarray}
First, the ROD is increasing because the spin-spin coupling is generated by the elimination of the hybridization. Without changing the reference state the flow equation 
(\ref{DEQ poor mans scaling with CUT Anderson}) leads to divergence, see dashed black line. 
At some point $l_1$ before the divergence occurs the  singlet formation becomes 
energetically favorable, see Eq.\ (\ref{ratio at which the singlet forms}), 
and the reference state is changed. 

The ROD shows a discontinuous behavior when the reference state is changed because we use the modified generator which includes additional terms. The reason why
the ROD is abruptly increasing is the same as in the case of the Kondo model, see Sect.\ 
\ref{ssec:modified_flow}. 
In contrast to the flow without change of reference state, the modified flow equation converges
and leads to a finite, well-defined effective Hamiltonian. 
Next, we analyze the same energy scales as for the Kondo model.


\subsection{Effective model for the Anderson impurity model}

Fig.\ \ref{fig:energy_scale_change_of_ref_state_N52_L2_V01414_l0_Anderson_alternative} shows the flow parameter $l_1$ at which the reference state is changed.
We again find discontinuities for the same reason as they occurred in Fig.\ 
\ref{fig:Tk_Anderson_N60_L2}. 
Between two discontinuities we find the exponential behavior characteristic 
of the Kondo energy scale in the Anderson impurity model
\begin{eqnarray}
l_{1}^{-1} \propto \exp\left(-A_{\Lambda} \frac{U}{8\rho_0V^2}\right).
\end{eqnarray}
Thus, for the Anderson impurity model the point where the reference state is changed is also given by the Kondo temperature $T_{K}$.


\subsection{Binding energy of the Kondo singlet}

\begin{figure}
\includegraphics[width=\columnwidth]{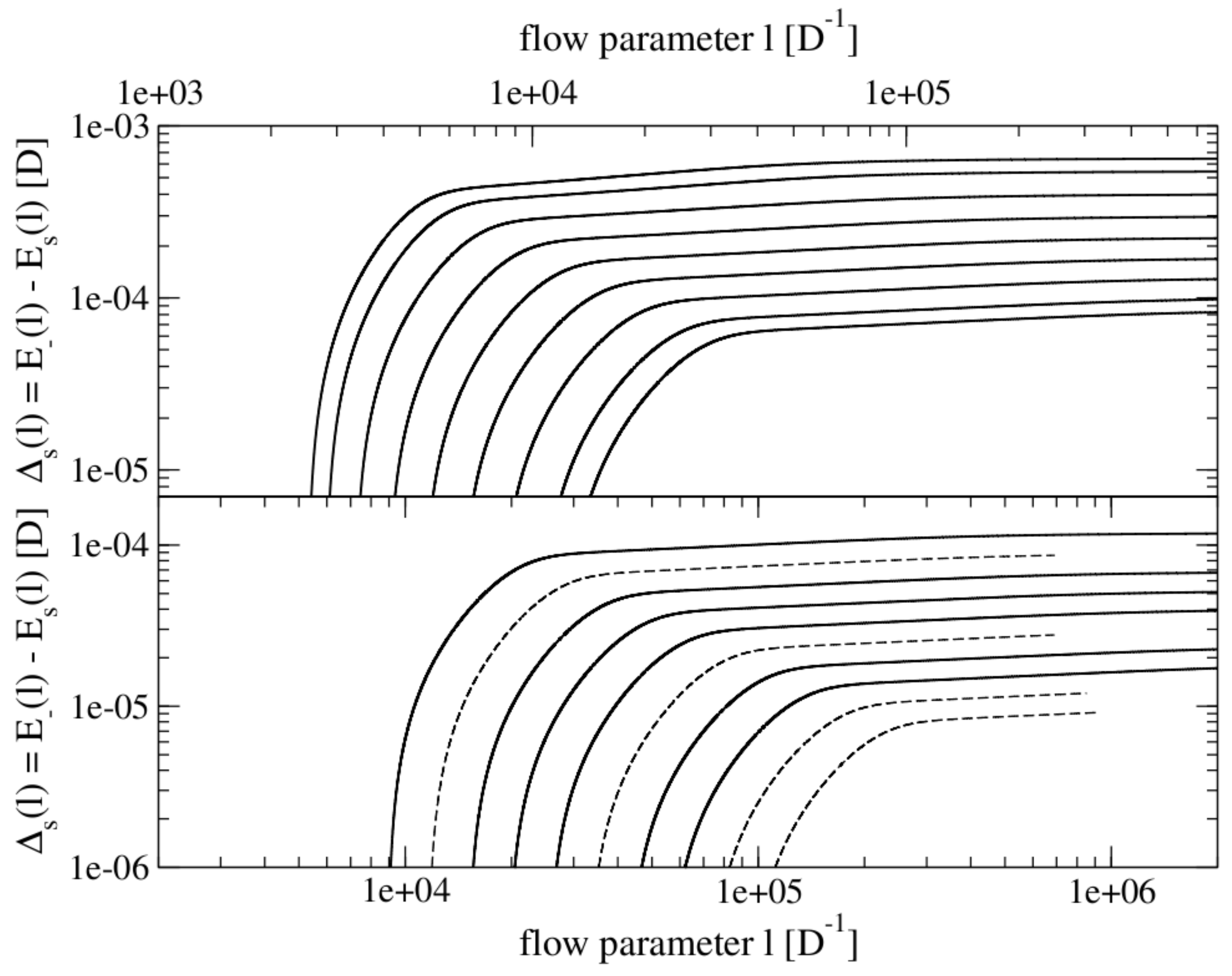}
\caption{(Color online) Flow of \mbox{$\Delta_{s}(l) = E_{-}(l) - E_{s}(l)$} where 
$E_{-}$ and $E_{s}$ result from 
(\ref{Chapter 7 effective Hamiltonian without interaction part energy levels}) and 
(\ref{change of reference state full model eigenvalues}) for 
the Anderson impurity model with \mbox{$N=52$}, \mbox{$\Lambda=2$} and \mbox{$V/D=0.01414$} and
various values of the interaction.
>From top to bottom: (upper panel) 
\mbox{$U/D\cdot10^{3}=1.464$}, $1.6$, $1.8$, $2$, $2.2$, $2.4$, $2.6$, $2.8$, $2.929$, 
(lower panel) 
\mbox{$U/D\cdot10^{3}=3.2, 3.4, 3.6, 3.8, 4, 4.2, 4.4, 4.6, 4.8, 5$}. 
The energy difference $\Delta_{s}$ converges to the binding energy of the Kondo singlet 
for \mbox{$l\rightarrow\infty$}. 
The flow for the different parameters is split into two plots
for clarity because each time when $U/2$ crosses $\epsilon_{n}$ $l_0$ jumps, cf.\ Fig.\ 
\ref{fig:E_singlett_m_E_FS_N52_L2_V01414_Anderson_vs_J}.}
\label{fig:E_singlett_m_E_FS_N52_L2_V01414_Anderson_U2to3}
\end{figure}

In Fig.\ \ref{fig:E_singlett_m_E_FS_N52_L2_V01414_Anderson_U2to3} the flow of 
$\Delta_s$ from (\ref{Delta_s}) is displayed. Discontinuities occur whenever $U/2$
an energy level $\epsilon_n$. 
For clarity the different regions between two consecutive  values of 
$\epsilon_n$ are depicted in two panels in 
Fig.\ \ref{fig:E_singlett_m_E_FS_N52_L2_V01414_Anderson_U2to3}. 
We find that $\Delta_s$ increases quickly and converges to the binding energy of the singlet
$\Delta_s\left(\infty\right)$.

\begin{figure}
\includegraphics[width=\columnwidth]{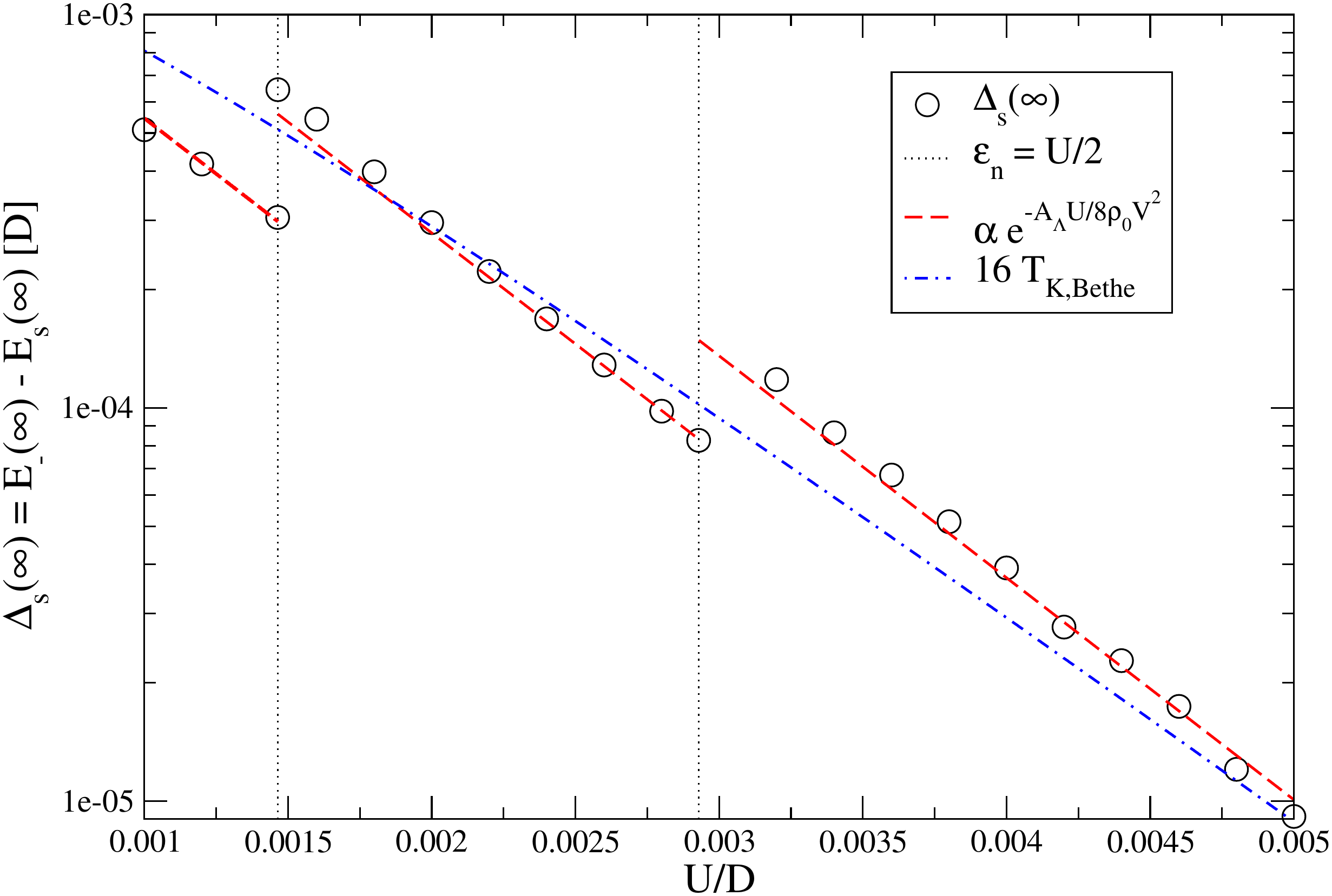}
\caption{(Color online) Binding energy of the singlet 
\mbox{$\Delta_{s}\left(\infty\right) = E_{-}\left(\infty\right) - E_{s}\left(\infty\right)$} with $E_{-}$ and $E_{s}$ from 
(\ref{Chapter 7 effective Hamiltonian without interaction part energy levels}) and (\ref{change of reference state full model eigenvalues}) 
for the Anderson impurity model with \mbox{$N=52$}, \mbox{$\Lambda=2$} and 
\mbox{$V/D=0.01414$}. 
The binding energy shows the exponential behavior 
\mbox{$\Delta_{s}\left(\infty\right)\propto
\exp\left(-A_{\Lambda}U/8\rho_{0}V^2\right)$}. 
The dashed-dotted curve (blue) depicts the Bethe ansatz result \eqref{eq:bethe_aim}.
The factor $A_{\Lambda}$ from (\ref{discretization effect on Tk}) takes the effects of  discretization into account.}
\label{fig:E_singlett_m_E_FS_N52_L2_V01414_Anderson_vs_J}
\end{figure}

The binding energy of the singlets 
is analyzed in Fig.\ \ref{fig:E_singlett_m_E_FS_N52_L2_V01414_Anderson_vs_J}. We again find the discontinuities already observed in Fig.\ \ref{fig:Tk_Anderson_N60_L2}.
Between these discontinuities the binding energy decreases according to
\begin{eqnarray}
\Delta_s\left(\infty\right) \propto \exp\left(-A_{\Lambda} \frac{U}{8\rho_0V^2}\right).
\end{eqnarray}
Thus, we again retrieve a singlet ground state with a binding energy 
 given by the Kondo temperature $T_{K}$.

\section{Summary}

\subsection{Conclusions}

To treat the exponentially small Kondo energy scale reliably is a key problem
in correlated fermionic systems.
We presented a way how to use CUTs in order to derive effective models for the Kondo and the  Anderson impurity model. The conventional
CUT approach with a fixed reference state leads to diverging flow equations \cite{kehre06}. 
We identified the origin of this divergence which lies in the inappropriate reference state.
We introduced a modified approach based on the change of the reference state 
for the Kondo and the Anderson impurity model which solves the problem of diverging couplings during the flow and results in a well-behaved effective low-energy model
with finite parameters at arbitrarily small energies. This is the main achievement of
the present work.
We find a singlet ground state with a binding energy that is given by the Kondo temperature $T_{K}$.
Our approach is able to capture the exponentially small Kondo energy scale.
The quantitative result for the Anderson impurity model compares well with the Bethe ansatz result \cite{tsvel83,hewso93}.

The ground state of the effective model obtained by the CUT is a singlet and a Fermi sea.
But we stress that the complete effective model also comprises couplings between the triplet 
states of the impurity and the fermions in the bath. This implies that even the effective model
represents a correlated problem with non-trivial properties.
Furthermore, interactions within the fermionic bath have not been 
tracked. For these reasons it is beyond
the scope of the present work to analyze other characteristic quantities such as the Wilson ratio
and the like.

Earlier approaches for the Kondo model result in diverging couplings at a characteristic flow parameter \cite{kehre06}, an effective model where the parameters still exhibit logarithmic infrared divergences  \cite{vogel05} or rely on a bosonized form of the Kondo model before applying the CUT \cite{hofst01a,lobas05, sleza03}.

In the case of the Anderson impurity model, 
only a few approaches based on CUTs were published 
\cite{kehre94,kehre96b,staub04} while none of them reveal the exponential character of the Kondo temperature $T_{\text{K}}$. Nevertheless, an important previous work is 
able to reconstruct the Schrieffer-Wolff transformation using CUTs \cite{kehre96b} and can be extended to other hosts \cite{zapal15}.
Our approach does not rely on a bosonized form and can be extended to the Anderson impurity model. 

Of course, there are  other methods which reliably provide the exponentially small Kondo energy. The first is the numerical renormalization group  \cite{wilso75,krish80a,bulla08} and the Bethe ansatz  solution \cite{tsvel83}. But the challenge to find a reliable RG approach
with convergent flow has continued to attract much attention. The functional RG approach 
yielded good results up to intermediate interactions for the Anderson impurity model \cite{hedde04}.
Many other studies \cite{karra08,barto09,isido10,freir12,kinza13}
improved the functional RG approach recently but did not capture the
strong coupling regime. Only in 2013, Streib and co-workers succeeded \cite{strei13}
exploiting a magnetic field as regulatory cutoff and conserved Ward identities 
similar to a renormalized perturbation theory developed by Hewson and his co-workers 
 \cite{hewso01,hewso06,edwar11}. The difficulties that these intricate approaches
had to face underlines impressively that the Kondo effect in the Anderson 
impurity problem represents a true challenge.

The modified approach based on CUTs advocated here has the merit to provide a
convergent, i.e., with finite coefficints,  effective low-energy models of the Anderson impurity model and the Kondo model. The key element is the 
change of the reference state capturing the exponential character of the Kondo temperature. 
This is the central finding of our study.

\subsection{Outlook}

Several extensions suggest themselves. One route is to extend the set of operators
to capture more than the leading processes in the two main parts of the
transformation: (i) the elimination of the hybridization governed by the
expansion in $V$ and (ii) the renormalization of the exchange couplings $J$
by eliminating the non-diagonal exchange couplings \cite{kehre06}. By such an extension,
higher order corrections beyond $V^2$ and $J^2$ can be addressed and
the results  for the Kondo energy scale can be improved quantitatively.

A second route is to further explore the properties of the obtained effective model.
For instance, it is interesting to compute explicitly the impurity contribution to the 
magnetic susceptibility $\chi$ and to the specific heat $C$. 
We stress, however, that the analysis of the effective model is not
straightforward because it still represents a correlated problem, for instance, due
to the interactions between the triplet states of the impurity and the fermions in the
bath. If $\chi$ and  $C$ are known quantitatively
the characteristic Wilson ratio is known which is an established measure for the
degree of correlation effects. The technical difficulty in the CUTs is to separate
the contribution of the impurity in the renormalization of the effective parameters.

A third route is to tackle the transformation of the observables as well.
Transforming the creation and annihilation operator of the impurity fermion will
allow us to compute the spectral densities which is a decisive quantity
in many applications \cite{hewso93,stewa84,rozen92,georg96,grobi07}.

A fourth extension is to address the case of the asymmetric Anderson impurity model
where the particle-hole asymmetry is broken \cite{hewso93} and fifth extension is
to address finite temperatures as well.

Finally, we think that the methodological progress developed for the treatment
of the Kondo problem by continuous unitary transformations will trigger improved
approaches to other strongly correlated problems in general.
Examples are extended correlated systems with massless excitations or the vicinities of
quantum phase transitions where the ground state has to be switched just
as the reference state has to be switched in the present study.

\begin{acknowledgments}
We thank Frithjof B.\ Anders, Nils A.\ Drescher and  Sebastian Schmitt
for many useful discussions. Financial support by the  Helmholtz 
Virtual Institute ``New states of matter and their excitations'' is acknowledged.
\end{acknowledgments}


\begin{thebibliography}{48}%
\makeatletter
\providecommand \@ifxundefined [1]{%
 \@ifx{#1\undefined}
}%
\providecommand \@ifnum [1]{%
 \ifnum #1\expandafter \@firstoftwo
 \else \expandafter \@secondoftwo
 \fi
}%
\providecommand \@ifx [1]{%
 \ifx #1\expandafter \@firstoftwo
 \else \expandafter \@secondoftwo
 \fi
}%
\providecommand \natexlab [1]{#1}%
\providecommand \enquote  [1]{``#1''}%
\providecommand \bibnamefont  [1]{#1}%
\providecommand \bibfnamefont [1]{#1}%
\providecommand \citenamefont [1]{#1}%
\providecommand \href@noop [0]{\@secondoftwo}%
\providecommand \href [0]{\begingroup \@sanitize@url \@href}%
\providecommand \@href[1]{\@@startlink{#1}\@@href}%
\providecommand \@@href[1]{\endgroup#1\@@endlink}%
\providecommand \@sanitize@url [0]{\catcode `\\12\catcode `\$12\catcode
  `\&12\catcode `\#12\catcode `\^12\catcode `\_12\catcode `\%12\relax}%
\providecommand \@@startlink[1]{}%
\providecommand \@@endlink[0]{}%
\providecommand \url  [0]{\begingroup\@sanitize@url \@url }%
\providecommand \@url [1]{\endgroup\@href {#1}{\urlprefix }}%
\providecommand \urlprefix  [0]{URL }%
\providecommand \Eprint [0]{\href }%
\providecommand \doibase [0]{http://dx.doi.org/}%
\providecommand \selectlanguage [0]{\@gobble}%
\providecommand \bibinfo  [0]{\@secondoftwo}%
\providecommand \bibfield  [0]{\@secondoftwo}%
\providecommand \translation [1]{[#1]}%
\providecommand \BibitemOpen [0]{}%
\providecommand \bibitemStop [0]{}%
\providecommand \bibitemNoStop [0]{.\EOS\space}%
\providecommand \EOS [0]{\spacefactor3000\relax}%
\providecommand \BibitemShut  [1]{\csname bibitem#1\endcsname}%
\let\auto@bib@innerbib\@empty
\bibitem [{\citenamefont {Kondo}(1964)}]{kondo64}%
  \BibitemOpen
  \bibfield  {author} {\bibinfo {author} {\bibfnamefont {J.}~\bibnamefont
  {Kondo}},\ }\href@noop {} {\bibfield  {journal} {\bibinfo  {journal} {Prog.
  Theor. Phys.}\ }\textbf {\bibinfo {volume} {32}},\ \bibinfo {pages} {37}
  (\bibinfo {year} {1964})}\BibitemShut {NoStop}%
\bibitem [{\citenamefont {Anderson}(1961)}]{ander61}%
  \BibitemOpen
  \bibfield  {author} {\bibinfo {author} {\bibfnamefont {P.~W.}\ \bibnamefont
  {Anderson}},\ }\href@noop {} {\bibfield  {journal} {\bibinfo  {journal}
  {Phys. Rev.}\ }\textbf {\bibinfo {volume} {124}},\ \bibinfo {pages} {41}
  (\bibinfo {year} {1961})}\BibitemShut {NoStop}%
\bibitem [{\citenamefont {Hewson}(1993)}]{hewso93}%
  \BibitemOpen
  \bibfield  {author} {\bibinfo {author} {\bibfnamefont {A.~C.}\ \bibnamefont
  {Hewson}},\ }\href@noop {} {\emph {\bibinfo {title} {The Kondo Problem to
  Heavy Fermions}}}\ (\bibinfo  {publisher} {Cambridge University Press},\
  \bibinfo {address} {Cambridge},\ \bibinfo {year} {1993})\BibitemShut
  {NoStop}%
\bibitem [{\citenamefont {Stewart}(1984)}]{stewa84}%
  \BibitemOpen
  \bibfield  {author} {\bibinfo {author} {\bibfnamefont {G.~R.}\ \bibnamefont
  {Stewart}},\ }\href@noop {} {\bibfield  {journal} {\bibinfo  {journal} {Rev.
  Mod. Phys.}\ }\textbf {\bibinfo {volume} {56}},\ \bibinfo {pages} {755}
  (\bibinfo {year} {1984})}\BibitemShut {NoStop}%
\bibitem [{\citenamefont {Rozenberg}\ \emph {et~al.}(1992)\citenamefont
  {Rozenberg}, \citenamefont {Zhang},\ and\ \citenamefont {Kotliar}}]{rozen92}%
  \BibitemOpen
  \bibfield  {author} {\bibinfo {author} {\bibfnamefont {M.~J.}\ \bibnamefont
  {Rozenberg}}, \bibinfo {author} {\bibfnamefont {X.~Y.}\ \bibnamefont
  {Zhang}}, \ and\ \bibinfo {author} {\bibfnamefont {G.}~\bibnamefont
  {Kotliar}},\ }\href@noop {} {\bibfield  {journal} {\bibinfo  {journal} {Phys.
  Rev. Lett.}\ }\textbf {\bibinfo {volume} {69}},\ \bibinfo {pages} {1236}
  (\bibinfo {year} {1992})}\BibitemShut {NoStop}%
\bibitem [{\citenamefont {Georges}\ \emph {et~al.}(1996)\citenamefont
  {Georges}, \citenamefont {Kotliar}, \citenamefont {Krauth},\ and\
  \citenamefont {Rozenberg}}]{georg96}%
  \BibitemOpen
  \bibfield  {author} {\bibinfo {author} {\bibfnamefont {A.}~\bibnamefont
  {Georges}}, \bibinfo {author} {\bibfnamefont {G.}~\bibnamefont {Kotliar}},
  \bibinfo {author} {\bibfnamefont {W.}~\bibnamefont {Krauth}}, \ and\ \bibinfo
  {author} {\bibfnamefont {M.~J.}\ \bibnamefont {Rozenberg}},\ }\href@noop {}
  {\bibfield  {journal} {\bibinfo  {journal} {Rev. Mod. Phys.}\ }\textbf
  {\bibinfo {volume} {68}},\ \bibinfo {pages} {13} (\bibinfo {year}
  {1996})}\BibitemShut {NoStop}%
\bibitem [{\citenamefont {Grobis}\ \emph {et~al.}(2007)\citenamefont {Grobis},
  \citenamefont {Rau}, \citenamefont {Potok},\ and\ \citenamefont
  {Goldhaber-Gordon}}]{grobi07}%
  \BibitemOpen
  \bibfield  {author} {\bibinfo {author} {\bibfnamefont {M.}~\bibnamefont
  {Grobis}}, \bibinfo {author} {\bibfnamefont {I.~G.}\ \bibnamefont {Rau}},
  \bibinfo {author} {\bibfnamefont {R.~M.}\ \bibnamefont {Potok}}, \ and\
  \bibinfo {author} {\bibfnamefont {D.}~\bibnamefont {Goldhaber-Gordon}},\
  }\href@noop {} {\emph {\bibinfo {title} {Kondo Effect in Mesoscopic Quantum
  Dots}}},\ edited by\ \bibinfo {editor} {\bibfnamefont {H.}~\bibnamefont
  {Kronm\"uller}}\ and\ \bibinfo {editor} {\bibfnamefont {S.}~\bibnamefont
  {Parkin}},\ \bibinfo {series} {Handbook of Magnetism and Advanced Magnetic
  Materials}, Vol.~\bibinfo {volume} {5}\ (\bibinfo  {publisher} {Wiley},\
  \bibinfo {year} {2007})\BibitemShut {NoStop}%
\bibitem [{\citenamefont {Wilson}(1975)}]{wilso75}%
  \BibitemOpen
  \bibfield  {author} {\bibinfo {author} {\bibfnamefont {K.~G.}\ \bibnamefont
  {Wilson}},\ }\href@noop {} {\bibfield  {journal} {\bibinfo  {journal} {Rev.
  Mod. Phys.}\ }\textbf {\bibinfo {volume} {47}},\ \bibinfo {pages} {773}
  (\bibinfo {year} {1975})}\BibitemShut {NoStop}%
\bibitem [{\citenamefont {Krishna-murthy}\ \emph {et~al.}(1980)\citenamefont
  {Krishna-murthy}, \citenamefont {Wilkins},\ and\ \citenamefont
  {Wilson}}]{krish80a}%
  \BibitemOpen
  \bibfield  {author} {\bibinfo {author} {\bibfnamefont {H.~R.}\ \bibnamefont
  {Krishna-murthy}}, \bibinfo {author} {\bibfnamefont {J.~W.}\ \bibnamefont
  {Wilkins}}, \ and\ \bibinfo {author} {\bibfnamefont {K.~G.}\ \bibnamefont
  {Wilson}},\ }\href@noop {} {\bibfield  {journal} {\bibinfo  {journal} {Phys.
  Rev. B}\ }\textbf {\bibinfo {volume} {21}},\ \bibinfo {pages} {1003}
  (\bibinfo {year} {1980})}\BibitemShut {NoStop}%
\bibitem [{\citenamefont {Bulla}\ \emph {et~al.}(2008)\citenamefont {Bulla},
  \citenamefont {Costi},\ and\ \citenamefont {Pruschke}}]{bulla08}%
  \BibitemOpen
  \bibfield  {author} {\bibinfo {author} {\bibfnamefont {R.}~\bibnamefont
  {Bulla}}, \bibinfo {author} {\bibfnamefont {T.~A.}\ \bibnamefont {Costi}}, \
  and\ \bibinfo {author} {\bibfnamefont {T.}~\bibnamefont {Pruschke}},\
  }\href@noop {} {\bibfield  {journal} {\bibinfo  {journal} {Rev. Mod. Phys.}\
  }\textbf {\bibinfo {volume} {80}},\ \bibinfo {pages} {395} (\bibinfo {year}
  {2008})}\BibitemShut {NoStop}%
\bibitem [{\citenamefont {Tsvelick}\ and\ \citenamefont
  {Wiegmann}(1983)}]{tsvel83}%
  \BibitemOpen
  \bibfield  {author} {\bibinfo {author} {\bibfnamefont {A.~M.}\ \bibnamefont
  {Tsvelick}}\ and\ \bibinfo {author} {\bibfnamefont {P.~B.}\ \bibnamefont
  {Wiegmann}},\ }\href@noop {} {\bibfield  {journal} {\bibinfo  {journal} {Adv.
  Phys.}\ }\textbf {\bibinfo {volume} {32}},\ \bibinfo {pages} {453} (\bibinfo
  {year} {1983})}\BibitemShut {NoStop}%
\bibitem [{\citenamefont {Hedden}\ \emph {et~al.}(2004)\citenamefont {Hedden},
  \citenamefont {Meden}, \citenamefont {Pruschke},\ and\ \citenamefont
  {Sch\"onhammer}}]{hedde04}%
  \BibitemOpen
  \bibfield  {author} {\bibinfo {author} {\bibfnamefont {R.}~\bibnamefont
  {Hedden}}, \bibinfo {author} {\bibfnamefont {V.}~\bibnamefont {Meden}},
  \bibinfo {author} {\bibfnamefont {T.}~\bibnamefont {Pruschke}}, \ and\
  \bibinfo {author} {\bibfnamefont {K.}~\bibnamefont {Sch\"onhammer}},\
  }\href@noop {} {\bibfield  {journal} {\bibinfo  {journal} {J. Phys.: Condens.
  Matter}\ }\textbf {\bibinfo {volume} {16}},\ \bibinfo {pages} {5279}
  (\bibinfo {year} {2004})}\BibitemShut {NoStop}%
\bibitem [{\citenamefont {Karrasch}\ \emph {et~al.}(2008)\citenamefont
  {Karrasch}, \citenamefont {Hedden}, \citenamefont {Peters}, \citenamefont
  {Pruschke}, \citenamefont {Sch\"onhammer},\ and\ \citenamefont
  {Meden}}]{karra08}%
  \BibitemOpen
  \bibfield  {author} {\bibinfo {author} {\bibfnamefont {C.}~\bibnamefont
  {Karrasch}}, \bibinfo {author} {\bibfnamefont {R.}~\bibnamefont {Hedden}},
  \bibinfo {author} {\bibfnamefont {R.}~\bibnamefont {Peters}}, \bibinfo
  {author} {\bibfnamefont {T.}~\bibnamefont {Pruschke}}, \bibinfo {author}
  {\bibfnamefont {K.}~\bibnamefont {Sch\"onhammer}}, \ and\ \bibinfo {author}
  {\bibfnamefont {V.}~\bibnamefont {Meden}},\ }\href@noop {} {\bibfield
  {journal} {\bibinfo  {journal} {J. Phys.: Condens. Matter}\ }\textbf
  {\bibinfo {volume} {20}},\ \bibinfo {pages} {345205} (\bibinfo {year}
  {2008})}\BibitemShut {NoStop}%
\bibitem [{\citenamefont {Bartosch}\ \emph {et~al.}(2009)\citenamefont
  {Bartosch}, \citenamefont {Freire}, \citenamefont {Cardenas},\ and\
  \citenamefont {Kopietz}}]{barto09}%
  \BibitemOpen
  \bibfield  {author} {\bibinfo {author} {\bibfnamefont {L.}~\bibnamefont
  {Bartosch}}, \bibinfo {author} {\bibfnamefont {H.}~\bibnamefont {Freire}},
  \bibinfo {author} {\bibfnamefont {J.~J.~R.}\ \bibnamefont {Cardenas}}, \ and\
  \bibinfo {author} {\bibfnamefont {P.}~\bibnamefont {Kopietz}},\ }\href@noop
  {} {\bibfield  {journal} {\bibinfo  {journal} {J. Phys.: Condens. Matter}\
  }\textbf {\bibinfo {volume} {21}},\ \bibinfo {pages} {305602} (\bibinfo
  {year} {2009})}\BibitemShut {NoStop}%
\bibitem [{\citenamefont {Isidori}\ \emph {et~al.}(2010)\citenamefont
  {Isidori}, \citenamefont {Roosen}, \citenamefont {Bartosch}, \citenamefont
  {Hofstetter},\ and\ \citenamefont {Kopietz}}]{isido10}%
  \BibitemOpen
  \bibfield  {author} {\bibinfo {author} {\bibfnamefont {A.}~\bibnamefont
  {Isidori}}, \bibinfo {author} {\bibfnamefont {D.}~\bibnamefont {Roosen}},
  \bibinfo {author} {\bibfnamefont {L.}~\bibnamefont {Bartosch}}, \bibinfo
  {author} {\bibfnamefont {W.}~\bibnamefont {Hofstetter}}, \ and\ \bibinfo
  {author} {\bibfnamefont {P.}~\bibnamefont {Kopietz}},\ }\href@noop {}
  {\bibfield  {journal} {\bibinfo  {journal} {Phys. Rev. B}\ }\textbf {\bibinfo
  {volume} {81}},\ \bibinfo {pages} {235120} (\bibinfo {year}
  {2010})}\BibitemShut {NoStop}%
\bibitem [{\citenamefont {Freire}\ and\ \citenamefont
  {Corr\^{e}a}(2012)}]{freir12}%
  \BibitemOpen
  \bibfield  {author} {\bibinfo {author} {\bibfnamefont {H.}~\bibnamefont
  {Freire}}\ and\ \bibinfo {author} {\bibfnamefont {E.}~\bibnamefont
  {Corr\^{e}a}},\ }\href@noop {} {\bibfield  {journal} {\bibinfo  {journal} {J.
  Low Temp. Phys.}\ }\textbf {\bibinfo {volume} {166}},\ \bibinfo {pages} {192}
  (\bibinfo {year} {2012})}\BibitemShut {NoStop}%
\bibitem [{\citenamefont {Kinza}\ \emph {et~al.}(2013)\citenamefont {Kinza},
  \citenamefont {Ortloff}, \citenamefont {Bauer},\ and\ \citenamefont
  {Honerkamp}}]{kinza13}%
  \BibitemOpen
  \bibfield  {author} {\bibinfo {author} {\bibfnamefont {M.}~\bibnamefont
  {Kinza}}, \bibinfo {author} {\bibfnamefont {J.}~\bibnamefont {Ortloff}},
  \bibinfo {author} {\bibfnamefont {J.}~\bibnamefont {Bauer}}, \ and\ \bibinfo
  {author} {\bibfnamefont {C.}~\bibnamefont {Honerkamp}},\ }\href@noop {}
  {\bibfield  {journal} {\bibinfo  {journal} {Phys. Rev. B}\ }\textbf {\bibinfo
  {volume} {87}},\ \bibinfo {pages} {035111} (\bibinfo {year}
  {2013})}\BibitemShut {NoStop}%
\bibitem [{\citenamefont {Streib}\ \emph {et~al.}(2013)\citenamefont {Streib},
  \citenamefont {Isidori},\ and\ \citenamefont {Kopietz}}]{strei13}%
  \BibitemOpen
  \bibfield  {author} {\bibinfo {author} {\bibfnamefont {S.}~\bibnamefont
  {Streib}}, \bibinfo {author} {\bibfnamefont {A.}~\bibnamefont {Isidori}}, \
  and\ \bibinfo {author} {\bibfnamefont {P.}~\bibnamefont {Kopietz}},\
  }\href@noop {} {\bibfield  {journal} {\bibinfo  {journal} {Phys. Rev. B}\
  }\textbf {\bibinfo {volume} {87}},\ \bibinfo {pages} {201107(R)} (\bibinfo
  {year} {2013})}\BibitemShut {NoStop}%
\bibitem [{\citenamefont {Hewson}(2001)}]{hewso01}%
  \BibitemOpen
  \bibfield  {author} {\bibinfo {author} {\bibfnamefont {A.~C.}\ \bibnamefont
  {Hewson}},\ }\href@noop {} {\bibfield  {journal} {\bibinfo  {journal} {J.
  Phys.: Condens. Matter}\ }\textbf {\bibinfo {volume} {13}},\ \bibinfo {pages}
  {10011} (\bibinfo {year} {2001})}\BibitemShut {NoStop}%
\bibitem [{\citenamefont {Hewson}(2006)}]{hewso06}%
  \BibitemOpen
  \bibfield  {author} {\bibinfo {author} {\bibfnamefont {A.~C.}\ \bibnamefont
  {Hewson}},\ }\href@noop {} {\bibfield  {journal} {\bibinfo  {journal} {J.
  Phys.: Condens. Matter}\ }\textbf {\bibinfo {volume} {18}},\ \bibinfo {pages}
  {1815} (\bibinfo {year} {2006})}\BibitemShut {NoStop}%
\bibitem [{\citenamefont {Edwards}\ and\ \citenamefont
  {Hewson}(2011)}]{edwar11}%
  \BibitemOpen
  \bibfield  {author} {\bibinfo {author} {\bibfnamefont {K.}~\bibnamefont
  {Edwards}}\ and\ \bibinfo {author} {\bibfnamefont {A.~C.}\ \bibnamefont
  {Hewson}},\ }\href@noop {} {\bibfield  {journal} {\bibinfo  {journal} {J.
  Phys.: Condens. Matter}\ }\textbf {\bibinfo {volume} {23}},\ \bibinfo {pages}
  {045601} (\bibinfo {year} {2011})}\BibitemShut {NoStop}%
\bibitem [{\citenamefont {Anderson}(1970)}]{ander70a}%
  \BibitemOpen
  \bibfield  {author} {\bibinfo {author} {\bibfnamefont {P.~W.}\ \bibnamefont
  {Anderson}},\ }\href@noop {} {\bibfield  {journal} {\bibinfo  {journal} {J.
  Phys. C}\ }\textbf {\bibinfo {volume} {3}},\ \bibinfo {pages} {2436}
  (\bibinfo {year} {1970})}\BibitemShut {NoStop}%
\bibitem [{\citenamefont {Kehrein}(2006)}]{kehre06}%
  \BibitemOpen
  \bibfield  {author} {\bibinfo {author} {\bibfnamefont {S.}~\bibnamefont
  {Kehrein}},\ }\href@noop {} {\emph {\bibinfo {title} {The Flow Equation
  Approach to Many-Particle Systems}}},\ \bibinfo {series} {Springer Tracts in
  Modern Physics}, Vol.\ \bibinfo {volume} {217}\ (\bibinfo  {publisher}
  {Springer},\ \bibinfo {address} {Berlin},\ \bibinfo {year}
  {2006})\BibitemShut {NoStop}%
\bibitem [{\citenamefont {Vogel}(2005)}]{vogel05}%
  \BibitemOpen
  \bibfield  {author} {\bibinfo {author} {\bibfnamefont {E.}~\bibnamefont
  {Vogel}},\ }\href@noop {} {\emph {\bibinfo {title} {Flussgleichungen f\"ur
  das Kondo-modell}}}\ (\bibinfo  {publisher} {Dissertation},\ \bibinfo
  {address} {Heidelberg},\ \bibinfo {year} {2005})\BibitemShut {NoStop}%
\bibitem [{\citenamefont {Hofstetter}\ and\ \citenamefont
  {Kehrein}(2001)}]{hofst01a}%
  \BibitemOpen
  \bibfield  {author} {\bibinfo {author} {\bibfnamefont {W.}~\bibnamefont
  {Hofstetter}}\ and\ \bibinfo {author} {\bibfnamefont {S.}~\bibnamefont
  {Kehrein}},\ }\href@noop {} {\bibfield  {journal} {\bibinfo  {journal} {Phys.
  Rev. B}\ }\textbf {\bibinfo {volume} {63}},\ \bibinfo {pages} {140402}
  (\bibinfo {year} {2001})}\BibitemShut {NoStop}%
\bibitem [{\citenamefont {Lobaskin}\ and\ \citenamefont
  {Kehrein}(2005)}]{lobas05}%
  \BibitemOpen
  \bibfield  {author} {\bibinfo {author} {\bibfnamefont {D.}~\bibnamefont
  {Lobaskin}}\ and\ \bibinfo {author} {\bibfnamefont {S.}~\bibnamefont
  {Kehrein}},\ }\href@noop {} {\bibfield  {journal} {\bibinfo  {journal} {Phys.
  Rev. B}\ }\textbf {\bibinfo {volume} {71}},\ \bibinfo {pages} {193303}
  (\bibinfo {year} {2005})}\BibitemShut {NoStop}%
\bibitem [{\citenamefont {Slezak}\ \emph {et~al.}(2003)\citenamefont {Slezak},
  \citenamefont {Kehrein}, \citenamefont {Pruschke},\ and\ \citenamefont
  {Jarrell}}]{sleza03}%
  \BibitemOpen
  \bibfield  {author} {\bibinfo {author} {\bibfnamefont {C.}~\bibnamefont
  {Slezak}}, \bibinfo {author} {\bibfnamefont {S.}~\bibnamefont {Kehrein}},
  \bibinfo {author} {\bibfnamefont {T.}~\bibnamefont {Pruschke}}, \ and\
  \bibinfo {author} {\bibfnamefont {M.}~\bibnamefont {Jarrell}},\ }\href@noop
  {} {\bibfield  {journal} {\bibinfo  {journal} {Phys. Rev. B}\ }\textbf
  {\bibinfo {volume} {67}},\ \bibinfo {pages} {184408} (\bibinfo {year}
  {2003})}\BibitemShut {NoStop}%
\bibitem [{\citenamefont {Kehrein}\ and\ \citenamefont
  {Mielke}(1994{\natexlab{a}})}]{kehre94}%
  \BibitemOpen
  \bibfield  {author} {\bibinfo {author} {\bibfnamefont {S.~K.}\ \bibnamefont
  {Kehrein}}\ and\ \bibinfo {author} {\bibfnamefont {A.}~\bibnamefont
  {Mielke}},\ }\href@noop {} {\bibfield  {journal} {\bibinfo  {journal} {J.
  Phys. A: Math. Gen.}\ }\textbf {\bibinfo {volume} {27}},\ \bibinfo {pages}
  {4259} (\bibinfo {year} {1994}{\natexlab{a}})}\BibitemShut {NoStop}%
\bibitem [{\citenamefont {Kehrein}\ and\ \citenamefont
  {Mielke}(1994{\natexlab{b}})}]{kehre94e}%
  \BibitemOpen
  \bibfield  {author} {\bibinfo {author} {\bibfnamefont {S.~K.}\ \bibnamefont
  {Kehrein}}\ and\ \bibinfo {author} {\bibfnamefont {A.}~\bibnamefont
  {Mielke}},\ }\href@noop {} {\bibfield  {journal} {\bibinfo  {journal} {J.
  Phys. A: Math. Gen.}\ }\textbf {\bibinfo {volume} {27}},\ \bibinfo {pages}
  {5705} (\bibinfo {year} {1994}{\natexlab{b}})}\BibitemShut {NoStop}%
\bibitem [{\citenamefont {Kehrein}\ and\ \citenamefont
  {Mielke}(1996)}]{kehre96b}%
  \BibitemOpen
  \bibfield  {author} {\bibinfo {author} {\bibfnamefont {S.~K.}\ \bibnamefont
  {Kehrein}}\ and\ \bibinfo {author} {\bibfnamefont {A.}~\bibnamefont
  {Mielke}},\ }\href@noop {} {\bibfield  {journal} {\bibinfo  {journal} {Ann.
  of Phys.}\ }\textbf {\bibinfo {volume} {252}},\ \bibinfo {pages} {1}
  (\bibinfo {year} {1996})}\BibitemShut {NoStop}%
\bibitem [{\citenamefont {Stauber}\ and\ \citenamefont
  {Guinea}(2004)}]{staub04}%
  \BibitemOpen
  \bibfield  {author} {\bibinfo {author} {\bibfnamefont {T.}~\bibnamefont
  {Stauber}}\ and\ \bibinfo {author} {\bibfnamefont {F.}~\bibnamefont
  {Guinea}},\ }\href@noop {} {\bibfield  {journal} {\bibinfo  {journal} {Phys.
  Rev. B}\ }\textbf {\bibinfo {volume} {69}},\ \bibinfo {pages} {035301}
  (\bibinfo {year} {2004})}\BibitemShut {NoStop}%
\bibitem [{\citenamefont {Zapalska}\ and\ \citenamefont
  {Doman\'nski}(2014)}]{zapal15}%
  \BibitemOpen
  \bibfield  {author} {\bibinfo {author} {\bibfnamefont {M.}~\bibnamefont
  {Zapalska}}\ and\ \bibinfo {author} {\bibfnamefont {T.}~\bibnamefont
  {Doman\'nski}},\ }\href@noop {} {\ ,\ \bibinfo {pages} {arxiv:1402.1291}
  (\bibinfo {year} {2014})}\BibitemShut {NoStop}%
\bibitem [{\citenamefont {Wegner}(1994)}]{wegne94}%
  \BibitemOpen
  \bibfield  {author} {\bibinfo {author} {\bibfnamefont {F.~J.}\ \bibnamefont
  {Wegner}},\ }\href@noop {} {\bibfield  {journal} {\bibinfo  {journal} {Ann.
  Physik}\ }\textbf {\bibinfo {volume} {3}},\ \bibinfo {pages} {77} (\bibinfo
  {year} {1994})}\BibitemShut {NoStop}%
\bibitem [{\citenamefont {G{\l}azek}\ and\ \citenamefont
  {Wilson}(1993)}]{glaze93}%
  \BibitemOpen
  \bibfield  {author} {\bibinfo {author} {\bibfnamefont {S.~D.}\ \bibnamefont
  {G{\l}azek}}\ and\ \bibinfo {author} {\bibfnamefont {K.~G.}\ \bibnamefont
  {Wilson}},\ }\href@noop {} {\bibfield  {journal} {\bibinfo  {journal} {Phys.
  Rev. D}\ }\textbf {\bibinfo {volume} {48}},\ \bibinfo {pages} {5863}
  (\bibinfo {year} {1993})}\BibitemShut {NoStop}%
\bibitem [{\citenamefont {G{\l}azek}\ and\ \citenamefont
  {Wilson}(1994)}]{glaze94}%
  \BibitemOpen
  \bibfield  {author} {\bibinfo {author} {\bibfnamefont {S.~D.}\ \bibnamefont
  {G{\l}azek}}\ and\ \bibinfo {author} {\bibfnamefont {K.~G.}\ \bibnamefont
  {Wilson}},\ }\href@noop {} {\bibfield  {journal} {\bibinfo  {journal} {Phys.
  Rev. D}\ }\textbf {\bibinfo {volume} {49}},\ \bibinfo {pages} {4214}
  (\bibinfo {year} {1994})}\BibitemShut {NoStop}%
\bibitem [{\citenamefont {Wegner}(2006)}]{wegne06}%
  \BibitemOpen
  \bibfield  {author} {\bibinfo {author} {\bibfnamefont {F.}~\bibnamefont
  {Wegner}},\ }\href@noop {} {\bibfield  {journal} {\bibinfo  {journal} {J.
  Phys. A: Math. Gen.}\ }\textbf {\bibinfo {volume} {39}},\ \bibinfo {pages}
  {8221} (\bibinfo {year} {2006})}\BibitemShut {NoStop}%
\bibitem [{\citenamefont {Duffe}\ and\ \citenamefont {Uhrig}(2011)}]{duffe11}%
  \BibitemOpen
  \bibfield  {author} {\bibinfo {author} {\bibfnamefont {S.}~\bibnamefont
  {Duffe}}\ and\ \bibinfo {author} {\bibfnamefont {G.~S.}\ \bibnamefont
  {Uhrig}},\ }\href@noop {} {\bibfield  {journal} {\bibinfo  {journal} {Eur.
  Phys. J. B}\ }\textbf {\bibinfo {volume} {84}},\ \bibinfo {pages} {475}
  (\bibinfo {year} {2011})}\BibitemShut {NoStop}%
\bibitem [{\citenamefont {Fischer}\ \emph {et~al.}(2010)\citenamefont
  {Fischer}, \citenamefont {Duffe},\ and\ \citenamefont {Uhrig}}]{fisch10a}%
  \BibitemOpen
  \bibfield  {author} {\bibinfo {author} {\bibfnamefont {T.}~\bibnamefont
  {Fischer}}, \bibinfo {author} {\bibfnamefont {S.}~\bibnamefont {Duffe}}, \
  and\ \bibinfo {author} {\bibfnamefont {G.~S.}\ \bibnamefont {Uhrig}},\
  }\href@noop {} {\bibfield  {journal} {\bibinfo  {journal} {New J. Phys.}\
  }\textbf {\bibinfo {volume} {10}},\ \bibinfo {pages} {033048} (\bibinfo
  {year} {2010})}\BibitemShut {NoStop}%
\bibitem [{\citenamefont {Fischer}\ \emph {et~al.}(2011)\citenamefont
  {Fischer}, \citenamefont {Duffe},\ and\ \citenamefont {Uhrig}}]{fisch11a}%
  \BibitemOpen
  \bibfield  {author} {\bibinfo {author} {\bibfnamefont {T.}~\bibnamefont
  {Fischer}}, \bibinfo {author} {\bibfnamefont {S.}~\bibnamefont {Duffe}}, \
  and\ \bibinfo {author} {\bibfnamefont {G.~S.}\ \bibnamefont {Uhrig}},\
  }\href@noop {} {\bibfield  {journal} {\bibinfo  {journal} {Europhys. Lett.}\
  }\textbf {\bibinfo {volume} {96}},\ \bibinfo {pages} {47001} (\bibinfo {year}
  {2011})}\BibitemShut {NoStop}%
\bibitem [{\citenamefont {Hamerla}\ \emph {et~al.}(2010)\citenamefont
  {Hamerla}, \citenamefont {Duffe},\ and\ \citenamefont {Uhrig}}]{hamer10}%
  \BibitemOpen
  \bibfield  {author} {\bibinfo {author} {\bibfnamefont {S.~A.}\ \bibnamefont
  {Hamerla}}, \bibinfo {author} {\bibfnamefont {S.}~\bibnamefont {Duffe}}, \
  and\ \bibinfo {author} {\bibfnamefont {G.~S.}\ \bibnamefont {Uhrig}},\
  }\href@noop {} {\bibfield  {journal} {\bibinfo  {journal} {Phys. Rev. B}\
  }\textbf {\bibinfo {volume} {82}},\ \bibinfo {pages} {235117} (\bibinfo
  {year} {2010})}\BibitemShut {NoStop}%
\bibitem [{\citenamefont {Uhrig}\ and\ \citenamefont
  {Normand}(1998)}]{uhrig98c}%
  \BibitemOpen
  \bibfield  {author} {\bibinfo {author} {\bibfnamefont {G.~S.}\ \bibnamefont
  {Uhrig}}\ and\ \bibinfo {author} {\bibfnamefont {B.}~\bibnamefont
  {Normand}},\ }\href@noop {} {\bibfield  {journal} {\bibinfo  {journal} {Phys.
  Rev. B}\ }\textbf {\bibinfo {volume} {58}},\ \bibinfo {pages} {14705(R)}
  (\bibinfo {year} {1998})}\BibitemShut {NoStop}%
\bibitem [{\citenamefont {Knetter}\ and\ \citenamefont
  {Uhrig}(2000)}]{knett00a}%
  \BibitemOpen
  \bibfield  {author} {\bibinfo {author} {\bibfnamefont {C.}~\bibnamefont
  {Knetter}}\ and\ \bibinfo {author} {\bibfnamefont {G.~S.}\ \bibnamefont
  {Uhrig}},\ }\href@noop {} {\bibfield  {journal} {\bibinfo  {journal} {Eur.
  Phys. J. B}\ }\textbf {\bibinfo {volume} {13}},\ \bibinfo {pages} {209}
  (\bibinfo {year} {2000})}\BibitemShut {NoStop}%
\bibitem [{\citenamefont {Krull}\ \emph {et~al.}(2012)\citenamefont {Krull},
  \citenamefont {Drescher},\ and\ \citenamefont {Uhrig}}]{krull12}%
  \BibitemOpen
  \bibfield  {author} {\bibinfo {author} {\bibfnamefont {H.}~\bibnamefont
  {Krull}}, \bibinfo {author} {\bibfnamefont {N.~A.}\ \bibnamefont {Drescher}},
  \ and\ \bibinfo {author} {\bibfnamefont {G.~S.}\ \bibnamefont {Uhrig}},\
  }\href@noop {} {\bibfield  {journal} {\bibinfo  {journal} {Phys. Rev. B}\
  }\textbf {\bibinfo {volume} {86}},\ \bibinfo {pages} {125113} (\bibinfo
  {year} {2012})}\BibitemShut {NoStop}%
\bibitem [{\citenamefont {Yang}\ and\ \citenamefont {Schmidt}(2011)}]{yang11a}%
  \BibitemOpen
  \bibfield  {author} {\bibinfo {author} {\bibfnamefont {H.-Y.}\ \bibnamefont
  {Yang}}\ and\ \bibinfo {author} {\bibfnamefont {K.~P.}\ \bibnamefont
  {Schmidt}},\ }\href@noop {} {\bibfield  {journal} {\bibinfo  {journal}
  {Europhys. Lett.}\ }\textbf {\bibinfo {volume} {94}},\ \bibinfo {pages}
  {17004} (\bibinfo {year} {2011})}\BibitemShut {NoStop}%
\bibitem [{\citenamefont {Mielke}(1998)}]{mielk98}%
  \BibitemOpen
  \bibfield  {author} {\bibinfo {author} {\bibfnamefont {A.}~\bibnamefont
  {Mielke}},\ }\href@noop {} {\bibfield  {journal} {\bibinfo  {journal} {Eur.
  Phys. J. B}\ }\textbf {\bibinfo {volume} {5}},\ \bibinfo {pages} {605}
  (\bibinfo {year} {1998})}\BibitemShut {NoStop}%
\bibitem [{\citenamefont {Knetter}\ \emph {et~al.}(2003)\citenamefont
  {Knetter}, \citenamefont {Schmidt},\ and\ \citenamefont {Uhrig}}]{knett03a}%
  \BibitemOpen
  \bibfield  {author} {\bibinfo {author} {\bibfnamefont {C.}~\bibnamefont
  {Knetter}}, \bibinfo {author} {\bibfnamefont {K.~P.}\ \bibnamefont
  {Schmidt}}, \ and\ \bibinfo {author} {\bibfnamefont {G.~S.}\ \bibnamefont
  {Uhrig}},\ }\href@noop {} {\bibfield  {journal} {\bibinfo  {journal} {J.
  Phys. A: Math. Gen.}\ }\textbf {\bibinfo {volume} {36}},\ \bibinfo {pages}
  {7889} (\bibinfo {year} {2003})}\BibitemShut {NoStop}%
\bibitem [{\citenamefont {Reischl}\ \emph {et~al.}(2004)\citenamefont
  {Reischl}, \citenamefont {M\"uller-Hartmann},\ and\ \citenamefont
  {Uhrig}}]{reisc04}%
  \BibitemOpen
  \bibfield  {author} {\bibinfo {author} {\bibfnamefont {A.}~\bibnamefont
  {Reischl}}, \bibinfo {author} {\bibfnamefont {E.}~\bibnamefont
  {M\"uller-Hartmann}}, \ and\ \bibinfo {author} {\bibfnamefont {G.~S.}\
  \bibnamefont {Uhrig}},\ }\href@noop {} {\bibfield  {journal} {\bibinfo
  {journal} {Phys. Rev. B}\ }\textbf {\bibinfo {volume} {70}},\ \bibinfo
  {pages} {245124} (\bibinfo {year} {2004})}\BibitemShut {NoStop}%
\bibitem [{\citenamefont {Schrieffer}\ and\ \citenamefont
  {Wolff}(1966)}]{schri66}%
  \BibitemOpen
  \bibfield  {author} {\bibinfo {author} {\bibfnamefont {J.~R.}\ \bibnamefont
  {Schrieffer}}\ and\ \bibinfo {author} {\bibfnamefont {P.~A.}\ \bibnamefont
  {Wolff}},\ }\href@noop {} {\bibfield  {journal} {\bibinfo  {journal} {Phys.
  Rev.}\ }\textbf {\bibinfo {volume} {149}},\ \bibinfo {pages} {491} (\bibinfo
  {year} {1966})}\BibitemShut {NoStop}%
\end{thebibliography}

%


\appendix

\section{Adapted operator basis for the Kondo and Anderson impurity model}
\label{Appendix Basis states for the new operator basis} 

We consider the Hamiltonian from Eq.\ (\ref{Chapter 7 H_r for the Kondo and Anderson model})
\begin{eqnarray}
\nonumber
  H_{\text{r}}&=&\sum_{\sigma} \epsilon_{r} \left(c^{\dagger}_{r \sigma}c^{\phantom{\dagger}}_{r \sigma} 
	- c^{\dagger}_{\bar{r} \sigma}c^{\phantom{\dagger}}_{\bar{r} \sigma}\right)
	\\
	\label{Appendix Hamiltonian part of sites included in the formation of the singlet}
	&+& J_{rr}\sum_{\mu}\sum_{\alpha,\beta}\sigma_{\alpha\beta}^{\mu}S_I^{\mu}
	\left(c^{\dagger}_{r \alpha}c^{\phantom{\dagger}}_{r \beta}
  +c^{\dagger}_{\bar{r} \alpha}c^{\phantom{\dagger}}_{\bar{r} \beta}\right) 
\end{eqnarray}
and its basis states. In the calculations presented in the main text we truncate basis states 
that are lying higher in energy than the triplet states.
All basis states are shown below in the  Tables \ref{tab:zero-part}, \ref{tab:one-part}
\ref{tab:two-part}, \ref{tab:three-part}, and \ref{tab:four-part}. 
The eigenstates are sorted by the number of fermions 
besides the impurity spin in the levels occuring in the Hamiltonian 
\eqref{Appendix Hamiltonian part of sites included in the formation of the singlet}. 
The arrow in the middle entry of the ket stands for the
state of the impurity. The arrow to the left for the state of the particle state
at $\epsilon_r > 0$; the arrow to the left for the state of the hole state
at $-\epsilon_r<0$. In addition, the total spin $S$ and the total z-component $S^{z}$ are given
for the states. The column `used' indicates whether or not the state
is considered in our calculations.

\begin{table}[htb]
\begin{tabular}{|c||c||c||c||c|}
\hline
states & energy $\epsilon_{i}$ & $S$ & $S^{z}$ & used  
\\\hline
$|a_1 \rangle = |0,\uparrow,0 \rangle$   
& $ \epsilon_{a_1} = 0$ 
& $\frac{1}{2}$
& $+\frac{1}{2}$
& \XSolidBrush
\\ \hline
$|a_2 \rangle = |0,\downarrow,0 \rangle$ 
& $\epsilon_{a_2} = 0$ 
& $\frac{1}{2}$
& $-\frac{1}{2}$
& \XSolidBrush
\\ \hline
\end{tabular}
\caption{States with zero fermion besides the impurity spin}
\label{tab:zero-part}
\end{table}

\begin{table}[htb]
\begin{tabular}{|l||c||c||c||c|}
\hline
states &  $\epsilon_{i}$ & $S$ & $S^{z}$  & used  
\\\hline
$|s^{-} \rangle = \frac{1}{\sqrt{2}}\left(|0,\uparrow,\downarrow \rangle - |0,\downarrow,\uparrow \rangle \right)$   
& $ -\frac{3J_{rr}}{2} - \epsilon_{r}$
& $0$
& $0$
& \Checkmark
\\ \hline
$|t^{-}_1 \rangle = |0,\uparrow,\uparrow \rangle$  
& $ \frac{J_{rr}}{2} - \epsilon_{r}$
& $1$
& $+1$
& \Checkmark
\\ \hline
$|t^{-}_2 \rangle = \frac{1}{\sqrt{2}}\left(|0,\uparrow,\downarrow \rangle + |0,\downarrow,\uparrow \rangle \right)$ 
& $ \frac{J_{rr}}{2} - \epsilon_{r}$
& $1$
& $0$
& \Checkmark
\\ \hline
$|t^{-}_3 \rangle = |0,\downarrow,\downarrow  \rangle$  
& $ \frac{J_{rr}}{2} - \epsilon_{r}$
& $1$
& $-1$
& \Checkmark
\\ \hline
$|a_3 \rangle = \frac{1}{\sqrt{2}}\left(|\!\uparrow,\downarrow,0 \rangle - |\!\downarrow,\uparrow,0 \rangle \right)$ 
& $ -\frac{3J_{rr}}{2} + \epsilon_{r}$
& $0$
& $0$
& \XSolidBrush
\\ \hline
$|a_4 \rangle = |\!\downarrow,\downarrow,0 \rangle $ 
& $ \frac{J_{rr}}{2} + \epsilon_{r}$
& $1$
& $-1$
& \XSolidBrush
\\ \hline
$|a_5 \rangle = \frac{1}{\sqrt{2}}\left(|\!\uparrow,\downarrow,0 \rangle + |\!\downarrow,\uparrow,0 \rangle \right)$ 
& $ \frac{J_{rr}}{2} + \epsilon_{r}$
& $1$
& $+1$
& \XSolidBrush
\\ \hline
$|a_6 \rangle = |\!\uparrow,\uparrow,0 \rangle $  
& $ \frac{J_{rr}}{2} + \epsilon_{r}$
& $1$
& $+1$
& \XSolidBrush
\\ \hline
\end{tabular}
\caption{States with one  fermion besides the impurity spin}
\label{tab:one-part}
\end{table}

\begin{table}[htb]
\begin{footnotesize}
\begin{tabular}{|l||c||c||c||c|}
\hline
states & $\epsilon_{i}$ & $S$ & $S^{z}$  & used 
\\\hline
$|\text{FS},\uparrow \rangle = |0,\uparrow,\uparrow\downarrow \rangle$ 
& $-2\epsilon_{r}$
& $\frac{1}{2}$
& $+\frac{1}{2}$
& \Checkmark
\\\hline
$|\text{FS},\downarrow \rangle = |0,\downarrow,\uparrow\downarrow \rangle$ 
& $-2\epsilon_{r}$
& $\frac{1}{2}$
& $-\frac{1}{2}$
& \Checkmark
\\ \hline
$|\tilde{\uparrow} \rangle = \frac{1}{\sqrt{6}}\left[
|\!\uparrow,\uparrow,\downarrow \rangle
-2 |\!\uparrow,\downarrow,\uparrow \rangle
+  |\!\downarrow,\uparrow,\uparrow \rangle\right]$  
& $-2J_{rr}$
& $\frac{1}{2}$
& $+\frac{1}{2}$
& \Checkmark
\\ \hline
$|\tilde{\downarrow} \rangle = \frac{1}{\sqrt{6}}\left[|\!\downarrow,\downarrow,\uparrow \rangle-2 |\!\downarrow,\uparrow,\downarrow \rangle+  |\!\uparrow,\downarrow,\downarrow \rangle\right]$ 
& $-2J_{rr}$
& $\frac{1}{2}$
& $-\frac{1}{2}$
& \Checkmark
\\ \hline
$|a_{11} \rangle =  \frac{1}{\sqrt{2}}\left[|\!\uparrow,\uparrow,\downarrow \rangle-|\!\downarrow,\uparrow,\uparrow \rangle\right]$
& $0$
& $\frac{1}{2}$
& $+\frac{1}{2}$
& \XSolidBrush
\\ \hline
$|a_{12} \rangle = \frac{1}{\sqrt{2}}\left[|\!\downarrow,\downarrow,\uparrow \rangle-|\!\uparrow,\downarrow,\downarrow \rangle\right]$ 
& $0$
& $\frac{1}{2}$
& $-\frac{1}{2}$
& \XSolidBrush
\\ \hline
$|a_{13} \rangle = \frac{1}{\sqrt{3}}\left[|\!\uparrow,\uparrow,\downarrow \rangle +|\!\uparrow,\downarrow,\uparrow \rangle+|\!\downarrow,\uparrow,\uparrow \rangle \right]$ 
& $J_{rr}$
& $\frac{3}{2}$
& $+\frac{1}{2}$
& \XSolidBrush
\\ \hline
$|a_{14} \rangle = \frac{1}{\sqrt{3}}\left[|\!\downarrow,\downarrow,\uparrow \rangle+|\!\downarrow,\uparrow,\downarrow \rangle+|\!\uparrow,\downarrow,\downarrow \rangle \right]$ 
& $J_{rr}$
& $\frac{3}{2}$
& $-\frac{1}{2}$
& \XSolidBrush
\\ \hline
$|a_{15} \rangle = |\!\uparrow,\uparrow,\uparrow \rangle$ 
& $J_{rr}$
& $\frac{3}{2}$
& $+\frac{3}{2}$
& \XSolidBrush
\\ \hline
$|a_{16} \rangle = |\!\downarrow,\downarrow,\downarrow \rangle$ 
& $J_{rr}$
& $\frac{3}{2}$
& $-\frac{3}{2}$
& \XSolidBrush
\\ \hline
$|a_{17} \rangle = |\!\uparrow\downarrow,\uparrow,0 \rangle$ 
& $2\epsilon_{r}$
& $\frac{1}{2}$
& $+\frac{1}{2}$
& \XSolidBrush
\\ \hline
$|a_{18} \rangle = |\!\uparrow\downarrow,\downarrow,0 \rangle$ 
& $2\epsilon_{r}$
& $\frac{1}{2}$
& $-\frac{1}{2}$
& \XSolidBrush
\\ \hline
\end{tabular}
\end{footnotesize}
\caption{States with two fermions besides the impurity spin}
\label{tab:two-part}
\end{table}

\begin{table}[htb]
\begin{footnotesize}
\begin{tabular}{|l||c||c||c||c|}
\hline
 states & $\epsilon_{i}$ & $S$ & $S^{z}$  & used 
\\\hline
$|s^{+} \rangle = \frac{1}{\sqrt{2}}\left(|\!\downarrow,\uparrow,\uparrow\downarrow \rangle - |\!\uparrow,\downarrow,\uparrow\downarrow \rangle \right) $ 
& $-\frac{3J_{rr}}{2} - \epsilon_{r}$
& $0$
& $0$
& \Checkmark
\\\hline
$|t^{+}_1 \rangle =|\!\uparrow,\uparrow, \uparrow\downarrow  \rangle$   
& $\frac{J_{rr}}{2} - \epsilon_{r}$
& $1$
& $+1$
& \Checkmark
\\ \hline
$|t^{+}_2 \rangle = \frac{1}{\sqrt{2}}\left(|\!\downarrow,\uparrow,\uparrow\downarrow \rangle + |\!\uparrow,\downarrow,\uparrow\downarrow \rangle \right) $ 
& $\frac{J_{rr}}{2} - \epsilon_{r}$
& $1$
& $0$
& \Checkmark
\\ \hline
$|t^{+}_3 \rangle =|\!\downarrow,\downarrow,\uparrow\downarrow \rangle$ 
& $\frac{J_{rr}}{2} - \epsilon_{r}$
& $1$
& $-1$
& \Checkmark
\\ \hline
$|a_7 \rangle = \frac{1}{\sqrt{2}}\left(|\!\uparrow\downarrow,\downarrow,\uparrow \rangle - |\!\uparrow\downarrow,\uparrow,\downarrow \rangle \right)$ 
& $-\frac{3J_{rr}}{2} + \epsilon_{r}$
& $0$
& $0$
& \XSolidBrush
\\ \hline
$|a_8 \rangle =|\!\uparrow\downarrow,\uparrow,\uparrow \rangle$ 
& $\frac{J_{rr}}{2}+\epsilon_{r}$
& $1$
& $+1$
& \XSolidBrush
\\ \hline
$|a_9 \rangle =|\!\uparrow\downarrow,\downarrow,\downarrow \rangle$ 
& $\frac{J_{rr}}{2}+\epsilon_{r}$
& $1$
& $-1$
& \XSolidBrush
\\ \hline
$|a_{10} \rangle = \frac{1}{\sqrt{2}}\left(|\!\uparrow\downarrow,\downarrow,\uparrow \rangle + 
|\!\uparrow\downarrow,\uparrow,\downarrow \rangle \right)$ 
& $\frac{J_{rr}}{2} + \epsilon_{r}$
& $1$
& $0$
& \XSolidBrush
\\ \hline
\end{tabular}
\end{footnotesize}
\caption{States with three fermions besides the impurity spin}
\label{tab:three-part}
\end{table}

\begin{table}[htb]
\begin{tabular}{|l||c||c||c||c|}
\hline
 states & $\epsilon_{i}$ & $S$ & $S^{z}$  & used 
\\\hline
$|4,\uparrow \rangle = |\!\uparrow\downarrow,\uparrow,\uparrow\downarrow \rangle$ 
& $0$
& $\frac{1}{2}$
& $+\frac{1}{2}$
& \XSolidBrush
\\\hline
$|4,\downarrow \rangle = |\!\uparrow\downarrow,\downarrow,\uparrow\downarrow \rangle$ 
& $0$
& $\frac{1}{2}$
& $-\frac{1}{2}$
& \XSolidBrush
\\ \hline
\end{tabular}
\caption{States with four fermions besides the impurity spin}
\label{tab:four-part}
\end{table}

\end{document}